\documentclass[eqsecnum,showpacs,twocolumn]{revtex4}

\usepackage{epsf}
\usepackage{amsmath}
\usepackage{graphicx}
\usepackage{bm}

\def\beq{\begin{equation}}
\def\eeq{\end{equation}}
\def\bea{\begin{eqnarray}}
\def\eea{\end{eqnarray}}

\def\bq{{\bf q}}
\def\bk{{\bf k}}
\def\dst{\displaystyle}

\newcommand{\gbar}{\bar{g}}
\newcommand{\tSigma}{\tilde{\Sigma}}

\newcommand{\f}[2]{{\ensuremath{\mathchoice%
        {\displaystyle{\frac{#1}{#2}}}
        {\displaystyle{\frac{#1}{#2}}}
        {\frac{#1}{#2}}
        {\frac{#1}{#2}}
        }}}

\begin{document}

\title{Quantum critical behavior in itinerant electron systems -- Eliashberg theory and instability of a ferromagnetic quantum-critical point. }

\author{J\'er\^ome Rech$^{1,2}$, Catherine P\'epin$^1$ and Andrey V. Chubukov$^3$ }
\affiliation{$^1$SPhT, L'Orme des Merisiers, CEA-Saclay, 91191 Gif-sur-Yvette, France\\
$^2$Center for Materials Theory, Rutgers University, Piscataway, NJ 08855, USA\\
$^3$Department of Physics, University of Wisconsin-Madison,
1150 Univ. Ave., Madison, WI 53706-1390, USA}
\date{\today}

\begin{abstract}
We consider the problem  of fermions interacting with gapless
long-wavelength collective bosonic modes. The theory
 describes, among other cases,  
 a ferromagnetic quantum-critical point (QCP) and a QCP towards nematic
 ordering.  We construct a controllable expansion at the QCP in two steps: 
 we first create a new, non Fermi-liquid ``zero-order'' Eliashberg-type 
theory, and then demonstrate  that the residual interaction effects are small.
We prove that this approach is justified under two
conditions:  the interaction should be smaller than
the fermionic bandwidth, and  either the band mass $m_B$
should be much smaller than $m = p_F/v_F$, or the number of fermionic
flavors $N$ should be large.  
 For an $SU(2)$ symmetric ferromagnetic QCP, 
we find that the Eliashberg theory itself includes a  set of singular
renormalizations which 
 can be understood as a consequence of an effective long-range dynamic
 interaction between quasi-particles, generated  by the Landau damping term.
These singular renormalizations give
 rise to a negative non-analytic $q^{3/2}$ correction to the static spin
 susceptibility, and destroy a ferromagnetic QCP.
We  demonstrate that this effect can be understood
in the framework of the $\phi^4$ theory of quantum-criticality.
We also show that the non-analytic $q^{3/2}$ correction to the bosonic propagator
is specific to the $SU(2)$ symmetric case.
For systems with a scalar order parameter, the $q^{3/2}$
contributions from individual diagrams cancel out in the full expression of the susceptibility, and the QCP remains stable.

\end{abstract}
\maketitle

\section{Introduction}

Quantum-critical behavior in two-dimensional (2D) 
systems with  continuous symmetry 
continues to attract  substantial interest from the condensed-matter community.
Near criticality, bosonic collective modes in either the spin or the charge channel
 (depending on the problem) are soft, and mutual feedback effects between
 bosonic and fermionic degrees of freedom lead to a rather peculiar behavior
 of both the fermionic and bosonic propagators.  In this paper, we study in detail 
 this behavior for two-dimensional (2D) systems.

For systems with continuous symmetry, the dynamics of low-energy bosons is
dominated by Landau damping, and the collective mode propagator dressed by
 a particle-hole bubble behaves as
\begin{equation}
\chi (q, \Omega_m) = \frac{\chi_0}{\xi^{-2} + q^2 + \gamma \frac{|\Omega_m|}{q}}\label{c_1}.
\end{equation}
At $\xi = \infty$, the bosonic propagator becomes massless, signaling an
 instability towards a particular ordering.
The dynamical exponent $z$, which measures how the frequency scales with momentum at criticality ($\omega \sim q^z$),  is  $z=3$.

Physically, the complexity of the problem resides in the presence of gapless
 fermions at the QCP.
In ordinary QCP in localized electron systems, 
one deals with only one type of massless modes, namely  
 a bosonic mode associated with the fluctuations of the order parameter.
 In itinerant electron QCP, 
 massless  bosonic modes interact with conduction electrons, which are gapless 
 at the Fermi surface, and this interaction affects both, electrons and bosons.
One can still reduce the problem to interacting bosons by 
 formally integrating the fermions out of the partition
function. It was originally conjectured~\cite{hertz} that 
 this leads to a conventional $\phi^4$ field theory with the bare 
propagator given by (\ref{c_1}), i.e., to a $\phi^4$ theory  
in an effective dimension $d+z$. 
Since $z=3$,  the bosonic sector is above its upper critical
dimension  for $d>1$, and  
the critical exponents have mean-field values. 

However, this description turns out to be oversimplified by two reasons. 
First, it does not address the issue of what happens to the fermions at the QCP. 
It turns out that for fermions, the upper critical dimension is $d_c^{+} =3$, such that in $d=2$ the fermionic self-energy is singular and critically affects the 
behavior of low-energy fermions. 
Second,  the  $\phi^4$ theory for bosons is actually rather 
 peculiar as the prefactors for the $\phi^4$ and higher-other terms are
 determined by low-energy fermions, and are sensitive to 
 to the behavior of these fermions near the QCP.

These two  arguments imply that at the QCP in itinerant electron systems, 
 fermions and bosons should be considered self-consistently and 
 on equal footing. The key theoretical challenge in this context is
to develop a controlled computational scheme to describe the correct behavior 
 of electrons interacting with gapless bosonic collective degrees of freedom.

The problem of fermions interacting with bosons with the propagator
(\ref{c_1}) at $\xi = \infty$
was previously  analyzed in the context of 2D fermions interacting
 with a singular gauge field~\cite{aim,khvesh,nayak},
 and was also applied to a gauge theory
 of high $T_c$ superconductors~\cite{ioffe_larkin},
and compressible Quantum Hall Effect ~\cite{Llevel}. 
Later, this problem was studied in the context of 2D
 fermions near a ferromagnetic instability~\cite{pairing,cfhm,gork_dzero,chubukov}, and, very recently, in
 the context of fermions near an instability towards a nematic-type
ordering with angular momentum $l=2$~\cite{fko,fradkin_2,metzner,kim}.
That last transition was argued by some studies~\cite{fk}
to be relevant to the cuprates.
Similar problems have been studied in the context of finite momentum 
 spin \cite{spin-fermion} and charge \cite{dicastro} ordering transitions in 
itinerant fermionic systems. 

An analytic treatment of the problem
 was originally carried out by Altshuler, Ioffe and Millis (AIM)~\cite{aim} for the interaction with the gauge-field. They showed that 
 the fermionic self-energy scales as $\Sigma (\omega_m) = i (|\omega_m|)^{2/3} \omega_0^{1/3} \text{sign } (\omega_m)$ to second-order in perturbation,
 pointing to a breakdown of the Fermi liquid behavior. They estimated higher-order terms and argued that the $\omega^{2/3}$ form of the self-energy survives
 to all orders in perturbation. On the other hand,
 non-perturbative eikonal expansion~\cite{khvesh}
 and closely related approach based on 2D bosonization~\cite{kwon,neto}
 yielded a different behavior, in which the fermionic Green's function
decays exponentially with coordinate and time. This
 would be consistent with a divergent perturbative expansion for the self-energy.
 Altshuler and collaborators argued that this last result only survives in the artificial limit of a vanishing number of fermionic flavors $N \rightarrow 0$: at any finite $N$ (including the physical case $N=1$),
 the finite curvature of the fermionic dispersion prevents the perturbation series for the self-energy to become singular (see also Ref.\cite{metzner_2}).

The discussion on the interplay between perturbative calculations and  2D bosonization re-emerged recently in the context of the quantum critical point for a Pomeranchuk instability towards nematic ordering.
 Metzner and collaborators~\cite{metzner} and, very recently, Khveshchenko and
 one of us~\cite{chub_khvesh} argued that $\Sigma (\omega) \propto \omega^{2/3}$ is the correct result at criticality, while Lawler et al~\cite{fradkin_2} argued, based on 2D bosonization, that non-perturbative effects change this
 behavior. Adding to the controversy,  Kopietz~\cite{kopietz} argued
 that higher-order corrections to the self-energy hold in powers of $\omega^{2/3} |\log \omega|^n$, where the geometrical series of logarithms gives rise to an extra power of frequency, such that $\Sigma \propto \omega^{2/3 -a}$.

In this paper, we re-analyze the problem.  We consider  a ferromagnetic QCP
($z=3$ with spin SU(2) symmetry), and a QCP towards nematic ordering, towards
 Ising-type ferromagnetism, and a gauge-field problem -- these three last problems are mathematically equivalent and correspond to $z=3$ and U(1) symmetry  of the order parameter.
 We construct a controllable expansion at  the  QCP by creating  a new,
non Fermi-liquid  ``zero-order'' theory by solving the set of coupled equations for the fermionic and bosonic propagators,
while neglecting the vertex corrections as well as the momentum dependence of the fermionic self-energy.
This procedure is often called Eliashberg theory because of its resemblance to the Eliashberg theory for the electron-phonon interaction~\cite{eliash}.
 We analyze the residual interaction effects using the new zero-order propagator instead of free fermions. We confirm an earlier result of \cite{aim}
that the residual interaction does not change the functional behavior of the self-energy.
We go beyond previous works and 
perform a careful analysis of the structure of the infra-red 
divergences in the theory. We
analyze in detail the
 vertex corrections at various momenta and frequencies,
 the interplay between the Migdal approximation and Ward identities,
 the role of the curvature of the fermionic dispersion
 and the interplay between a direct perturbation theory for free fermions and
 an effective perturbation theory in which one expands around the Eliashberg
 solution. We also obtain the leading correction to the
 fermionic density of states (DOS).

A generic condition for the validity of the Eliashberg theory
 is that bosons should be slow modes compared to fermions (i.e., for a fixed frequency, the bosonic momentum should be larger than the fermionic one).
Then fermions, forced by the interaction to vibrate at frequencies near the bosonic pole, are far from  their own resonance and thus have a small spectral weight, giving rise to only a small correction to electron-boson vertex (Migdal theorem).
Typical bosonic momenta in Eq. (\ref{c_1}) scale as $\omega^{1/3}$ and are
 obviously slower than free fermions whose momenta scale as $\omega$.
The situation becomes less clear once the fermionic self-energy is included. We show that the correction to the static fermion-boson vertex is determined by 
  frequencies at which the fermionic self-energy is of order of a bare $\omega$, 
 and the static vertex correction is small, for a fermion-boson coupling 
 smaller than the fermionic bandwidth.  This coincides with the 
generic condition for the validity of the low-energy description since
 otherwise the physics is not restricted to the vicinity of the Fermi surface.

This generic condition, however, is not a sufficient one at the 
QCP of spatially isotropic systems  -- we show that there are corrections 
to the Eliashberg theory which  come from the
scattering process in which one component of the bosonic momentum is near the bosonic mass shell,
while the other is near the fermionic mass shell. We find that such
 corrections are dangerous if the expansion 
around the Eliashberg solution holds in powers of terms of order one. The same  expansion around 
free fermions yields terms which formally diverge
 as powers of $\omega^{-1/3}$ if one neglects the curvature of the Fermi surface~\cite{fradkin_2,chub_khvesh}.
 We show, in agreement with \cite{aim}, that in this situation, the
 way to construct a fully controllable perturbation expansion around the 
 Eliashberg theory at the QCP 
 is to either assume that the curvature of the Fermi surface 
  is large, or extend the theory to a large number of fermionic flavors, $N$.
 In this case,  the self-energy diagrams with vertex insertions are all small, and the theory is under control.  

We emphasize that this smallness does not imply that the theory is in the weak coupling limit -- the Eliashberg self-energy (given by the one-loop diagram)  
  doesn't ``feel'' the curvature and
 diverges as $\omega^{2/3}$ leading to non-Fermi liquid physics at the QCP. 
 Another exception is  the pairing  vertex,
which doesn't feel  the curvature as well, and is of order one.

In the second part of the paper we show that there exists a  third singular 
scattering process in  which both
 fermionic and bosonic momenta vibrate near the fermionic mass-shell.
This third process is qualitatively different from the first two
 scattering processes in which at least one component of the bosonic momentum is near the bosonic mass shell.

We show that this process (which by virtue of scattering near the fermionic mass shell is within the Eliashberg theory) gives rise to a
non-analytic momentum expansion of the static vertex.
We show that for a ferromagnetic $SU(2)$ symmetric QCP, 
this  non-analyticity eventually gives rise 
 to a non-analytic and negative correction to the static spin susceptibility.
This correction exceeds the $q^2$ term in (\ref{c_1}) and 
makes a ferromagnetic QCP unstable. 

The issue of 
 whether the QCP is internally stable has been the subject of 
 numerous discussions in the recent literature. 
This work was pioneered by Belitz, Kirkpatrick and Vojta~\cite{bkv}
 who found that in a generic 3D Fermi liquid far from the QCP, the static spin susceptibility
$\chi_s (q)$ has a negative non-analytic momentum dependence,
leading to a minimum of $\chi_s^{-1} (q)$ located at some incommensurate momentum,
rather than at $q=0$. The same result was later obtained for 2D systems~\cite{chub_maslov}.
If one were to formally extend the Fermi liquid results to the quantum critical region,
 one would obtain that the continuous QCP becomes unstable~\cite{belitz_rmp}.
 It was a priori unclear, however, whether this extension procedure is 
justified, since the Fermi liquid behavior does not seem to survive as one approaches the QCP.

To address this issue we
  explicitly compute the static spin susceptibility at criticality,
and show that it is  negative and non-analytic at the smallest momenta.
This implies that a ferromagnetic QCP is indeed unstable, as the Fermi liquid analysis suggests.
 We argue that the non-analyticity in $\chi_s (q)$ is associated with the Landau damping term,
which gives rise to an effective long-range dynamic interaction between quasiparticles both away from,
 and at the QCP. The singular fermionic
self-energy at criticality only modifies, in not a very essential way,
 the functional form of the non-analyticity compared to that in a Fermi liquid.

We also discuss the emergence of the non-analytic term in the
static spin susceptibility in the framework of the Hertz-Millis-Moriya (HMM)
$\phi^4$ theory of quantum-criticality~\cite{hertz}. As we said above, this
 theory assumes that there exists a
regular expansion of the effective action in powers of the order
parameter field $\phi$. We show that this is actually not the
case, and the pre-factor
 of the  $\phi^4$ term is non-analytic and depends on the ratio between typical momenta and frequencies.
We show that this non-analyticity feeds back as a non-analytic $|q|^{3/2}$ 
correction of the quadratic term in $\phi$.
We study how the  non-analyticity in $\chi_s (q)$ affects
 the fermionic self-energy and show that it gives rise to 
 self-energy terms larger that $\omega^{2/3}$ (beginning at the three-loop order).
The series of such terms eventually leads to a breakdown of the Eliashberg theory
for the fermionic self-energy at the energy scale related to the typical
momentum scale at which $\chi_s (q)$ becomes negative.

We show that the non-analytic corrections to the bosonic
propagator and the divergent corrections to the fermionic
self-energy are specific to the SU(2) spin symmetric case. For a
nematic instability, as well as for a ferromagnetic instability in
systems with Ising symmetry the dangerous terms in the bosonic
propagator and the fermionic self-energy cancel out, and the
QCP is stable.

The paper is organized as follows. In Sec. II we discuss the model.
In Sec. III we present a quick
 analysis of the self-energies, justifying all at once, the Eliashberg-like treatment, and the need to include the curvature of the fermionic dispersion. In Sec IV, we discuss the Eliashberg theory near quantum criticality.
In Sec. V, we analyze in detail the conditions one has to impose in order for the Eliashberg theory to be valid. This includes the computation of all vertex corrections, as well as the momentum-dependent self-energy at the two-loop level.
 We also compare the results obtained by strict perturbative expansion using free fermions, and the Eliashberg-type calculations.

In Sec. VI  we address the issue of the stability of a
ferromagnetic QCP. 
We  revisit the scaling arguments that a 
 ferromagnetic QCP must be stable in dimension $D>1$ and 
show that the prefactor for the $\phi^4$ term is actually a nonanalytic function of the ratio of frequency and momentum. We 
 show that  this non-analyticity feeds back as a 
 non-analytic correction to the static spin susceptibility. 
We  explicitly compute  the  momentum-dependent term in 
 $\chi_s (q)$ at the two-loop order, 
 both in the Fermi liquid regime away from a ferromagnetic QCP 
 and at criticality. 
 In both cases, we find that the dominant term at small $q$ is 
 negative and non-analytic.
We also show  that the instability of a
ferromagnetic QCP can be also detected by computing the fermionic
self-energy which at the three-loop order 
acquires extra singular terms because of the singularity  
 in the static susceptibility. 
In Sec. VII, we evaluate the charge susceptibility at two-loop and
three-loop
 orders and show  that it remains analytic -- all non-analytic contributions from individual diagrams cancel out.
 Finally, in Sec. VIII, we present our conclusions and discuss the consequences of the instability of a continuous QCP towards ferromagnetic ordering.
 Technical details are presented in the Appendices A-F.

A short version addressing part of the results of section VI has been presented in ~\cite{our_prl}.

\section{The model}

The model we consider describes low-energy fermions interacting with Landau-overdamped
collective bosonic excitations which are either gapless by symmetry reasons
(as it is the case for the interaction with a gauge field), or become gapless at the quantum critical point
(for the nematic and the ferromagnetic problems).

The underlying lattice models may be quite different for these three cases, but
the low-energy models are very similar, the only difference being that in the case
of the ferromagnetic QCP the gapless bosonic excitations are in the spin channel, whereas in the nematic case and the gauge field problem
 they are in the charge channel.

The general strategy to derive the low-energy model is 
to start with a model with fermion-fermion interaction, 
 assume that there is only one low-energy collective degree of freedom 
 near the QCP, decouple the four-fermion interaction term using the
critical bosonic field as an Hubbard-Stratonovich field,  and 
 integrate out
of the partition function all high-energy degrees of freedom, with
energies between the fermionic bandwidth $W$ and some cutoff
$\Lambda$~\cite{spin-fermion,prr}.

If this procedure was performed completely we would obtain a full
Renormalization Group treatment of the problem. Unfortunately, there is no controllable
scheme to perform this procedure. It is widely believed, though
 that although the propagators of the remaining low-energy
modes  possess some memory of the physics at high energies, the
 integration of high-energy fermions  does
 not give rise to anomalous dimensions for the bare fermionic 
and bosonic propagators in the low-energy model.
In practical terms, this  assumption 
implies  that the bare propagator of the relevant collective mode 
 is an  analytic function of
momentum and frequency, and the fermionic propagator has the  Fermi liquid 
form:
\begin{equation}
G (k, \omega) = \frac{Z_0}{i\omega - \epsilon_k} ,
\label{c_1_1}
\end{equation}
where $Z_0 <1$ is a constant, and $\epsilon_k$ is the renormalized band dispersion. Near the Fermi surface,
\begin{equation}
\epsilon_k = v_F k_{\perp} + \frac{k^2_\parallel}{2m_B}.
\label{c_1_2}
\end{equation}
Here ${\bf k}$ is the momentum deviation from ${\bf k_F}$, the parallel and perpendicular components are with respect to the direction along the Fermi surface at ${\bf k_F}$, $m_B$ is the band mass, the Fermi velocity $v_F = k_F/m$, and for a circular Fermi surface one has $m = m_B$.

One can then re-cast the original model of  fermion-fermion
 interaction into an effective low-energy
fermion-boson model. Consider for definiteness
 that the system is close to a ferromagnetic QCP. Then the low-energy degrees
 of freedom are fermions (with the propagator given by (\ref{c_1_1})) and long-wavelength collective spin excitations whose propagator (the spin susceptibility) is {\it analytic} near $q=0$ and $\Omega =0$:
\begin{equation} \label{chibare}
\chi_{s,0} (q, \Omega) = \frac{\chi_0}{\xi^{-2} + q^2 + A \Omega^2 + O(q^4, \Omega^4)}.
\end{equation}
Here $A$ is a constant, and $\xi$ is the correlation length, which becomes infinite at the QCP. We prove in the next section that the $\Omega^2$ term does not play any role in our analysis, and we therefore neglect it from now and approximate the above bare propagator by the static one $\chi_{s,0} (q)$. The model can then be described by the phenomenological spin-fermion Hamiltonian:
\begin{eqnarray} \label{sfermion}
H_{sf}  & =  & \sum_{k  , \alpha  } \epsilon_k  c^\dagger_{k, \alpha}
c_{k, \alpha}  + \sum_q  \chi^{-1}_{s,0} (q) {\bf  S}_q {\bf  S}_{-q }
\nonumber  \\ & &  + g  \sum_{k, q  , \alpha  , \beta  } c^\dagger_{k,
\alpha }  {\bf \sigma }_{\alpha \beta }  c_{k + q, \beta  } \cdot {\bf
S}_q .
\end{eqnarray}
Here ${\bf S}_q$ with $q < \Lambda/v_F$ are vector bosonic variables, and $g$ is the effective fermion-boson interaction. For convenience, we incorporated the fermionic residue $Z_0$ into $g$.

To illustrate how this effective Hamiltonian can, in principle, be
derived from the microscopic model of interacting conduction
electrons, we consider a model in which the electrons interact
with a short range four-fermion interaction $U(q)$ and assume that
only the forward scattering is relevant ($U(0) = U$): \bea
\label{start}
 H & = & \sum_{k , \alpha } \epsilon_k c^\dagger_{k, \alpha} c_{k, \alpha} \nonumber \\
  & + & \frac{1}{2} \sum_q U
 \sum_{k,k^\prime, \alpha , \beta }  c^\dagger_{k, \alpha} c_{k + q,
\alpha }  c^\dagger_{k^\prime \beta } c_{k^\prime - q , \beta }  ,
\eea In this situation, the interaction is renormalized
independently in the spin and in the charge channels~\cite{cmgg}.
Using the identity for the Pauli matrices ${\bf \sigma}_{\alpha
\beta}\cdot{\bf \sigma}_{\gamma \delta}= -\delta_{\alpha \beta}
\delta_{\gamma \delta}+ 2\delta_{\alpha \delta}\delta_{\beta
\gamma}$, one can demonstrate~\cite{cmgg} that in each of the
channels, the RPA summation is exact, and the fully renormalized
four-fermion interaction $U^{full}_{\alpha \beta,\gamma \varepsilon}
(q)$ is given by:
\begin{equation}
U^{full}_{\alpha \beta ,\gamma \varepsilon} (q) = U \left[\delta _{\alpha \gamma }\delta
_{\beta \varepsilon } \left( \frac{1}{2}+\mathcal{G}_{\rho }\right) +
\sigma
_{\alpha \gamma }^{a}\sigma _{\beta \varepsilon }^{a} \left( \frac{1}{2}+%
\mathcal{G}_{\sigma }\right)\right],  \label{b4}
\end{equation}
where $\sigma _{\alpha \beta }^{a}$ are Pauli matrices ($a=x,y,z)$, and
\begin{equation}
\mathcal{G}_{\rho }\equiv \frac{1}{2}\frac{1}{1-U\Pi \left(
q\right) }; \qquad \mathcal{G}_{\sigma }\equiv-\frac{1}{2}\frac{1}{1+U\Pi \left( q\right) },
\label{f8}
\end{equation}
with $\Pi (q) = -\frac{m}{2\pi} (1 - a^2 (q/k_F)^2)$, $a >0$.

For positive values of $U$ satisfying $m U/2\pi \approx 1$,
the interaction in the spin channel is much larger
than the one in the charge channel. Neglecting then the
interaction in the charge channel, we can simplify the Hamiltonian (\ref{start}):
\beq \label{start2}
\begin{array}{c}
{\displaystyle H = \sum_{k , \alpha
} \epsilon_k c^\dagger_{k, \alpha} c_{k, \alpha}} +
\frac{1}{2} \sum_q U_{eff} (q) \\ \\
{\displaystyle \times \sum_{k, k^\prime,\alpha , \beta, \gamma, \delta }
c^\dagger_{k, \alpha } {\bf \sigma }_{\alpha \beta } c_{k + q, \beta }
\cdot  c^\dagger_{k^\prime \gamma } {\bf \sigma }_{\gamma \delta }
c_{k^\prime - q , \delta }}   .
\end{array}
\eeq
where $U_{eff} (q) =  (1/2) U^2 \Pi (q)/(1 + U \Pi (q))$.
Performing a Hubbard-Stratonovich decomposition in the three fields ${\bf S}_q$,
one recasts (\ref{start2}) into  Eq. (\ref{sfermion}) with:
\beq
\left\{
\begin{array}{lcl}
g & = & U\frac{a}{2} \\
\chi_0 &= & 2 \frac{k^2_F}{U a^2} \\
{\bar g} &=& g^2 \chi_0 = (U/2) k^2_F \\
\xi^{-2} &=& \frac{k^2_F}{a^2} \left( \frac{2 \pi}{m U}-1\right)
\end{array}
\right.
\eeq
The QCP is reached when $m U /2\pi =1$, i.e., $\xi^{-2} =0$.
This coincides with the Stoner criterion for a ferromagnetic instability~\cite{stoner}.

We emphasize that the bosonic propagator in  Eq. (\ref{chibare})
does not contain the Landau damping term.
This is because we only integrated out the high-energy fermions,
whereas the Landau damping of a collective
mode of energy $\Omega$ comes from fermions of energy $\omega < \Omega$,
and can only be generated within the low-energy theory.
The dynamics of both the bosonic fields ${\bf S}_q$ and
the fermionic $c$ and $c^\dagger$ is determined
self-consistently by treating both fluctuations on equal footing.

To put under control the computations carried out later in the paper, it is necessary to extend the model by introducing $N$ identical fermion species, while keeping the $SU(2)$ spin symmetry. The Hamiltonian (\ref{sfermion}) can then be rewritten as:
\bea \label{start5}
H_{sf} & = & H_f + H_b + H_{int} \quad \mbox{where} \nonumber \\
H_f &=& \sum_{k,j,\alpha} \epsilon_k c^\dagger_{k, j, \alpha} c_{k,j, \alpha} \nonumber \\
 H_b &=& \sum_q  \chi^{-1}_{s,0} (q)  {\bf  S}_q \cdot {\bf  S}_{-q}  \nonumber \\
 H_{int} &=& g \sum_{k, q ,j,\alpha,\beta} c^\dagger_{k,j,\alpha} {\bf \sigma}_{\alpha \beta} c_{k + q, j,\beta}\cdot{\bf  S}_q ,
\eea
where the index $j=1 \ldots N$ labels the fermionic species.

We use the spin-fermion Hamiltonian of Eq. (\ref{start5}) as the
starting point of our analysis. In the case of a QCP in the charge
channel, or a ferromagnetic instability with Ising symmetry, the
bosonic vector field ${\bf S}$ becomes a scalar field designated
as $\phi$. The interacting term is also modified, the Pauli
matrices being replaced by $\delta_{\alpha \beta}$ for the
interaction with charge fluctuations, and by $\sigma^Z_{\alpha
\beta}$ in the Ising case. The corresponding interaction
Hamiltonians are: \beq \left\{
\begin{array}{lcl}
H_{int}^{Charge} & = & g \sum_{k,q,j,\alpha,\beta} c^\dagger_{k,j,\alpha} \delta_{\alpha \beta} c_{k + q, j,\beta} \phi_q \\
H_{int}^{Ising} & = & g \sum_{k,q,j,\alpha,\beta} c^\dagger_{k,j,\alpha} \sigma^Z_{\alpha \beta} c_{k + q, j,\beta} \phi_q
\end{array}
\right.
\eeq

\begin{figure}[tbp]
\centerline{\includegraphics[width=3.4in]{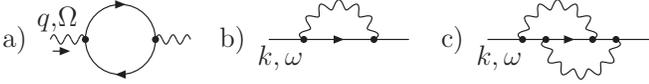}}
\caption{a) Polarization bubble b) One-loop fermionic self-energy c) Two-loop fermionic self-energy.}
\label{self}
\end{figure}

\section{Direct perturbation theory}
\label{direct}

In this section we compute the fermionic and
bosonic self-energies for the model presented in
Eq. (\ref{sfermion}) using a  perturbation expansion around non-interacting fermions.
Our goal here is three-fold:
to relate the Landau damping coefficient to the fermion-boson coupling constant $g$,
to distinguish between
$\Sigma (\omega)$ and $\Sigma (k)$ and to
demonstrate the importance of the curvature of the Fermi surface.

We evaluate the self-energy in this section
up to two-loop order. We  verified that to this order,
there is no qualitative difference between the quantum critical
point in the spin or in the charge channel. We then restrict our presentation to the
 spin-fermion model near a ferromagnetic QCP.

\subsection{Bosonic self-energy: the Landau damping term}

The full bosonic propagator depends on the self-energy $\Pi (q, \Omega)$ according to:
\begin{equation} \label{chi_1}
\chi_{s} (q, \Omega) = \frac{\chi_0}{\xi^{-2} + q^2 + \Pi (q, \Omega)}.
\end{equation}

At the lowest order in the spin-fermion interaction, the bosonic
self-energy is given by the first diagram in Fig. \ref{self}, and reads:
\begin{equation}
\Pi (\bq,\Omega) = 2N \bar{g} \int \f{d^{2}k~ d\omega}{(2\pi)^3}~
G(\bk,\omega)~ G(\bk+\bq,\omega+\Omega).
\end{equation}

The curvature of the fermionic dispersion does
not affect much the result of this computation
as it only leads to small corrections in $q/{m_B v_F}$.
Neglecting the quadratic term in the fermionic propagators,
we introduce the angle
$\theta$ defined as $\epsilon_{\bf k+q}=\epsilon_{ \bf k} + v_F q \cos \theta$
and perform the integration over $\epsilon_{ \bf k}$, which gives us:
\bea
\Pi (\bq,\Omega)& = &  i \f{N\bar{g} m}{2 \pi^2} \int_{- \infty}^{+
\infty} d\omega~\left( \theta(\omega+\Omega) -
\theta(\omega) \right) \nonumber \\
& & \qquad \times \int_0^{2 \pi} d\theta~ \f{1}{i\Omega -v_F q \cos \theta} \nonumber \\
& =&  \f{Nm \bar{g}}{\pi} \f{|\Omega|}{\sqrt{(v_F q)^2+\Omega^2}}.
\eea

At the QCP, the bosonic mass-shell corresponds to the region of
momentum and frequency space for which the terms in the inverse
propagator are of the same order, i.e. near a mass shell
$q$ and $\Omega$ satisfies $\Pi(q,\Omega) \sim q^2$. It follows that,
at the QCP, near the bosonic mass shell, $v_F q/\Omega  \sim v_F (m \gbar v^2_F/\Omega^2)^{1/3} \gg1$ at small enough frequency, so that $v_F q$ is the largest term in the denominator of $\Pi(q,\Omega)$. The expression of the bosonic self-energy then reduces to:
\beq \label{ex_6}
\Pi (\bq,\Omega) =  \gamma~ \f{|\Omega |}{q}  ,
\eeq
where $\gamma = \f{Nm \bar{g}}{\pi v_F}$.

This expression describes the Landau damping with a pre-factor
depending on the fermion-boson coupling constant. This term
 is larger than a regular $O(\Omega^2)$ term, and fully
 determines the dynamics of the collective bosonic mode.

\subsection{One-loop fermionic self-energy}

We now turn to the fermionic self-energy, right at the QCP,
where $\xi^{-1}=0$. To the lowest order in the interaction,
the fermionic self-energy contains one bosonic line,
as represented in Fig.\ref{self}b, and its analytic form writes:
\begin{equation}
\Sigma^{\rm free}_1 ({\bf k},\omega)  =  3 i g^2 \int \f{d^2q~d\Omega}{(2 \pi)^3}
 G_0(\bk+\bq,\omega+\Omega) \chi(\bq,\Omega) .
\label{ex_1}
\end{equation}
The superscript ``free'' implies that we use the free fermionic $G_0 (k, \omega)$ in the integral for the self-energy.

In a direct perturbation theory in $g$,
one would have to use the bare form of the bosonic propagator,
Eq. (\ref{chibare}), which leads to $\Sigma (\omega) \propto \log \omega$.
However, this result is useless, as we already know that the Landau
damping term completely overshadows a regular frequency dependence
in (\ref{chibare}). It makes more sense then to estimate
 the perturbative self-energy using the full bosonic propagator
 Eq. (\ref{chi_1}). This is {\it not} a fully self-consistent procedure,
 but we use it here to estimate the functional forms of the self-energy
 at various orders in perturbation around free fermions.

It is instructive to distinguish
 between $\Sigma ({\bf k_F},\omega)=\Sigma(\omega)$ and $\Sigma({\bf k},0)=\Sigma({\bf k})$.

Substituting the renormalized bosonic propagator with the Landau damping term into
(\ref{ex_1}), the frequency-dependent self-energy reads:
\begin{eqnarray}
\Sigma^{\rm free}_1 (\omega)  &=&  \f{3
 i \gbar}{(2 \pi)^3} \int d\Omega~q dq~d\theta ~
 \f{1}{q^2+\gamma \frac{| \Omega |}{q}} \nonumber \\ & &
 \times \f{1}{i(\omega+\Omega)-v_F q \cos \theta} .
\label{ex_2}
\end{eqnarray}
Here $\theta$ is the angle between ${\bf k_F}$ and $\bq$, and we linearized the fermionic dispersion. Evaluating the integral over the angle, and using that the typical internal bosonic momentum $q \sim \Omega^{1/3}$ is much larger than  $\Omega \sim \omega$, we obtain
\bea \label{sigres}
\Sigma^{\rm free}_1 (\omega) & = &  \f{3 \gbar}{2 \pi^2 v_F} \int_0^{\omega} d\Omega~\int
\f{d q q }{q^3 + \gamma| \Omega |} \nonumber \\
 & = & \omega^{1/3}_0 \omega^{2/3} ,
\eea
where
\begin{equation}
\omega_{0}  =  \frac{3 \sqrt{3}  \gbar^3}{8 \pi^3 v^3_F\gamma} = \frac{3\sqrt{3} \gbar^2}{8 \pi^2 N m v^2_F} .
\label{omega0}
\end{equation}
This result has been first obtained in Ref. \cite{ioffe_larkin}. It shows that in $D=2$,  the interaction between bare fermions and critical bosons leads to a breakdown of the Fermi-liquid behavior: at low energies, the $\omega^{2/3}$ term in (\ref{sigres}) is larger than the bare $\omega$ in the fermionic propagator. This obviously makes one wonder if higher order insertions lead to even more singular contributions.

We next  compute the one-loop  self-energy $\Sigma({\bf k})$, given by:
\begin{eqnarray}
\Sigma^{\rm free}_1 ({\bf k}) & = & \f{3 i \gbar}{(2 \pi)^3} \int d\Omega~q dq~d\theta ~
 \f{1}{q^2+\gamma \frac{| \Omega |}{q}} \nonumber \\
 & & \qquad \times \f{1}{i\Omega- \epsilon_k - v_F q \cos \theta} \nonumber \\
& = & \f{3 \gbar}{(2 \pi)^2} \epsilon_{\bf k} \int \f{d\Omega~dq}{q^2+\gamma \frac{| \Omega |}{q}} \f{q |\Omega|}{(\Omega^2 +  (v_F q)^2)^{3/2}} .  \qquad
\label{ex_3}
\end{eqnarray}
One can make sure that the  integral is infrared-convergent, i.e. $\Sigma^{\rm free} (\epsilon_{\bf k}) \propto \epsilon_{\bf k}$, with an interaction dependent prefactor, which also depends on the upper cutoff of the theory, $\Lambda$.
This suggests that the momentum dependent part of the fermionic self-energy is regular at the QCP and only leads to a finite mass renormalization.

\subsection{Two-loop fermionic self-energy}

We next calculate the contribution to the fermionic self-energy $\Sigma (\omega)$ from diagrams at the two-loop level. For illustrative purposes, we consider the diagram presented in Fig. \ref{self}c, which writes:
\beq \label{2lse}
\begin{array}{l}
{\displaystyle \Sigma^{\rm free}_2 (\omega) \sim \gbar^2 \int d\omega_1 d^2 q_1 \int d\omega_2 d^2 q_2 G({\bf k_F + q_1},\omega+\omega_1) } \\ \\
{\displaystyle \times~ G({\bf k_F+q_1+q_2},\omega+\omega_1+\omega_2)~ G({\bf k_F+q_2},\omega+\omega_2)} \\ \\
{\displaystyle \times ~\chi_s({\bf q_1},\omega_1)~ \chi_s({\bf q_2},\omega_2)   }    ,
\end{array}
\eeq
where we use the full bosonic propagator, the free fermionic one, and we restrict ourselves to the frequency dependence.

We first compute this integral expanding the dispersion of the internal fermions to linear order, since the quadratic term was not significant in the computation of the one-loop bosonic and fermionic self-energy. Choosing the $x$ axis along the external ${\bf k} = {\bf k_F}$ and  integrating over $q_1^x$ and $q_2^x$, we are left with:
\bea
\Sigma^{\rm free}_2 (\omega) & \sim & \frac{\gbar^2}{v_F^2 \omega} \int_0^{\omega} \frac{d\omega_1 dq_{1y}}{q_{1y}^2+\frac{\gamma|\omega_1|}{q_{1y}}} \int_{\omega - \omega_1}^\omega \frac{d\omega_2 d q_{2y}}{q_{2y}^2+\frac{\gamma|\omega_2|}{q_{2y}}} \nonumber \\
& \sim & \omega_0^{2/3} \omega^{1/3}  .
\eea
where $\omega_0$ is defined in (\ref{omega0}). At low-energy, this two-loop self-energy diverges faster than the one-loop self-energy obtained in (\ref{sigres}). Estimating higher-order diagrams, we find that they form a series in powers of $(\omega_0/\omega)^{1/3}$, such that the perturbative expansion around free fermions breaks up at $\omega \sim \omega_0$. This result is in line with the one obtained by Lawler et al.~\cite{fradkin_2} using a two-dimensional bosonization scheme. The scale $\omega_0$ is related by $\omega_0 = v_F /x_0$ to the spatial scale at which the equal time fermionic Green's function $G(x)$, obtained from bosonization, begins decaying exponentially ($G(x) \propto e^{-(x/x_0)^{1/3}}$).

However, the divergence of the perturbation theory can be cured once the curvature of the fermionic dispersion is included, as we now show. We re-evaluate the two-loop self-energy (\ref{2lse}), using now the full fermionic dispersion, Eq. (\ref{c_1_2}). After integrating over the momentum component $q_1^x$ and $q_2^x$, one has:
\bea
\Sigma^{\rm free}_2 (\omega) & \sim & \frac{\gbar^2}{v_F^2} \int_0^{\omega} \frac{d\omega_1 dq_{1y}}{q_{1y}^2+\frac{\gamma|\omega_1|}{q_{1y}}} \int_{\omega - \omega_1}^\omega \frac{d\omega_2 d q_{2y}}{q_{2y}^2+\frac{\gamma|\omega_2|}{q_{2y}}} \nonumber \\
& & \qquad \qquad \times \frac{1}{i\omega-\frac{q_{1y}q_{2y}}{m}}  \nonumber \\
 & \sim & \frac{m^2 \gbar^2}{\gamma^2 v_F^2}~ \omega \log^2 \omega .   \label{ex_10}
\eea

This result agrees with~\cite{aim}. We see that, when the curvature of the fermionic dispersion is included, that the two-loop self-energy turns out to be small compared to its one-loop counterpart, at low energy. In a separate study~\cite{chub_khvesh}, one of us (A.C.) and D. Khveshchenko reconsidered the bosonization procedure in the presence of the curvature and obtained the same results as in (\ref{ex_10}).

\subsection{Summary}

As a conclusion, this first approach suggests that both
the fermionic and the bosonic self-energies are important at the QCP.
The bosonic self-energy sets the dynamics of the bosons,
while the fermionic self-energy is non-analytic and parametrically
larger than the bare $\omega$ term at low energy, which implies a breakdown
of the Fermi-liquid behavior at criticality.

We also found that only the frequency-dependent part of the self-energy matters,
the momentum-dependent one only leads to a regular renormalization of the effective mass.
Finally, we found that the curvature of the Fermi surface
plays an important role in regularizing the perturbation expansion.

The full account of these effects cannot be obtained from this simple
analysis and one has to develop a controllable way to
treat the bosonic and fermionic self-energies on equal footing.
Since we found that only the frequency-dependent $\Sigma (\omega)$
is relevant, a way to proceed is to verify whether an Eliashberg-like theory,
similar to the one developed in the context of phonon
superconductivity\cite{eliash}, may be such a controllable approximation.

\section{Eliashberg theory}

The Eliashberg procedure allows to compute the fermionic
self-energy $\Sigma (\omega)$ and the bosonic polarization $\Pi
(q,\Omega)$, by solving the self-consistent set of coupled Dyson's
equations, neglecting all
 contributions coming from the vertex corrections and
 the momentum-dependent fermionic self-energy.

Specifically the Eliashberg theory follows three steps:
\begin{itemize}

\item neglect both the vertex corrections and the momentum dependent part of the fermionic self-energy, i.e., approximate:
\begin{eqnarray}
&&\Sigma  ({\bf  k},\omega_n)  =  \Sigma(\omega_n) \nonumber \\
&& g_{\text{Tot}}  =  g + \Delta g =  g
\end{eqnarray}

\item construct the set of self-consistent Dyson's equations:
\begin{eqnarray}
&&  G^{-1} (k,\omega_n)  =   i\omega_n -v_F
(k-k_F)  + i\Sigma  (\omega_n)  \nonumber \\
&&  \chi_s  (q, \Omega_m)  \f{\chi_0}{\xi^{-2}+q^2+\Pi (q, \Omega_m)} ,
\label{eq:propag}
\end{eqnarray}
with the following fermionic and bosonic self-energies:
\begin{eqnarray} \label{selfcath}
i\Sigma (\omega_n) & = &
\parbox[c]{2cm}{\includegraphics[width=2cm]{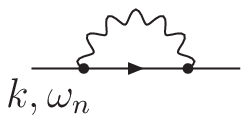}}  \nonumber  \\
\chi^{-1}_0 \Pi (q, \Omega_m) & = &
\parbox[c]{2cm}{\includegraphics[width=2cm]{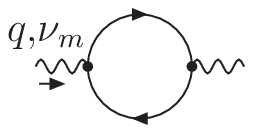}}
\label{eq:selfs}
\end{eqnarray}
The fermionic Green's functions in (\ref{eq:selfs}) are full (they are represented diagrammatically by a straight  line) and  $\chi_s (q,\Omega_m)$ is the full bosonic propagator (represented by a wavy line).

\item check a posteriori that the neglected terms $\Delta g$ and
$\Sigma (\bk)$, are all parametrically small.

\end{itemize}

The evaluation of the momentum integral for the fermionic self-energy 
 in the Eliashberg theory requires care. Since fermions are faster than bosons,
 the leading contribution to $\Sigma (\omega)$ is obtained 
if one integrates over the momentum component transverse to the Fermi surface only
 in the fermionic propagator and sets this component to zero in the bosonic
propagator (this implies that the momentum integral is factorized). 
One can show that the corrections that arise from keeping the transverse component
 of momentum in the bosonic propagator are small to the same parameter 
 as  $\Delta g/g$ and should therefore be neglected, as keeping them 
 would be beyond the accuracy of the theory. 
 The factorization of the
 momentum integration is what distinguishes the Eliashberg theory
 from the FLEX approximation. In the FLEX approximation,
  one also neglects vertex corrections, but 
does not factorize the momentum integral in (\ref{eq:selfs}).

The evaluation of the bosonic  and fermionic 
self-energies within the Eliashberg theory
is presented in Appendices \ref{app:boson} and \ref{app:fermion}.
We list here the main results.

At large but finite correlation
length $\xi$ and for a bosonic momentum and
frequency satisfying $v_F q \gg \Sigma (\Omega)$ we obtain:
\beq \label{sols}
\left \{
\begin{array}{l}
\Pi (q, \Omega) = \gamma \f{| \Omega  | }{q} \ , \mbox{and} \\ \\ \Sigma
(\omega)  =  \lambda \omega  F \left( \gamma \omega \xi^3 \right)  ,
\end{array} \right.
\eeq
where $F(x \ll 1)  = 1 +O(x)$, and $F( x \gg 1) = \frac{2}{\sqrt{3}} x^{-1/3}$.
The parameter $\gamma$ is the same as for free fermions,
\begin{equation}
\gamma  = \f{N m \gbar }{ \pi v_F },
\label{param_1}
\end{equation}
and $\lambda$ is the dimensionless coupling
\beq \label{param}
\begin{array}{cc}
{\displaystyle \lambda= \frac{3 \gbar }{4 \pi v_F \xi^{-1} } } & \qquad {\bar  g } =g^2 \chi_0
\end{array}
\eeq At finite $\xi^{-1}$ and vanishing $\omega$, the self-energy
has a Fermi liquid form: \beq  \label{self_fl} \Sigma(\omega ) = \lambda \omega . \eeq The Fermi liquid theory is stable, and the
low-energy quasi-particles have a finite effective mass: \beq
\label{m*} m^* =  m (1  +\lambda)  . \eeq The effective mass
diverges proportionally to $\xi$ in the vicinity of the QCP.

At $\omega \gg (\gamma \xi^3)^{-1}$, however, the system
is in the quantum-critical regime. Here the Fermi liquid theory
breaks down in the sense that the quasi-particles damping becomes
comparable to its energy. We have:
\bea \label{defs}
&&\Sigma( \omega ) = {\omega }_0^{1/3}  | \omega |^{2/3} {\rm sign}( \omega )  ,
\eea
where $\omega_0 = 3\sqrt{3} {\gbar^2} /(8 \pi^2 N m v^2_F)$ is the same as in (\ref{omega0}).

At the QCP, $\xi^{-1} =0$, the region of Fermi-liquid behavior collapses,
and the $\omega^{2/3}$ dependence of the self-energy extends down
to the lowest frequencies. The expression for $\Sigma (\omega)$
 is valid for all frequencies below the cutoff $\Lambda$.
However, only frequencies $\omega \leq \omega_0$ are actually relevant as
 at higher frequencies the system behaves as a nearly ideal Fermi gas.
 Note that the curvature of the fermionic dispersion is unimportant
 here and only accounts for small corrections containing higher powers of frequencies.

Comparing (\ref{sols}) and (\ref{defs}) with (\ref{ex_6}) and (\ref{sigres}),
we see that the self-energies in the Eliashberg theory coincide
with the one-loop perturbative results around free fermions.
This arises from the fact that the momentum integration is factorized in
the Eliashberg theory, and the full fermionic propagator appears
in both self-energies only via the fermionic density of states (DOS):
\beq
N(\omega) = \frac{i}{\pi} \int d \epsilon_k  \frac{1}{\omega + \Sigma (\omega) - \epsilon_k} .
\eeq
This DOS reduces to $N(\omega) = sign (\omega)$, independently on the self-energy:
it remains the same as for free fermions. Note that Eq. (\ref{sols})
 for the bosonic self-energy is only valid as long as the interplay
 between the external momentum and frequency is such that $v_F q \gg \Sigma (\Omega)$.
 In the opposite limit, the vertex corrections cannot be neglected as
 we argue in the next section.

For further analysis of the corrections to the Eliashberg theory,
it is then convenient to rescale $N$ out of the formulas for
$\Sigma(\omega)$ and for $\Pi(q,\Omega)$.
 This can be done by
rescaling $m$ and $k_F$ leaving $v_F$ intact: \beq m \rightarrow
m/N \qquad k_F \rightarrow k_F/N  . \label{rescaling} \eeq  We
emphasize that the fermionic self-energy $\Sigma \propto
\omega^{2/3}$ and the Landau damping term in the bosonic
propagator do not contain $N$ after rescaling, and therefore must
be included into the new ``zero-order'' theory about which we then
expand using $1/N$ as a small parameter. This zero-order theory
includes: \beq \left\{
\begin{array}{lcl}
{\dst \chi (q, \Omega)} & = & \dst{\frac{\chi_0}{\xi^{-2} + q^2 + \gamma |\Omega|/q}} \\
{\dst G(k, \omega)} & = & \dst{\frac{1}{i (\omega + \Sigma (\omega)) - \epsilon_k}} .
\end{array}
\right. \label{apr25_1} \eeq At the QCP, this theory  becomes
completely parameter-free once we measure the frequency in units
of $\omega_0$ and rescale the bosonic momentum.

We show in the next two sessions that
 the corrections to the Eliashberg theory are small in
 powers of two parameters, one of which vanishes in the limit $N \to \infty$,
 the other being parametrically small.
 Accordingly, the Eliashberg theory becomes exact at $N = \infty$,
 however the last limit has to be taken at a finite curvature.

For the $SU(2)$ ferromagnetic case, we will show below
that Eq. (\ref{apr25_1}) is incomplete. There are extra contributions to
 both $\chi (q, \Omega)$ and $\Sigma (\omega)$ which also belong to
 the Eliashberg theory, showing up at three-loop level and higher for the fermionic self-energy.
 These terms, however,
 appear due to rather specific reasons related to the
 presence of an effective dynamic long-range interaction in the theory
 and all cancel out for the gauge field problem, at a nematic QCP and
 for an Ising ferromagnetic QCP.
We consider these extra terms in the next section and proceed
here without taking them into consideration.

We can reformulate the Eliashberg theory by introducing the
following effective Lagrangian describing the fermion-boson interaction:
\beq
L =  L_F + L_B  + L_{int} \  , \mbox{with} \nonumber
\eeq
\begin{align} \label{eff}
L_B =& T \sum_{q,n} {\bf S}_{q,n} \chi^{-1} (q, \Omega_n) {\bf  S}_{-q,-n}   , \nonumber \\
L_F =&  T \sum_{k,n,\alpha,j} c^\dagger_{k,j,n,\alpha} G^{-1} (k, \omega_n)
c_{k,j,n,\alpha}   ,  \nonumber \\
L_{int} =&  g T^2 \sum_{n,m,k,q} {\bf S}_{q,m}
  c^\dagger_{k,j,n \alpha} {\bf \sigma}_{\alpha \beta} c_{k-q, j, n-m,\beta}  ,
\end{align}
where $n, m$ number Matsubara frequencies, $\alpha,\beta$ are spin indices, and $j$ is a flavor index. The upper limit of the frequency summation is the cutoff $\Lambda$.

We emphasize that there is no double counting in the bosonic
propagator (\ref{eff}). The integration of high-energy fermions,
above the cutoff $\Lambda$ leads to the momentum
dependence of the static bosonic propagator, whereas the interaction
at frequencies below $\Lambda $ gives rise to the Landau damping,
without affecting the static part.

To summarize, the Eliashberg-type theory at the QCP contains a non-analytic
fermionic self-energy which scales as $\omega^{2/3}$ and breaks down the Fermi liquid description
 of fermions. At the same time, the bosonic propagator is regular -- the only
 effect of the interaction with low-energy fermions is the appearance of
 a Landau damping.

 We will see below that for a charge QCP ( and for a spin QCP with a scalar order parameter) this Eliashberg theory
 is entirely stable. For a ferromagnetic $SU(2)$ symmetric QCP,
 the Eliashberg theory has to be extended, as we will see in Sec VI,
 to incorporate extra singular terms associated with the existence
 of long-range dynamic interaction between quasi-particles.

We emphasize 
 that even for a charge QCP
the fully renormalized {\it bosonic} susceptibility  does not necessary
 coincide with the one in (\ref{apr25_1}) and  may, in particular,
acquire an anomalous dimension~\cite{ab_chub_prl_05} if $d$ is low enough.
  However,
 this can only be due to infra-red divergent corrections to Eliashberg theory, 
which are fully captured by the
 effective low-energy model of Eq. (\ref{eff}). The anomalous dimension emerges
 at the antiferromagnetic QCP in $d=2$ (Ref. \cite{spin-fermion,ab_chub_prl_05}), 
but not in  our case.

\begin{widetext}
\begin{center}
\begin{table}[hbp]
\renewcommand\arraystretch{1.8}
\begin{tabular}{|r@{$=$}l|p{9cm}|c|}
\hline
\multicolumn{2}{|c|}{Expression} & Definition  & Eq. \\
\hline
\multicolumn{2}{|c|}{${ v_F}$} & {Fermi velocity} & (\ref{c_1_2}) \\
\multicolumn{2}{|c|}{${ m}$}  & {bare quasiparticle mass, $m = k_F/v_F$}
 & (\ref{c_1_2}) \\
\multicolumn{2}{|c|}{${ m^*}$}  & {effective (renormalized) quasiparticle mass}
 & (\ref{m*}) \\
\multicolumn{2}{|c|}{${ m_B}$}  & {band mass, determines the curvature of the Fermi surface} & (\ref{c_1_2}) \\
\multicolumn{2}{|c|}{${ g}$} & {spin-fermion coupling constant} & (\ref{sfermion}) \\
\multicolumn{2}{|c|}{${ \xi}$} & {ferromagnetic correlation length} & (\ref{chibare}) \\
\multicolumn{2}{|c|}{$\chi_0 \xi^2$} & {value of the spin susceptibility at $q=0$} & (\ref{chibare})\\
\multicolumn{2}{|c|}{$N$} & {number of fermionic flavors} & \\
\hline
${ \gbar}$ & ${ g^2 \chi_0}$ & {effective four-fermion interaction} & (\ref{param}) \\
${ \gamma}$ & ${ \frac{N m \gbar}{\pi v_F}}$ & {Landau damping coefficient} & (\ref{param_1}) \\
${ \lambda}$ & ${ \frac{3 \gbar \xi}{4 \pi v_F}}$ & {dimensionless coupling constant, it measures the  mass enhancement: $\lambda = \frac{m^*}{m} -1$}  & (\ref{param})  \\
${ \omega_0}$ & ${ \frac{3\sqrt{3}\gbar^3}{8\pi^3 \gamma v_F^3}
\sim \frac{{\bar g}^2}{N E_F}}$ & {frequency up to which $\Sigma(\omega)$ dominates over $\omega$ in the fermionic propagator} & (\ref{defs})  \\
$\quad { \omega_{\rm max}}$ & ${ \sqrt{\gamma v_F^3} \sim \sqrt{N {\bar g} E_F}}\quad$ & {frequency up to which the fermionic and the bosonic mass-shells are well separated} &  (\ref{wmax}) \\
\hline
${ \alpha}$ & ${ \frac{\gbar^2}{\gamma v_F^3} \sim \frac{\bar g}{N E_F}}$ & {small parameter, measuring the slowness of the bosonic modes compared to the fermionic ones; the same small parameter justifies the low-energy description} &  (\ref{alpha}) \\
${ \beta}$ & ${ \frac{m_B}{m N}}$ & {small parameter related to the curvature of the fermionic dispersion}  &  (\ref{beta})  \\
\hline
\end{tabular}
\caption{List of the various parameters used in the text, their expression before the rescaling in $N$, and the reference equation where it is defined in the text.}
\end{table}
\end{center}
\end{widetext}

\section{Validity of the Eliashberg theory}

The essential part of the Eliashberg procedure is an a posteriori
verification that the neglected terms in the self-energies are small.
Quite generally, the validity of this procedure is based on the idea that the fermions are fast excitations compared to the bosons,
and hence the fermionic and bosonic mass shells are well separated in energy.

When scattering off physical mass-shell bosons,
the fermions are forced to vibrate on the bosonic mass shell,
which is far away from their own. The electronic spectral
function near the bosonic mass-shell is then small and
this reduces the self-energy that arises from true fermion-boson scattering.
In the case of the electron-phonon interaction, this is known as the Migdal theorem.

The computation of the fermionic self-energy $\Sigma (\omega)$ gives us
an idea of what the typical intermediate momenta and frequencies
are in the problem. One can make sure that at criticality the typical
fermionic momenta $k-k_F$ are of order
$\Sigma (\omega)/v_F = \omega^{1/3}_0 \omega^{2/3}/v_F$.
On the other hand, the typical bosonic momenta $q_{\perp}$ along
the direction of ${\bf k}_F$ (i.e., transverse to the Fermi surface)
are of the same order as the typical fermionic momenta,
while the  momenta $q_{\parallel}$ transverse to ${\bf k}_F$
(i.e., along the Fermi surface) are of order
$(\gamma \omega)^{1/3} \gg \omega^{1/3}_0 \omega^{2/3}/v_F$.
We see that for a given frequency $\omega$,
the typical $|q|=\sqrt{q^2_{\perp} + q^2_{\parallel}}$
are much larger than $k-k_F$, i.e. the effective boson
velocity is much smaller than $v_F$. One then expects that the Migdal theorem holds.

The ratio of the typical fermionic $k-k_F$ and bosonic $|q|$ at
the same frequency $\omega$ is $(\omega_0 \omega/\gamma v^2_F)^{1/3}$.
At $\omega \sim \omega_0$, this ratio then becomes:
\beq
\alpha = \left(\frac{\omega^2_0}{\gamma v^3_F}\right)^{1/3} \sim \frac{\bar g}{ N E_F} ,
\label{alpha}
\eeq
and the slowness of the bosonic mode is then ensured in the
quantum critical regime provided that $\alpha \ll 1$.
This condition coincides, in our case, with the condition
that the interaction should be smaller than the bandwidth.
This is a necessary condition for the effective low-energy model
to be valid, for if it is not satisfied,
one cannot distinguish between contributions coming from low and from high energies.

However, the smallness of $\alpha$ is not sufficient.
In the integral for the fermionic self-energy, 
only one component of the bosonic momentum, $q_{\parallel}$,
is much larger than $k-k_F$, the other one is comparable: $q_{\perp}\sim k - k_F$.
One needs to check whether the corrections to the Eliashberg theory scale
as the ratio of $k-k_F$ to the modulus of $|q|$ or one of its components.

To address these issues, we explicitly compute the vertex corrections
and the fermionic self-energy at the two-loop level.

\subsection{Vertex corrections}

We consider the vertex corrections due to the insertion of
one bosonic propagator (three-leg vertex) and two bosonic propagators (four-leg vertex).
The behavior of these vertex corrections strongly depends on
the interplay between the internal and external momenta and frequencies. We present the results
 below and discuss technical details of the calculations in Appendix C.

\subsubsection{Three-leg vertex with zero external momentum and frequency}

We begin with the simplest 3-leg vertex, with strictly
zero incoming frequency $\Omega$ and momentum $q$. The one-loop vertex renormalization diagram contains one bosonic line and is presented in Fig. \ref{vertices}a. In analytic form, it writes:
\bea \label{migdal1}
&&\left. \frac{\Delta g }{ g }\right|_{q=\Omega =0} \sim g^2 \int d \omega d^2 p~ G({\bf k_F},\omega)^2~ \chi({\bf p},\omega) \nonumber \\
& \sim & \gbar \int \frac{ d \omega d^2 p } { \frac{\gamma | \omega | }{  p } + p^2 } \ \frac{1}{
\left ( i \tilde{\Sigma} (\omega ) - v_F p_x - \frac{p_y^2 }{ 2m_B  }\right  )^2 }   , \nonumber
\label{apr25_3}
\eea
where we defined $\tilde{\Sigma}(\omega) = \omega + \Sigma(\omega)$ and we have chosen ${\bf k}_F$ along the $x$ axis, so that $p_x = p_{\perp}$ and $p_y = p_{\parallel}$.

Since the poles coming from the fermionic Green's functions are
in the same half plane, the integral over $q_x$ is only non-zero
because of the branch cut in the bosonic propagator.
At the branch cut $p_x \sim p_y$, so that we can safely
drop the quadratic term in the fermionic propagators.
Introducing polar coordinates, and integrating successively over
the angle between ${\bf k_F}$ and ${\bf p}$, then over the modulus $p$, we obtain:
\bea
\left. \frac{\Delta g }{ g }\right|_{q=\Omega =0}  &\sim& \frac{\gbar}{\gamma v_F^3} \int_0^{\omega_{\rm max}} d\omega \f{\tSigma(\omega)}{\omega},
\label{migdal2}
\eea
where $\omega_{\rm max}$ is the frequency up to which  bosons are
slow modes compared to fermions, i.e. up to which
$\tSigma (\omega)/v_F < (\gamma \omega)^{1/3}$.  This frequency
exceeds $\omega_0$, so to find it
we have to use $\tSigma (\omega) \approx \omega$. We then obtain
\beq \label{wmax}
\omega_{\rm max} = \sqrt{\gamma v^3_F} \sim \sqrt{N \bar g E_F} .
\eeq
Note that for small values of $\alpha$, $\omega_{\rm max} \gg \omega_0$,
and the maximum frequency up to which the bosons can be treated as slow
modes well exceeds the upper limit of the quantum-critical behavior.

Substituting ${\tilde \Sigma}$ and $\omega_{\rm max}$ into (\ref{migdal2}), we obtain
\beq
\left. \frac{\Delta g }{ g }\right|_{q=\Omega =0}  \sim \sqrt{\alpha}
\label{apr25_2}
\eeq
This correction to the vertex can then
be neglected provided that $\alpha$ is small.

\subsubsection{Three-leg vertex with finite external momentum}

We now turn to the three-leg vertex with zero external frequency but a finite
bosonic momentum ${\bf q}$. The one-loop renormalization is given in Fig.\ref{vertices},
and its analytic form writes:
\bea
\left. \frac{\Delta g }{ g }\right|_{q, \Omega=0} & \sim & g^2 \int d \omega d^2 p~ G({\bf k_F+p+q},\omega) \nonumber \\
& & \qquad G({\bf k_F+p},\omega) \chi({\bf p},\omega) \nonumber \\
& \sim & \gbar \int \frac{ d \omega d^2 p } { \frac{\gamma | \omega | }{  p } + p^2  } \
\frac{1}{i {\tilde \Sigma} (\omega ) - v_F p_x - \frac{p_y^2 }{ 2m_B }} \nonumber  \\
& &  \times \frac{1}{i{\tilde \Sigma}( \omega ) - v_F q_x - v_F p_x - \frac{p_y^2}{2 m_B} -\frac{q_y p_y}{m_B}  }  ,  \nonumber \label{stat1}
\eea
where $p_x$ is the projection of ${\bf p}$ along ${\bf k_F}$.

As before, the integral over $p_x$ is
 can be reduced to the contribution from only the branch cut in
 the bosonic propagator. At the branch cut, $p_x \sim p_y$,
 hence we can neglect the quadratic term in the fermionic Green's function,
 which then allows us to integrate over $p_y$. Expanding in $q_x$, and
 subtracting the constant term at zero momentum calculated in (\ref{migdal2}),
 we obtain (detail in the Appendix C)
\bea
\left. \frac{\Delta g }{ g }\right|_{q, \Omega=0} - \left. \frac{\Delta g }{ g }\right|_{q=\Omega=0}  & \sim &  i \f{q_x}{k_F} \int_{|v_F q_x|} \f{d\omega}{|\omega|} \log [i \tSigma(\omega)] \nonumber \\
& \sim & \f{q_x}{k_F} \log |q_x|  , \label{stat4}
\eea

When not only the bosonic momentum $q$ is finite but also the external fermionic momentum for the static three-leg vertex is away from $k_F$, the $q \times \log$ dependence of the static $ \Delta g/g$ survives, but the argument of the logarithm is the 
 maximum of bosonic $q$ and fermionic $k-k_F$ (we directed both along $x$).

In general, the typical value of $q_x$ is much smaller than $k_F$,
hence the momentum dependent part of this vertex correction is small.
However, we argue in the next sections that because of the logarithmic
term in (\ref{stat4}), the insertion of this vertex correction into the $SU(2)$
 static susceptibility gives rise to a non-analytic $q^{3/2}$ term.

We also emphasize that although the calculations
of $\left. \frac{\Delta g }{ g }\right|_{q=\Omega=0}$ and
$\left. \frac{\Delta g }{ g }\right|_{q, \Omega=0}$ look similar,
the characteristic bosonic momenta are different for the two cases.
At $q=\Omega =0$, typical bosonic momenta in (\ref{apr25_3}) are of
order $(\gamma \omega)^{1/3}$, i.e.,
 near the bosonic mass shell. On the other hand,  typical bosonic momenta in
(\ref{stat1}) are of order $\tSigma (\omega)/v_F$, i.e., near the fermionic mass shell.

\subsubsection{Generic three-leg vertex}

\begin{figure}[tbp]
\centerline{\includegraphics[width=3.4in]{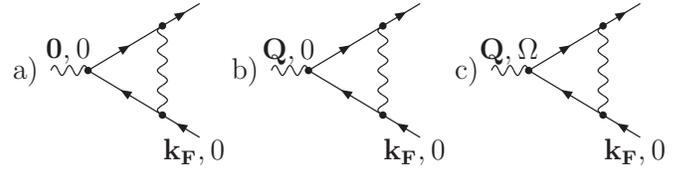}}
\caption{Three-leg vertices: a) zero external momentum and frequency b) finite momentum c) generic vertex.}
\label{vertices}
\end{figure}

We next consider the same vertex with small but finite
external momentum $q$ and frequency $\Omega$. This diagram,
presented in Fig. \ref{vertices}b, reads:
\bea
\left. \frac{\Delta g }{ g }\right|_{q, \Omega} & \sim & g^2 \int d \omega d^2 p~ G({\bf k_F+p+q},\omega+\Omega) \nonumber \\
& & \qquad G({\bf k_F+p},\omega) ~ \chi({\bf p},\omega) \nonumber \\
& \sim & \gbar \int \frac{ d \omega d^2 p } { \frac{\gamma | \omega | }{  p } + p^2  } \
\frac{1}{i {\tilde \Sigma} (\omega ) - v_F p_x - \frac{p_y^2 }{ 2m_B }} \nonumber  \\
& & \hspace{-0.5cm} \times \frac{1}{i{\tilde \Sigma}( \omega +\Omega) - v_F q_x - v_F p_x - \frac{p_y^2}{2 m_B} -\frac{q_y p_y}{m_B}  }  ,  \qquad \label{dyn1}
\eea
where we have chosen ${\bf k}_F$ along the $x$ axis,
so that $q_x = q_{\perp}$ and $q_y = q_{\parallel}$.

Integrating over $p_x$ first, one obtains two contributions:
one arising from the poles in the fermionic Green's functions (which  now can be
in different half-planes since $\Omega$ is finite),
and the other from the branch cut in the bosonic propagator.
The latter leads to the same result as (\ref{migdal2}), up to small
corrections due to the finiteness of the external $q$ and $\Omega$.
Focusing on the other contribution, one has:
\bea
\left. \frac{\Delta g }{ g }\right|_{q, \Omega} & \sim &  i \frac{\gbar}{v_F} \int_0^{\Omega} d\omega \int dp_y  \frac{|p_y|}{\gamma | \omega | + |p_y|^3 }  \nonumber \\
&& \hspace{-0.5cm} \times \frac{1}{i {\tilde \Sigma} (\Omega -\omega) + i {\tilde \Sigma} (\omega) -
 v_F q_x - \frac {q_y p_y }{  m_B  }}  , \quad
\eea
where the simplification of the frequency integral comes from the pole structure in $p_x$.

This generic correction to the vertex strongly depends
on the interplay between the external $q_x$, $q_y$, and $\Omega$.

When $q=0$, but $\Omega$ is finite, the integration over $p_y$ gives:
\bea
\left. \frac{\Delta g }{ g }\right|_{\rm q=0, \Omega} & \sim & \frac{\gbar}{v_F \gamma^{1/3}} \int_0^{\Omega} \frac{d\omega}{\omega^{1/3}}  \frac{1}{{\tilde \Sigma} (\Omega -\omega) + {\tilde \Sigma} (\omega)} \nonumber \\
 & \sim & \frac{\gbar}{v_F \gamma^{1/3} \omega_0^{1/3}} =O(1) .
\label{ex_8} \eea We see that the finite external frequency leads
to a vertex correction that is not small, and one can make sure
that higher-order corrections are of the same order. The series of
vertex corrections have been summed up in \cite{chubukov_ward}
(for a charge vertex), where $\Delta g/g$
 was proved to diverge like $(\omega_0/\Omega)^{1/3}$ at low frequency,
 as expected from the Ward identity.

When $q$ and $\Omega$ are both finite,
\beq
\left. \frac{\Delta g }{ g }\right|_{q, \Omega} =  {\cal F} \left( \frac{v_F q}{\Sigma(\Omega)}  \right) ,
\label{apr25_4}
\eeq
where 
\beq
{\cal F} (x) = \int_0^1 \frac{dz}{z^{1/3}} \frac{1}{(1-z)^{2/3}+z^{2/3}+ix}
\eeq
 has the following asymptotic behavior:
\beq
\left\{
\begin{array}{lcl}
{\cal F}(x\ll 1) & = & O(1) \\
{\cal F}(x\gg 1) & = & O \left( \frac{1}{x} \right)
\end{array}
\right.
\eeq
If  typical $q$ is on the bosonic mass shell,
then $q \sim (\gamma \Omega)^{1/3}$, and one has:
\beq \label{dynmassshell}
\left. \frac{\Delta g }{ g }\right|_{q, \Omega}
\sim \frac{\Sigma(\Omega)}{v_F q_x} \sim ~\sqrt{\alpha}~
\left( \frac{\Omega}{\omega_{\rm max}} \right)^{1/3}.
\eeq
This is obviously small in $\alpha$.

It turns out that the behavior of the vertex correction is
 more complex and the result for $\Delta g/g$  strongly
 depends on the direction of $q$ compared to the
 direction of fermionic ${\bf k}_F$.
 This directional dependence is important for our purposes
 as  we know from self-energy calculations that in the integral for the self-energy only
$y$ component of the bosonic momentum is near a bosonic mass shell and
scales as $(\gamma \Omega)^{1/3}$,
 the $x$ component of bosonic momentum is actually of
 the order of ${\tilde \Sigma} (\omega)/v_F$, i.e., is near a fermionic
 mass shell and is much
smaller.  We therefore take a more careful look at the vertex correction.

For the case where $q_x \sim \tSigma(\Omega)/v_F$ and
$q_y \sim (\gamma \Omega)^{1/3}$, one would argue from (\ref{dynmassshell})
that the vertex correction now becomes of order $O(1)$ and is no longer
parametrically small. However, the computation that lead to (\ref{dynmassshell})
cannot be extended to the strongly anisotropic case as for
 external $q_x \sim \tSigma(\Omega)/v_F$ and $q_y \sim (\gamma \Omega)^{1/3}$,
 the curvature of the fermionic dispersion becomes relevant and changes the result.
 The full dependence on $q_x$ and $q_y$ is rather complex and we restrict
 ourselves to the case when
\beq
\beta = \frac{1}{N} \frac{m_B}{m} \ll 1  .
\label{beta}
\eeq
In this situation, $q^2_y/m_B \sim (v_F q_x)/\beta  \gg v_F q_x$, i.e., the
 the quadratic term in fermionic propagator dominates.
Performing the integration, we then find that:
\bea
\left. \frac{\Delta g }{ g }\right|_{q, \Omega} & \sim &
\beta^2 \left( \frac{\gamma \Omega}{q_y^3} \right)^{2/3}
\log^2 \left[   \beta \frac{(\gamma \Omega)^{1/3}}{q_y} \right] \nonumber \\
 & \sim & \beta^2 \log^2 \beta  ,  \label{dyn4}
\eea

It follows that even when only one component of the bosonic momentum
 is near a bosonic mass shell, the vertex correction
 is small if $\beta$ is small. This is the second condition
 for the Eliashberg theory to be controllable at criticality.

The smallness of $\beta$ can be ensured by either extending the
theory to large values of $N$, or by considering a very strong
curvature of the Fermi surface which implies that $m_B \ll m$.
Even though the latter can hardly be satisfied for realistic Fermi
surfaces, we emphasize that the curvature of the dispersion plays
a crucial role in the theory, for even in the case of $N \gg 1$,
the vertex correction is of order $O(1)$ without this curvature.

\subsubsection{Pairing vertex}

By contrast to the previous vertices we studied, the pairing vertex
in the Cooper channel is not sensitive to the curvature of the Fermi surface.
This leads to a vertex of order $O(1)$ even in the large-N limit and the pairing problem
 then has to be carried out exactly  within the Eliashberg theory.

This vertex renormalization is presented in Fig. \ref{coop4leg} and its analytic form is given by:
\bea
\left. \frac{\Delta g }{ g }\right|_{\rm Cooper} & \sim & g^2 \int d \omega d^2 q~ \chi({\bf q},\omega)~G({\bf k_F+q},\omega) \nonumber \\
&  & \qquad \qquad \times G({\bf -k_F-q},-\omega-\Omega) \nonumber \\
& \sim & \gbar \int \frac{d\omega d^2 q}{\frac{\gamma |\omega|}{q}+q^2}\frac{1}{i\tSigma(\omega)-v_F q_x-\frac{q_y^2}{2m_B}} \nonumber \\
& &  \times  \frac{1}{-i\tSigma(\omega+\Omega) - v_F q_x - \frac{q_y^2}{2 m_B}}  .  \label{Coop1}
\eea

Integrating over $q_x$, restricting ourselves to the contribution
from the fermionic poles (the one from the branch cut can be proved to be smaller),
we find that the quadratic terms cancel out, leaving us with:
\beq \label{Coop2}
\left. \frac{\Delta g }{ g }\right|_{\rm Cooper}  \sim \frac{\gbar}{v_F} \int_{|\Omega|}^{D} \frac{d\omega}{\tSigma(\omega+\Omega)+\tSigma(\omega)} \int_0^{\infty} \frac{dq_y ~q_y}{\gamma \omega+q_y^3}  .
\eeq

Performing the remaining integral, the pre-factor simplifies and
we obtain: \bea
\left. \frac{\Delta g }{ g }\right|_{\rm Cooper}  &\sim &  \frac{{\bar g}}{\gamma^{1/3} \omega^{1/3}_0 v_F}   \log \left|  \f{\Omega}{D} \right| \nonumber \\
& \sim & \log \left|  \f{\Omega}{D} \right|    ,   \label{Coop3}
\eea
where we assumed that we were in the quantum critical regime, i.e. $|\Omega|<\omega_0$.

We emphasize that the  pre-factor of the $\log$ in (\ref{Coop3})
is $O(1)$,
 even when one takes into account the curvature of the fermionic dispersion.
 The result of Eq. (\ref{Coop3}) confirms previous studies~\cite{cfhm} advocating
 that the system at a ferromagnetic QCP is unstable towards pairing.

\subsubsection{Four-leg vertex}

\begin{figure}[tbp]
\centerline{\includegraphics[width=3.4in]{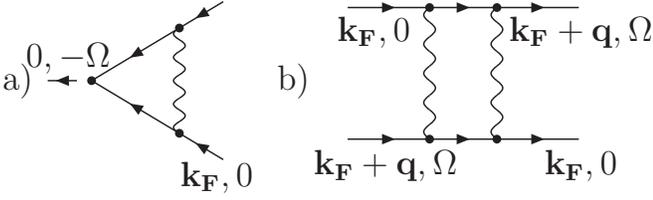}}
\caption{a) Cooper pairing vertex b) Four-leg vertex}
\label{coop4leg}
\end{figure}

We consider now higher-order corrections to the vertex through
the example of a four-leg vertex correction with two crossed bosonic lines,
also called a Cooperon insertion. Analytically, the
expression for this renormalized vertex, presented diagrammatically
in Fig. \ref{coop4leg}, writes:
\beq \label{cross1}
\begin{array} {l}
{\displaystyle \Gamma_2 (q,\Omega) \sim \gbar^2 \int d\omega d^2 p \ \chi_s \left(\frac{\Omega+\omega}{2},\frac{\bf p+q}{2} \right) } \\ \\
{\displaystyle \times  \chi_s \left(\frac{\Omega-\omega}{2},\frac{\bf q-p}{2} \right) G \left(\frac{\Omega+\omega}{2},{\bf k_F}+\frac{\bf p+q}{2}\right) } \\ \\
{\displaystyle \times G \left(\frac{\Omega-\omega}{2},{\bf k_F}+\frac{\bf q-p}{2} \right) } .
\end{array}
\eeq
After integrating over $p_x$ (projection of ${\bf p}$
along ${\bf k_F}$), and $\omega$, were are left with:
\beq \label{cross2}
\begin{array} {l}
{\displaystyle \Gamma_2 (q,\Omega) \sim \frac{\gbar^2}{v_F} \frac{\Omega}{|q_y|^3} \int dz \frac{\sqrt{\left( z^2 +\frac{q^2}{q_y^2} \right)^2 - 4z^2}}{2i\Sigma \left( \frac{\Omega}{2} \right) -v_F q_x -\frac{q_y^2}{4m} (1+z^2)} } \\ \\
{\displaystyle \times \frac{1}{\left( z^2 +2z+ \frac{q^2}{q_y^2} \right)^{3/2} + \frac{\gamma |\Omega|}{|q_y|^3}} \frac{1}{\left( z^2 - 2z+ \frac{q^2}{q_y^2} \right)^{3/2} - \frac{\gamma |\Omega|}{|q_y|^3}}  } ,
\end{array}
\eeq
where we have changed $p_y$ into $z=p_y/|q_y|$.

This renormalized 4-leg vertex $\Gamma_2 (q, \Omega)$ has
to be compared with the bare four-leg vertex, whose analytic form
is  given by the bosonic propagator multiplied by $g^2$:
\beq \label{Gamma1}
\Gamma_1 (q,\Omega) \sim \frac{\gbar}{q^2 + \frac{\gamma |\Omega|}{q}} .
\eeq

In the case of a typical external momentum $q_x \sim q_y \sim (\gamma \Omega)^{1/3}$,
the ratio $\Gamma_2/\Gamma_1$ is of order $\alpha$.
For a typical
$q_x \sim {\tilde \Sigma} (\omega)/v_F$ and $q_y \sim (\gamma \omega)^{1/3}$,
 we  obtain, to logarithmic accuracy,
\beq \label{cross4}
\frac{\Gamma_2}{\Gamma_1}  \sim \beta \ll 1 ,
\eeq
 This last result again critically depends on the curvature of the
 Fermi surface: neglecting the quadratic terms in the fermionic propagators,
 one would obtain $\Gamma_2/\Gamma_1 = O(1)$.
We see that likewise to the three-loop vertices,  the smallness of the crossed
vertex $\Gamma_2 (q,\Omega )$ requires both $\alpha$ and $\beta$  to be small.

\subsection{Self-energy corrections}

\subsubsection{Corrections to the self-energy at the two-loop level}

We found in our analysis of the vertex corrections that the result depends
on the interplay between the typical momentum and frequency. In our estimates,
we considered two regions of
 external $q$ and $\Omega$, namely $q_x \sim q_y \sim (\gamma \Omega)^{1/3}$
 and $q_x \sim \tSigma (\omega)/v_F$, $q_y \sim (\gamma \Omega)^{1/3}$.
In both cases, we found that vertex corrections are small.
 We verify here that the two-loop self-energy, obtained by inserting
 vertex corrections into one-loop self-energy diagram, is also small.

 The two-loop self-energy diagram is presented in fig. \ref{self}. We have:
\begin{align} \label{sig2loop1}
\Sigma_2 (\omega) & \sim \gbar^2 \int d\omega_1 d^2 q_1 \int d\omega_2 d^2 q_2 \chi ({\bf q_1},\omega_1)  \chi({\bf q_2},\omega_2) \nonumber \\
&  \qquad \times G({\bf k_F+q_1},\omega+\omega_1) G({\bf k_F+q_2},\omega+\omega_2) \nonumber \\
& \qquad \times G({\bf k_F+q_1+q_2},\omega+\omega_1+\omega_2) \nonumber \\
& \sim \gbar^2 \int d\omega_1 d^2 q_1 \int d\omega_2 d^2 q_2 \frac{q_1}{\gamma |\omega_1|+ q_1^3} \frac{q_2}{\gamma |\omega_2|+q_2^3} \nonumber \\
& \times \frac{1}{i\tilde{\Sigma}(\omega+\omega_1)-v_F q_{1x}} \frac{1}{i\tilde{\Sigma}(\omega+\omega_2)-v_F q_{2x}-\frac{q_{2y}^2}{2m_B}} \nonumber \\
& \times \frac{1}{i \tilde{\Sigma}(\omega+\omega_1+\omega_2)-v_F q_{1x}-v_F q_{2x}-\frac{q_{1y}^2}{2m}-\frac{q_{2y}^2}{2m_B}} ,
\end{align}
where we recall $\tilde{\Sigma}(\omega) = \omega+\Sigma(\omega)$.

Integrating successively over $q_{1x}$ and $q_{2x}$, and
rescaling the remaining momentum components by introducing
$x=q_{1y}/(\gamma |\omega_1|)^{1/3}$ and $y=q_{2y}/(\gamma
|\omega_2|)^{1/3}$, we obtain: \beq \label{sig2loop2}
\begin{array}{l}
{\displaystyle \Sigma_2 (\omega) \sim \frac{m_B \gbar^2}{v_F^2} \int_{0}^{\omega} d\omega_2 \int_{\omega-\omega_2}^{\omega} d\omega_1 \frac{1}{(\gamma^2 \omega_1 \omega_2)^{2/3}} }\\ \\
{\displaystyle \qquad \times \int_{-\infty}^{\infty} \frac{dx dy}{xy + i\zeta} \frac{|x y|}{\left( 1+ |x|^3  \right) \left( 1+ |y|^3    \right)}} ,
\end{array}
\eeq
where $\zeta = m_B \frac{\tilde{\Sigma}(\omega+\omega_1)+\tilde{\Sigma}(\omega+\omega_2)-\tilde{\Sigma}(\omega+\omega_1+\omega_2)}{\left( \gamma^2 \omega_1 \omega_2  \right)^{1/3}}$.

As the typical frequencies $\omega_1$ and $\omega_2$
are of order $\omega$, the typical value of $\zeta$ is of order
$\beta \ll 1$. Expanding then in (\ref{sig2loop2}) to first order in $\zeta$
and performing the remaining integrals, we obtain in the quantum-critical regime:
\bea \label{sig2loop3}
\Sigma_2 (\omega) &\sim &\frac{m \gbar^2}{v_F^2} \int_0^{\omega} d\omega_2 \int_{\omega-\omega_2}^{\omega} d\omega_1 \frac{\zeta \log^2 \zeta}{(\gamma^2 \omega_1 \omega_2)^{2/3}} \nonumber \\
& \sim & \Sigma(\omega) \beta^2 \log^2 \beta ,
\eea
where $\Sigma(\omega) = \Sigma_1 (\omega) = \omega^{1/3}_0 \omega^{2/3}$ is the self-energy in the Eliashberg theory.

This result agrees with  the one obtained in \cite{aim},
 and shows that $\Sigma_2(\omega) \sim \Sigma_1(\omega) \times \left. \frac{\Delta g}{g} \right|_{\rm q, \Omega}$
 where $\left. \frac{\Delta g}{g} \right|_{\rm q, \Omega}$ is given by
(\ref{dyn4}).
This last result implies that  typical internal $q$ and $\Omega$ for
the Eliashberg self-energy and for $\Sigma_2 (\omega)$ are the same.

It is also instructive to compare these two-loop results, obtained
as an expansion around the Eliashberg solution, to the
perturbation expansion around free fermions. In the latter, we
found in Eq. (\ref{ex_10}) that $\Sigma_2 (\omega) \sim \omega
\log^2 \omega$ whereas in the former we have $\Sigma_2 (\omega)
\propto \beta (\omega^{1/3}_0 \omega^{2/3}) \log^2\beta$, Eq.
(\ref{sig2loop3}). The free-fermion result can be reproduced if we
neglect the self-energy in (\ref{sig2loop2}). We see that the
expansion around free fermions does not reproduce the correct
frequency dependence of $\Sigma_2 (\omega)$. This obviously
implies that if one expands around free fermions, there exist
higher-order terms associated with insertions of the self-energy
$\Sigma(\omega)$ into the internal fermionic lines, which may
overshadow the two-loop result around free fermions. Accordingly,
 near the QCP, the expansion around free fermions does not converge,
 even if the curvature of the fermionic dispersion is included.
 On the other hand, the expansion around the Eliashberg solution is
 regular and holds in powers of the small parameters $\alpha$ and $\beta$.

\subsubsection{Momentum dependence of the self-energy and the density of states}

Along with the vertex corrections, we also neglected the momentum
dependence of the fermionic self-energy in order to proceed with
the Eliashberg scheme. We now verify whether the momentum dependent
part of the self-energy $\Sigma (\bk,\omega=0)=\Sigma(\bk)$ remains
small when evaluated with the full fermionic propagator.
 The $k$ dependent self-energy is given by:
\bea
\Sigma({\bf k},0) & = & 3 i g^2 \int \f{d^2q~d\Omega}{(2 \pi)^3}
 G(\bk+\bq,\Omega) \chi(\bq,\Omega) \nonumber \\
 & = &\frac{3i  \gbar}{(2 \pi)^3} \int \frac{d\Omega d^2q}{i{\tilde \Sigma}(\Omega)-\epsilon_{k+q}} \frac{q}{\gamma |\Omega|+q^3} .
\label{may2_2}
\eea
Defining the angular variable $\theta$ as
$\epsilon_{k+q} = \epsilon_k + v_F q \cos \theta$,  integrating
over it,  and expanding to linear order in $\epsilon_{\bf k}$ we obtain:
\bea
&& \Sigma({\bf k},0) =-3i \epsilon_k \times \frac{{\bar g}}{2 \pi^2}
\int_0^\infty d\Omega ~{\tilde \Sigma} (\Omega) \times \nonumber \\
&&\int \frac{q^2 dq }{\left(q^{3} +
(\gamma |\Omega|)\right)\left((v_F q)^2 + {\tilde \Sigma} (\Omega)^2\right)^{3/2}}.
\eea

A simple experimentation with the integrals shows that the
integration over momentum is confined to $q \sim {\tilde \Sigma} (\Omega)/v_F$,
while the frequency integral
 is confined to $\Omega < \omega_{max}$, where $\omega_{max}$,
defined in (\ref{wmax}), is the scale where
$(\gamma \Omega)^{1/3} = {\tilde \Sigma} (\Omega)/v_F$.
The computation of $\Sigma({\bf k})$  is given in the Appendix B, and the result is
\beq
 \Sigma({\bf k},0) = -i \epsilon_k \times  \frac{3 \times 1.3308}{2\sqrt{2} \pi^{3/2}} \sqrt{\alpha} = -i \epsilon_k ~0.253 \sqrt{\alpha}
\label{sigma2k2_2}
\eeq
Using our definition of $\Sigma$, we then obtain that $\Sigma({\bf k},0)$
 gives rise to a small, regular correction to the quasiparticle mass $\epsilon_k -i \Sigma ({\bf k},0) = \epsilon^*_k = v^*_F (k-k_F)$. 
\beq
v^*_F = v_F~ \left( 1 - 0.253 \sqrt{\alpha}\right)
\eeq  

Like the vertex correction at zero external bosonic momentum and frequency,
this small correction is of order $O(\sqrt{\alpha})$ and comes from frequencies
of order $\omega_{\rm max}$.

The momentum dependent self-energy, unlike the frequency dependent part,
generally gives rise to corrections to the fermionic density of states (DOS)
\beq
N(\omega) = - N_0  \int \frac{d \epsilon_k}{\pi} Im G(\epsilon_k, \omega). 
\eeq
where $N_0$ is the DOS of free fermions.

 Assuming that $\Sigma (k)$ is small and expanding in it in the 
fermionic propagator, we obtain, in real  frequencies  
\beq
\frac{N(\omega)}{N_0}  = 1-  Im \left( \left. \frac{\Sigma (\bk)}{\epsilon_k}\right|_{\epsilon_k = i \tSigma (-i\omega)} \right),
\label{may2_1}
\eeq
Sunstituting (\ref{sigma2k2_2}) into (\ref{may2_1}, we find that the density of states just shifts by a constant.  In order to extract the frequency dependence of the density of states,
one has to evaluate the momentum-dependent self-energy to next order in $\epsilon_k$ and on the mass shell,  where $\epsilon_k = i\tSigma (\omega)$.

The evaluation of the self-energy near a mass shell generally
 requires extra caution as the self-energy may possess mass-shell
singularities~\cite{cmgg}. 
 We, however,  have checked in Appendix  \ref{app:mass_shell} that in
 our case the self-energy does not possess any mass-shell singularity, and 
 the self-energy remains finite on the mass shell. 

The calculation of the self-energy to order $\epsilon^2_k$ is
 displayed in the Appendix B, and the result is
\beq
\Sigma({\bf k},\omega) = i \epsilon_k \times \frac{0.45}{8 \pi N}
 \frac{|\Sigma (\omega)|}{E_F}
 \log {\frac{\omega_1}{|\omega|}}
\label{sigma2k3}
\eeq  
Substituting this self-energy  into the expression
for the DOS and converting to real frequencies, we find at small $\omega$:
\beq
N(\omega) \sim N_0 \left( 1 - \left(\frac{\omega}{\omega_1}\right)^{2/3}
 \log \frac{\omega_{max}}{\omega}\right) ,
\label{sigma2k3_2}
\eeq
where we explicitly defined 
\beq
\omega_1 = \frac{128 \pi^{5/2}}{0.45^{3/2}~ 3^{3/4}} \frac{N^2 E^2_F}{{\bar g}}. 
\eeq

The correction to the fermionic DOS was earlier computed
by Lawler et al~\cite{fradkin_2} using the bosonization technique. 
They obtained the same $\omega^{2/3} \log \omega$ dependence as in (\ref{sigma2k3_2}. The  agreement
 with Ref.~\cite{fradkin_2} is, however,  likely  
 accidental as they evaluated $N(\omega)$  in the expansion around free 
fermions, while we obtained $N(\omega)$ by evaluating
 the $k-$dependent self-energy at $\epsilon_k = i \tSigma (-i\omega)$.
If we  neglected
 the Eliashberg self-energy (i.e., expanded around free fermions), we 
 would obtain $\omega \log \omega$ correction to the DOS. This 
last result agrees with the one obtained by 
\cite{chub_khvesh} using the same technique as in~\cite{fradkin_2}.

\subsection{Summary}

We have shown in this section that there are two conditions for the validity
of the Eliashberg theory that one can recast as the smallness of two parameters:
\beq \label{alphabeta}
\alpha \sim \frac{\gbar^2}{\gamma v_F^3} \sim \frac{\gbar}{N E_F} \ll 1 \qquad \beta \sim \frac{m \gbar}{\gamma v_F} \sim \frac{m_B}{N m}\ll 1 .
\eeq

The first condition is quite generic for a low-energy theory since it requires
that the fermion-fermion interaction mediated by the exchange of a boson should
be smaller than the Fermi energy. Otherwise, the physics is not restricted to the
vicinity of the Fermi surface anymore.
The parameter $\alpha$ plays the same role as the Migdal parameter for
the electron-phonon interaction: it sets the condition that fermions are
fast excitations compared to bosons. In the scattering processes that are small
in $\alpha$, fermions are forced by the interaction to vibrate at frequencies
near the bosonic mass shell.They are then far from their own resonance and
thus have a small spectral weight.

However, the condition $\alpha \ll 1$ is not sufficient to construct a fully
controllable perturbation expansion around the non-Fermi liquid state at the QCP.
 In spatially isotropic systems there exist vertex corrections for which
 the external momentum has a component on the fermionic mass-shell.
These corrections don't contain $\alpha$. However, these
corrections are sensitive to the curvature of the Fermi surface,
and are small if $\beta$ is small which can be achieved
 either by imposing $m_B \ll m$ or by extending the theory
 to a large number N of fermionic flavors.

A word of caution. In evaluating the renormalization of the static vertex, 
 we silently  assumed  that $\sqrt{\alpha} \ll \beta$, i.e., 
${\bar g}/E_F < (m_B/m)^2 /N$. At very large $N$, this is no longer valid. 
For this situation, i.e., when $\beta \ll \sqrt{\alpha}$, our estimates show that  the static vertex is even smaller than $\sqrt{\alpha}$.

Finally, the pairing vertex
in the Cooper channel stays of order $O(1)$, signaling the
possibility of a pairing
 instability close to the quantum critical point.
 Nevertheless, we assume, based on explicit calculations
 worked out in \cite{cfhm}, that the quantum critical behavior
 extends in the parameter space to a region where the superconductivity is not present.

\section{Instability of the ferromagnetic quantum critical point}

We found that the Eliashberg theory for fermions interacting with gapless
long-wavelength bosons is stable and controlled by two small parameters.
We verified this by calculating the fermionic self-energy in a two-loop
 expansion around the Eliashberg solution

One may wonder whether the same conclusions hold for the bosonic self-energy as well.
In particular, what are the corrections to the static susceptibility
$\chi_s (q,0)$? Naively one could assume
 that they are unimportant and do not change the bare $q^2$ behavior
 of the inverse bosonic propagator at the QCP.

For a ferromagnetic $SU(2)$ QCP, for which the massless bosons are
spin fluctuations, we show in this section that the corrections to
the static spin susceptibility are non-analytic: they scale like
$q^{3/2}$, and {\it do not} contain any pre-factor except for a
proper power of $k_F$. Such terms obviously overshadow the regular
$q^2$ of the bare susceptibility at small enough momenta.
 These terms  therefore belong
 to Eliashberg theory, which has to be extended to incorporate them.

 The physics behind the $q^{3/2}$ term in $\chi (q, 0)$
at a ferromagnetic QCP is, by itself, not directly related to
criticality: far away from the QCP, the spin susceptibility also
contains negative, non-analytic $|q|$ term \cite{bkv,millis,chub_maslov,cmm,gal,bet}.
 This term gradually evolves as the correlation length $\xi$ increases,
 and transforms into the $q^{3/2}$ term at the QCP. Both these non-analyticities,
 at and away from the QCP, emerge because the boson-mediated
 interaction between fermions contain a long-range dynamic component,
 generated from the Landau damping.

 For charge fluctuations, the $q^{3/2}$ terms appear in the individual diagrams
 for the susceptibility but
 cancel out in the full formula for $\chi (q,\Omega)$.
We discuss the physical origin of the difference between spin and
charge
 susceptibilities in the next section.

One of the reasons why the non-analyticity of the static spin
susceptibility at the QCP has not been analyzed much earlier is
because it was widely believed that an itinerant fermionic system
near a ferromagnetic QCP is adequately described by a
phenomenological $2+1$D $\phi^4$ bosonic theory (in our case, the
role of $\phi$ is played by the vector field ${\bf S}$) with the
dynamic exponent $z=3$ and a constant pre-factor for the $\phi^4$
term~\cite{hertz}. In dimensions $d\geq 4-z =1$, the model lies
above its upper critical dimension and the $\phi^4$ term is simply
irrelevant.

In this section, we derive the effective $\phi^4$ theory from the
spin-fermion model Hamiltonian, and show that it contains two new
elements absent from the phenomenological approach. First, the
pre-factor of the $\phi^4$ term strongly depends on the ratio
between the external momenta and frequencies, and contains a
non-analytic term in addition to the constant one. Second, there
also exists a cubic $\phi^3$ term whose pre-factor, although
vanishing in the static limit, also strongly depends on the
interplay between the external momenta and frequencies. We can
recast the non-analytic $q^{3/2}$ term in the static spin
susceptibility as arising from these cubic and quartic terms in
$\phi$.

We also prove that the non-analyticity appears in the
temperature-dependent uniform static susceptibility $\chi_s (T)$.
e show below that $\chi^{-1}_s (T) \propto T |\log T|$, again with
a negative pre-factor.

Finally, we show that the instability of the ferromagnetic QCP
can also be seen from the fermionic self-energy, which acquires
singular terms beginning at the three-loop order in the case of spin fluctuations.

\begin{figure}[tbp]
\centerline{\includegraphics[width=3.4in]{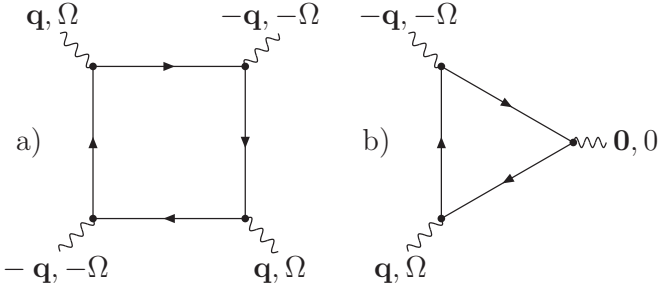}}
\caption{``$\phi^4$'' and ``$\phi^3$'' type diagram}
\label{phi4}
\end{figure}

\subsection{Hertz-Millis-Moriya theory revisited }

In order to derive a quantum critical $\phi^4$ model, one
has to integrate the fermions out of the partition function,
noticing that the Lagrangian of the spin-fermion model is quadratic in the fermions.

Expanding then in the number of bosonic fields ${\bf S}$,
the quartic term in the effective action reads:
\beq  \label{hertzrev1}
\begin{array}{l}
{\displaystyle \int  \frac{d^2  q d^2  p  d^2   p^\prime~  d  \Omega d  \nu  d
\nu^\prime }{(  2 \pi )^9}~ A  ({\bf p},{\bf p}^{\prime},{\bf q}, \nu,\nu^{\prime},\Omega) } \\ \\
{\displaystyle \qquad \times~ \left[ {\bf S}_{p+q/2}\cdot{\bf S}_{q/2-p} ~ {\bf S}_{p^\prime-q/2}\cdot{\bf S}_{-p^\prime-q/2}  \right.} \\ \\
{\displaystyle \qquad ~+~ {\bf S}_{p+q/2}\cdot{\bf S}_{-p^{\prime}-q/2} ~ {\bf S}_{q/2-p}\cdot{\bf S}_{p^\prime-q/2}}  \\ \\
{\displaystyle \qquad ~-~ \left. {\bf S}_{p+q/2}\cdot{\bf S}_{p^{\prime}-q/2} ~ {\bf S}_{-q/2-p^\prime}\cdot{\bf S}_{q/2-p}  \right]} ,
\end{array}
\eeq
where $p, p^\prime, q$ are bosonic momenta, and $\nu, \nu^\prime, \Omega$ are
 bosonic frequencies.

Our goal here is to prove that the pre-factor $A$ is not a regular
function of momenta and frequencies. To simplify the presentation,
we choose  to study only the dependence of $A$ on $q$ and $\Omega$
 and set ${\bf p},{\bf p^\prime}, \nu, \nu^\prime$ to zero (see Fig~\ref{phi4}).

The analytic form of this prefactor then writes:
\bea \label{hertzrev2}
A(q,\Omega) & \sim & N g^4 \int d \omega\ \int d^2 k
\ G({\bf k},\omega)^2 \qquad \qquad \nonumber \\
 & & \qquad \times ~ G({\bf k+q},\omega+\Omega)^2 .
\eea

Defining  $\theta$ as the angle between ${\bf k}$ and ${\bf q}$, and performing the integration over $\epsilon_k$,and the angular variable, we find:
\bea \label{hertzrev3}
A(q,\Omega) & \sim & \frac{Nmg^4}{\omega_0 \Omega} \int_0^{1} dz \nonumber \\
& & \f{\left[ \left(\frac{v_F q}{\Sigma(\Omega)}\right)^2 -2 \left( z^{2/3} + (1-z)^{2/3}\right)^2 \right]}{\left[ \left(\frac{v_F q}{\Sigma(\Omega)}\right)^2 +  \left(  z^{2/3} + (1-z)^{2/3}\right)^2 \right]^{5/2}} ,  \qquad
\eea
where we defined $z=\omega/\Omega$, and neglected at this stage a regular part that comes from large values of $\epsilon_k$ and for which the curvature is relevant.

We see that $A(q, \Omega)$ depends on the interplay between
the momentum and frequency. We can identify two regimes
\begin{itemize}
\item If $|q| \sim (\gamma |\Omega|)^{1/3}$, i.e. if the bosonic momenta
 are near the bosonic mass shell,  the self-energy in
 the denominator can be neglected, the frequency factors in
 the numerator and the denominator are
 canceled out, and we obtain:
\beq \label{hertzrev4}
A (\Omega) \sim \frac{N m g^4}{\gamma v_F^3} \sim \frac{1}{\chi^2_0}~
\alpha m \ ,
\eeq
We see that $A$  can be safely replaced by a constant $O(\alpha)$. This is
 consistent with the previous works~\cite{hertz}.
 The agreement is not surprising as the relation $q \sim (\gamma |\Omega|)^{1/3}$
 is assumed in
power counting based on $z=3$ scaling.
 Note that the condition  $|q| \sim (\gamma |\Omega|)^{1/3}$ only
 specifies the magnitude of $q$, one of its components (e.g., $q_x$) can be much smaller.

\item If  $q \sim {\bar \Sigma} (\omega) \sim \omega_0^{1/3} \Omega^{2/3}  v_F$ (at $\omega < \omega_0$), i.e. when a boson resonates near a
 fermionic mass shell, the $z=3$ scaling arguments are not applicable. We have in this regime
\beq
A(q,\Omega) \sim \frac{Nmg^4}{\omega_0 \Omega} \sim \frac{N m}{\chi_0^2} ~ \frac{1}{\sqrt{\alpha}}\frac{\omega_{\rm max}}{\Omega}.
\eeq
In this case, $A(\Omega)$ is a singular  function of frequency,
and cannot be replaced by a constant.
\end{itemize}

We see therefore that the pre-factor of the $\phi^4$ term is
actually singular
 outside the  scaling regime of a $z=3$ theory.

In the similar spirit, one can construct a cubic term in the
bosonic fields. \beq \label{hertzrev1_1} \int  \frac{d^2  q d^2  p
d  \Omega d  \nu }{(  2 \pi )^9}~ B  ({\bf p},{\bf q},
\nu,\Omega)~ {\bf S}_{p}\cdot( {\bf S}_{q-p/2} \times {\bf
S}_{-q-p/2}) , \eeq where the pre-factor $B$ is a convolution of
three fermionic Green's functions as presented in Fig.\ref{phi4},
and is given by: \bea \label{hertzrev2_1} && B({\bf q},{\bf p},
\Omega, \nu) \sim  N g^3 \int d \omega\ \int d^2 k
\ G({\bf k-p},\omega - \nu)  \nonumber \\
&& \qquad \times G({\bf k-p/2-q},\omega-\nu/2-\Omega)
~  G({\bf k},\omega) .
\eea

Proceeding as for the quartic term, we set for
simplicity ${\bf p} = {\bf 0}$, $\nu =0$, and integrate
over $\epsilon_k$ and the angular variable, leading to:
\bea \label{hertzrev3_1}
B(q,\Omega) & \sim & \frac{Nmg^3}{\omega_0^{2/3} \Omega^{1/3}} \int_0^{1} dz \nonumber \\
& & \f{\left( z^{2/3} + (1-z)^{2/3}\right)}{\left[ \left(\frac{v_F
q}{\Sigma(\Omega)}\right)^2 +  \left(  z^{2/3} +
(1-z)^{2/3}\right)^2 \right]^{3/2}}  , \qquad \eea where we again
introduced the rescaled frequency $z=\omega/\Omega$.

We can again identify two regimes.
\begin{itemize}
\item
In the $z=3$ regime where $q\sim (\gamma \Omega)^{1/3}$, one
can expand for large $\frac{v_F q}{\Sigma(\Omega)}$, which leads to:
\beq
B(q,\Omega) \sim \frac{Nmg}{\chi_0} ~ \alpha \left( \frac{\Omega}{\omega_{\rm max}} \right)^{2/3} ,
\eeq
where $\omega_{max}$ is given by (\ref{wmax}). This term is small
in the quantum critical regime where $\Omega < \omega_0 \sim \omega_{max} \alpha^{3/2}$, and
can be safely neglected.

\item
For $q \sim \Sigma(\Omega)$, the remaining integral is of order $O(1)$ and the result writes:
\beq
B(q,\Omega) \sim \frac{Nmg}{\chi_0} ~ \frac{1}{\sqrt{\alpha}} \left( \frac{\omega_{\rm max}}{\Omega} \right)^{1/3} ,
\eeq
which is large and cannot be neglected.
\end{itemize}

We demonstrate in the next section how the singular behavior of the
$\phi^3$ and $\phi^4$ terms leads to a non-analytic contribution to the static spin susceptibility.

\subsection{Non-analyticity in the static spin susceptibility}

\begin{figure}[tbp]
\centerline{\includegraphics[width=3.4in]{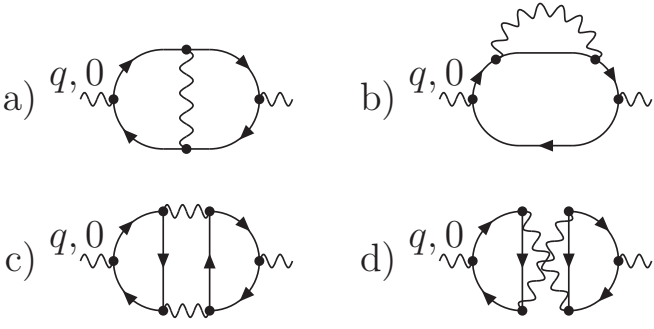}}
\caption{Corrections to the polarization bubble from diagrams with one and two extra bosonic lines}
\label{2loop}
\end{figure}

We  now  estimate  the  effect  of these  singularities  on  physical quantities.
Both the $\phi^3$ and $\phi^4$ terms in the effective action give rise
to corrections to the $\phi^2$ term, i.e. to the spin susceptibility. These
corrections are obtained diagrammatically by contracting the external legs
of the $\phi^3$ and $\phi^4$ terms, as shown in Fig. \ref{2loop}.  The computations are
described in detail Appendix \ref{app:chi}.

The contributions from the $\phi^4$ terms have been considered in our
short publication~\cite{our_prl}. The contributions from cubic terms were missed,
and were first considered in ~\cite{ch_masl_latest} in the analysis of the spin
susceptibility in the paramagnetic phase, away from a QCP.

In analytic form, we have, using $\chi (q,0) = \chi_0 /(\xi^{-2} + q^2 + \Pi (q,0))$,  and $\Pi = \Pi_1 (q,0) + \Pi_2 (q,0) + \Pi_3 (q,0) + \Pi_4 (q,0)$:

\begin{widetext}
\bea \label{pi12}
\Pi_{1} (q, 0) & = &  \Gamma^{S}_{1} \f{N \gbar^2}{(2 \pi)^6 \chi_0} \int d^2 k d \omega d^2  l d\Omega~ \chi_s (l,\Omega)~ G(\omega,k)~G(\omega+\Omega,k+l)~ G(\omega+\Omega,k+q+l)~ G(\omega,k+q) \nonumber \\
\Pi_{2} (q, 0) & = &  \Gamma^{S}_{2} \f{N \gbar^2}{(2 \pi)^6 \chi_0} \int d^2 k d \omega d^2 l d\Omega~ \chi_s (l,\Omega)~ G(\omega,k)^2~ G(\omega+\Omega,k+l)~ G(\omega,k+q)  \nonumber \\
\Pi_{3} (q, 0) & = &  \Gamma^{S}_{3} \f{N^2 \gbar^3}{(2 \pi)^9 \chi^2_0} \int d^2 k d \omega d^2 k^\prime d \omega^\prime d l d\Omega~ \chi_s (l,\Omega) ~\chi_s (q+l,\Omega)~ G(\omega,k) ~ G (\omega, k+q) \nonumber \\
&& \qquad \qquad G(\omega+\Omega,k+l+q)~ G(\omega^\prime ,k^\prime)~ G(\omega^\prime ,k^\prime +q)~ G (\omega^\prime + \Omega, k^\prime + l +q) \nonumber \\
\Pi_{4} (q, 0) & = & \Gamma^{S}_{4} \f{N^2 \gbar^3}{(2 \pi)^9 \chi^2_0} \int d^2 k d \omega d^2 k^\prime d \omega^\prime d l d\Omega~ \chi_s (l,\Omega) ~\chi_s (l,\Omega)~ G(\omega,k)~ G (\omega, k+q) \nonumber \\
&& \qquad \qquad G(\omega+\Omega,k+l+q)~ G(\omega^\prime ,k^\prime) ~G(\omega^\prime +\Omega,k^\prime +l)~ G (\omega^\prime + \Omega, k^\prime + l + q)
\eea
\end{widetext}
The factors of $N$ come from the fermionic loops and the numerical
pre-factors from the following spin summations:
\begin{equation} \label{gammaspin}
\left\{
\begin{array}{lcl}
{\dst \Gamma^{S}_1} & = & {\dst \sum_{\alpha,\beta,\gamma,\delta} \sigma^Z_{\alpha \beta}{\bm \sigma}_{\beta \gamma} \sigma^Z_{\gamma \delta} {\bm \sigma}_{\delta \alpha}}  =  -2 \\  \\
{\dst \Gamma^{S}_2} & = & {\dst \sum_{\alpha,\beta,\gamma,\delta} \sigma^Z_{\alpha \beta}{\bm \sigma}_{\beta \gamma} \cdot {\bm \sigma}_{\gamma \delta} \sigma^Z_{\delta \alpha}}  =  6 \\ \\
{\dst \Gamma^{S}_3} & = & {\dst \sum_{\alpha,\beta,\gamma,\delta,\epsilon, \zeta}  \sigma^Z_{\alpha \beta} \left({\bm \sigma}_{\beta \gamma} \cdot {\bm \sigma}_{\delta \epsilon}\right) \left( {\bm \sigma}_{\gamma \alpha} \cdot {\bm \sigma}_{\zeta \delta} \right) \sigma^Z_{\epsilon \zeta}}  =  8 \\ \\
{\dst \Gamma^{S}_4} & = & {\dst \sum_{\alpha,\beta,\gamma,\delta,\epsilon,\zeta} \sigma^Z_{\alpha \beta}\left({\bm \sigma}_{\beta \gamma} \cdot {\bm \sigma}_{\epsilon \delta } \right)  \left({\bm \sigma}_{\gamma \alpha} \cdot {\bm \sigma}_{\delta \zeta} \right) \sigma^Z_{\zeta \epsilon}}  =  - 8
\end{array} \right.
\end{equation}

These four diagrams are related by pairs. To verify this,
it is useful to expand the products of Green's functions according to:
\begin{equation}
G(\omega_1,k_1)    G(\omega_2,k_2)     =    \frac{G(\omega_1,k_1)    -
G(\omega_2,k_2)}{G^{-1}(\omega_2,k_2) - G^{-1}(\omega_1,k_1)}.
\end{equation}

Applying this to $\Pi_{2} (q,0)$, we find that it splits into two parts.
In one part, the poles in $\epsilon_k$ are located in the same  half-plane,
leading to a vanishing contribution. The  remaining term in $\Pi_2 (q,0)$ is
related  to $\Pi_1 (q,0)$ in such  a way that:
\begin{equation} \label{relation_pi}
\Pi_{1}(q,0) = -\frac{2\Gamma_1}{\Gamma_2} \Pi_{2}(q,0).
\end{equation}
(see Appendix \ref{app:chi} for details).

Similarly, $\Pi_3 (q,0)$ and $\Pi_4 (q,0)$ are related as
\begin{equation} \label{relation_pi_1}
\Pi_{3}(q,0) = -\frac{\Gamma_3}{\Gamma_4} \Pi_{4}(q,0).
\end{equation}

Collecting all four contributions and using the relations between
pre-factors, we obtain:
\begin{eqnarray} \label{total_pi}
\Pi(q,0) &  = & \Pi_{A}(q,0)  + \Pi_{B}(q,0) \ ,  \\
\Pi_{A}(q,0) & = & \Pi_{1}(q,0)  + 2 \Pi_{2}(q,0) \nonumber \\
 & =  & 16 \f{N \gbar^2}{(2  \pi)^6} \int d^2 K  d \omega d^2  l d\Omega~ \chi_s (l,\Omega)  G(\omega,k)^2  \nonumber   \\
 &  &  \qquad \quad \times G(\omega+\Omega,k+l)  G(\omega,k+q) \ , \nonumber \\
\Pi_{B}(q,0) & = & \Pi_{3}(q,0)+ \Pi_{4}(q,0) \nonumber \\
& =  & 16 \f{N^2 \gbar^3}{(2 \pi)^9 \chi^2_0} \int d^2 k d \omega d^2 k^\prime d \omega^\prime d l d\Omega~ \chi_s (l,\Omega) \nonumber \\
&& \quad \times \chi_s (q+l,\Omega)~ G(\omega,k) ~ G (\omega, k+q) \nonumber \\
&& \quad \times G(\omega+\Omega,k+l+q)~ G(\omega^\prime ,k^\prime +q) \nonumber \\
&& \quad \times G(\omega^\prime ,k^\prime)~ G (\omega^\prime + \Omega, k^\prime + l +q)
\end{eqnarray}

\subsubsection{Fermi-liquid regime}

Away from criticality, the correlation length $\xi$ is finite, and at
low frequency, the system is in the Fermi-liquid regime. The fermionic
self-energy is $\Sigma (\omega) = \lambda \omega$, Eq. (\ref{self_fl}).

The spin susceptibility in this regime has been analyzed in
\cite{millis,chub_maslov,cmm,gal,bet,ch_masl_latest}. It was shown there that to the
lowest order in the interaction, $\Pi_B (q,0) = \Pi_A (q,0)$,
i.e., $\Pi (q,0) = 2 \Pi_A (q,0)$. Beyond leading order, $\Pi_B
(q,0)$ and $\Pi_A (q,0)$ are not equivalent but are of the same
sign and of comparable magnitude. For simplicity, we assume that
the relation $\Pi_B (q,0) = \Pi_A (q,0)$ holds in the whole Fermi
liquid regime. We will see below that even at the QCP, $\Pi_B
(q,0)$ and $\Pi_A (q,0)$ are quite similar (at QCP $\Pi_B (q,0)
\approx 1.3 \Pi_A (q,0)$).

Introducing $\cos \theta =\frac{\bf k \cdot l}{|{\bf k}| |{\bf l}|}$
and $\cos \theta' = \frac{\bf k \cdot q}{|{\bf  k}| |{\bf q}|}$, and
successively integrating over $|{\bf k}|$, $\omega$ and $\theta'$, (\ref{total_pi}) can
be rewritten as:
\beq \begin{array}{l}
{\displaystyle \Pi (q,0)   =   \frac{8  \gbar |q|}{\pi^3 (1+\lambda)v_F} \int_0^{\infty}  dz \int_0^{\frac{\pi}{2}} d\phi  \int_0^{\pi} d\theta } \\ \\
{\displaystyle \qquad \frac{1}{\frac{1}{\tilde{\gamma}   \xi^2}  +  \tan   \phi}~ \frac{\cos   \phi  \sin   \phi}{\left(  i\sin \phi-\cos \theta \cos  \phi \right)^2} } \\ \\
{\displaystyle \qquad \times~ \frac{z}{\sqrt{1+z^2 (\sin \phi
+ i \cos \phi \cos \theta)^2}}}  ,
\end{array}
\eeq
where we defined $\tilde{\gamma} = \frac{\gamma v_F}{1+\lambda}$, and
introduced the new variables $z$ and $\phi$,
which satisfy $z \cos \phi =  \frac{l}{q}$ and $z \sin \phi = \frac{(1+\lambda)\Omega}{v_F q}$.

The universal part of $\Pi (q,0)$ can be isolated by subtracting
from it the constant part $\Pi (0,0)$. The integral over $z$ then becomes convergent.
Integrating successively over $z$, $\phi$ and $\theta$, we obtain:
\begin{equation} \label{mod_q1}
 \Pi (q,0) = -\frac{4}{\pi^2} \frac{\gbar}{v_F (1+\lambda)} |q| \
{\cal H} \left( \frac{1 + \lambda}{\tilde{\gamma} \xi^2} \right) ,
\end{equation}
where ${\cal H}(0) = \frac{1}{3}$, and ${\cal H}(x \gg 1) \approx 2/(3x^2)$
We do recover the non-analytic $|q|$ correction to the static spin susceptibility
in $D=2$, as obtained in earlier studies \cite{bkv,millis,chub_maslov,cmm}.

Note that Eq. (\ref{mod_q2}) does not contradict the Fermi liquid
relation  $\chi_s (q \to 0,\omega=0) \propto (1 + F_{1,s})/(1 + F_{0,a})$,
where $F_{1,s}$ and $F_{0,a}$ are Landau parameters. The Fermi liquid theory
only implies that the static spin susceptibility saturates to a constant
value as $q \to 0$, but does not impose any formal constraint on the
$q-$dependence of $\chi_s(q,\omega)$.

As one gets closer to the QCP, $\lambda = 3 {\bar g}/(4\pi v_S \xi^{-1})$
 diverges and the pre-factor of the $|q|$ term vanishes as:
\begin{equation} \label{mod_q2}
 \Pi (q,0) = -\frac{16}{9 \pi} \xi^{-1} |q|.
\end{equation}
This is not surprising since the Fermi liquid regime extends
on a region of the phase diagram that shrinks and ultimately
vanishes as one approaches the QCP.

Now, two different scenarios are possible:
\begin{itemize}
\item  the divergence of $\xi$ at the QCP completely eliminates
the non-analyticity and the expansion of $\Pi (q,0)$ begins as $q^2$,
like in a bare spin susceptibility,
\item the self-energy $\Sigma (\omega) \propto \omega^{2/3}$ at
the QCP still leads to a non-analytic term $\Pi (q,0)\propto |q|^{\delta}$,
with  $1 < \delta <2$, which dominates over the bare $q^2$ term.
\end{itemize}
We show in the next subsection that the second scenario is realized,
and at the QCP, one has $\Pi (q,0)\propto |q|^{3/2}$.

\subsubsection{At criticality}

At the QCP, we have to take into account two new elements:
the bosonic propagator is massless ($\xi^{-1}=0$) and the fermionic
self-energy is no longer Fermi-liquid-like, it is given by (\ref{defs}).

The full calculation of $\Pi_A (q,0)$ and $\Pi_B (q,0)$ is presented in Appendix \ref{app:chi}. We just outline here the main steps of the computation and show where the $q^{3/2}$ term comes from.

Consider first $\Pi_A (q,0)$. Using (\ref{total_pi}) as a starting point,
and substituting the full form of the spin susceptibility, Eq. (\ref{specchi}),
 we then expand  $\epsilon_{{\bf k} +{\bf l}}$ and $\epsilon_{{\bf k} + {\bf q}}$,
 and integrate successively over $l_y$ (projection of ${\bf l}$
 perpendicular to ${\bf k_F}$) and $\epsilon_k$, leading to:
\beq \label{pi_qcp}
\begin{array}{l} {\displaystyle
\Pi_A (q,0)  = 16i \frac{m\gbar^2}{(2 \pi)^5}  \int_{0}^{2 \pi}
d\theta  \int_{-\infty}^{+\infty} dl_x  \int_{-\infty}^{+\infty} d
\Omega \int_{-\Omega}^{0} d \omega } \\ \\ {\displaystyle
~\frac{1}{\left( i\Sigma(\omega+\Omega)-i\Sigma(\omega)-v_F l_x
\right)^2}  h \left(
\frac{\sqrt{l^2_x+ c^2 \Sigma^2 (\Omega)}}{(\gamma |\Omega|)^{1/3}}\right) } \\ \\ { \displaystyle
\times \frac{1}{(\gamma |\Omega|)^{1/3}}~\frac{1}{i\Sigma(\omega+\Omega)-i\Sigma(\omega)-v_F l_x +v_F q
\cos \theta} } ,
\end{array} \eeq
where $c \simeq 1.19878$ (see \ref{apr25_7}), and $h(x)$ is
the spin susceptibility $\chi_(l, \Omega)/\chi_0$ integrated over
the momentum component $l_y$. The asymptotic behavior of $h(x)$ is given by:
\begin{equation} \label{hofx}
\begin{cases}
h(x \gg 1)= \frac{\pi}{x}  &  \\ h(x \ll 1)= \frac{4 \pi}{3 \sqrt{3}}
+ (\log 2 -1) x^2 - \frac{x^2 \log x^2}{2} &
\end{cases}
\end{equation}

Since the integrand in (\ref{pi_qcp}) has poles in $l_x$ located in
the same half-plane, the only non-vanishing contributions to $\Pi$
comes from the non-analyticities in $h(x)$.

There are two non-analyticities in $h(x)$.
The first one comes from the $1/x$ behavior at large $x$,
which extends to $x \sim 1$. This is a conventional non-analyticity
associated with bosons vibrating near their own mass shell,
since at  $x \sim 1$, $l_x \sim l_y \sim (\gamma |\Omega|)^{1/3}$.

Subtracting the universal $\Pi_A^{(a)}(0,0)$,
expanding in $q$ in (\ref{pi_qcp}) and substituting
$l_x \sim (\gamma |\Omega|)^{1/3}$, we find for this contribution to $\Pi$:
\begin{eqnarray} \label{nonan2}
\delta \Pi_A^{(a)} & \propto & q^2 \frac{m\gbar^2}{v_F^3
 \gamma^{5/3}} \int_0^{\Omega_{\rm max}} \frac{d\Omega}{\Omega^{2/3}}
 \nonumber \\ & \propto & \sqrt{\alpha} q^2  ,
\end{eqnarray}
where $\Omega_{\rm max} \sim \sqrt{\gamma v_F^3}$.

We see that the integration over the momentum range relevant to the
$z=3$ scaling regime  yields a regular $q^2$ contribution to the static
susceptibility. Not only this contribution  is regular in $q$,
but it is also small in $\alpha$. This result is  similar to the
one we obtained in (\ref{migdal2}) for the static vertex at a vanishing momentum.

However, Eq. (\ref{hofx}) shows that $h(x)$ has a non-analytic
$x^2 \log x$ term already at small $x$, i.e. far from the bosonic mass shell:
the branch cut associated with the logarithmic term emerges
at $v_F l_x \sim \Sigma (\omega)$. The typical value of $l_y$ in
the integral that leads to this $x^2 \log x$ term is also of the same order,
although larger in the logarithmic sense. This implies that this second
non-analyticity comes from a process in which the bosons are  vibrating
near a fermionic mass shell and far from their own.

Furthermore, this logarithmic term in (\ref{hofx}) comes
exclusively from the Landau damping term in the bosonic propagator --
the $q^2$ term in $\chi (l, \Omega)$ can be safely omitted. Indeed, one has:
\begin{equation} \label{cflandau}
\int_{-\Lambda}^{\Lambda} dl_y \f{1}{\frac{\gamma
    |\Omega|}{\sqrt{l_x^2+l_y^2}}} = f(\Lambda) - \f{l_x^2 \log
  l_x^2}{2 \gamma \Omega} .
\end{equation}

\begin{figure}[tbp]
\centerline{\includegraphics[width=3.4in]{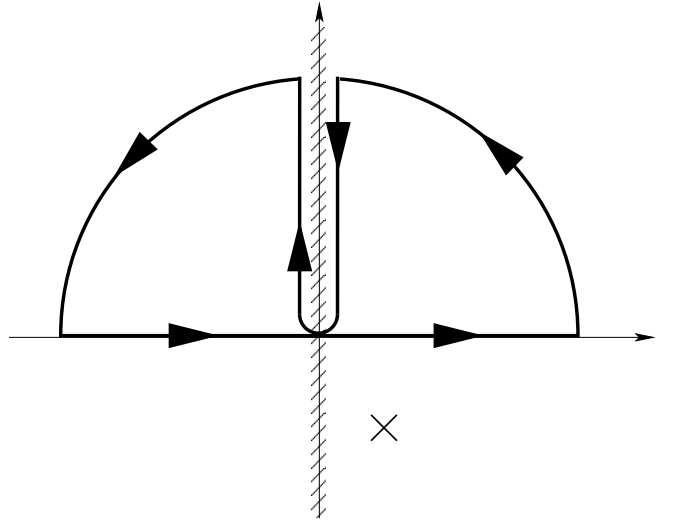}}
\caption{Integration contour: the hatched region stands for a branch cut, and the cross for a pole.}
\label{contour}
\end{figure}

Substituting the logarithmic term from (\ref{hofx}) into (\ref{pi_qcp}),
we subtract the non-universal constant term $\Pi_A^{(b)} (0,0)$, which
makes the integral over $l_x$ convergent. Introducing $z=l_x/(c \Sigma(\Omega))$,
one can perform the integral over $z$ over the contour of Fig. \ref{contour},
which leads to the following contribution to $\Pi$:
\begin{align}
&\Pi_A^{(b)} (q, 0) - \Pi_A^{(b)} (0, 0) =  \f{4 N m \gbar^2}{c^2 \pi^4 \gamma v_F} q^2 \int_0^{\pi/2}
d\theta \int_{1}^{+\infty} d y \nonumber \\
& \quad \int_{0}^{+\infty} \f{d\omega}{\Sigma(\Omega)^2} \int_0^{1} dw \f{\cos^2\theta}{\left[c^{-1}\left((1-w)^{2/3}+w^{2/3}\right)+y\right]^3} \nonumber \\
 & \qquad  \times ~\f{1-y^2}{\left[y+c^{-1}\left((1-w)^{2/3}+w^{2/3}\right)\right]^2+\left(\frac{v_F q \cos \theta}{c\Sigma(\Omega)}\right)^2} , \label{pi_qcp2}
\end{align}
where we defined $w=\omega/\Omega$.

Introducing the new variables $t=\left( \frac{c \Sigma(\Omega)}{v_F q \cos \theta} \right)^{3/2}$ and $v=t\left[ y+c^{-1} \left( (1-w)^{2/3}+w^{2/3}  \right)\right]$, it becomes possible to separate the integrals, leading to the following final result:
\bea
\Pi_A^{(b)} (q, 0) - \Pi_A^{(b)} (0, 0) & = & - 0.8377   q^{3/2} \f{Nm\gbar^2}{\pi^4 \gamma v_F^{3/2} \omega_0^{1/2}} \nonumber\\
& =& -0.1053~ \sqrt{k_F}~ q^{3/2} . \label{pi_qcp3}
\eea
We emphasize that this dependence comes from bosonic modes
vibrating at the fermionic mass-shell. This explains why
the result of (\ref{pi_qcp3}) is not small in $\alpha$, as this small parameter measures the softness of
the mass-shell bosons compared to the mass-shell fermions.

The integrals for $\Pi_B (q, 0)$ cannot be exactly evaluated analytically, but
 an approximate calculation in Appendix E
 yields
\begin{equation} \label{pi_qcp_2}
\Pi_B^{(b)} (q,0) - \Pi_B^{(b)} (0,0)  = - 0.14 q^{3/2} \sqrt{p_F}  ,
\end{equation}
such that the total
\begin{equation} \label{pi_qcp_3}
\Pi (q,0) - \Pi (0,0) = - 0.25 q^{3/2} \sqrt{p_F} .
\end{equation}

We see that $\Pi (q,0)$ is still non-analytic in $q$ and the
prefactor is negative. At small $q$, the negative $q^{3/2}$ term
well exceeds the regular $q^2$ term. This implies that {\it  the static spin susceptibility is negative at small momenta, i.e. a ferromagnetic QCP is unstable}. We discuss the consequences of this instability in the concluding section. The momentum $q_{min}$ below which $\chi_s$ is negative is determined by
\begin{equation} \label{pi_qcp_4}
\chi^{-1} (q_{min})  \propto  q_{min}^2 - 0.25 q_{min}^{3/2} \sqrt{p_F} =0,
\end{equation}
which gives $q_{min} = 0.06 k_F$. Parametrically, $q_{min}$ is
of order $k_F$, which is the largest momentum scale in our problem.
Strictly speaking, this suggests that the whole quantum-critical theory
for the ferromagnetic case is not valid, for the quantum
critical behavior extends up to energies of order $\omega_0$, i.e. up to
momenta of order $q \leq \omega_0/v_F \sim \alpha^2 k_F \ll k_F$.
Numerically, however, $q_{min}$ is much smaller than $k_F$. This implies
that for reasonable values of $\alpha$, there exists an intermediate momentum
and energy range  $q_{min} < q < \omega_0/v_F$ where the system is in the
quantum-critical regime, but the static spin susceptibility  is still positive.

The $q^{3/2}$ non-analyticity can also be viewed as emerging from
$q \log q$ momentum dependence of the static vertex. Using
 Eq. (\ref{stat4}), one can rewrite:
\begin{align}
\Pi (q,0) \sim N\gbar \int d^2 k d\omega \frac{1}{i\Sigma(\omega)-\epsilon_k}
\frac{\left. \frac{\Delta g}{g} \right|_{q,\Omega=0}}{i\Sigma(\omega)-\epsilon_{k+q}}
\end{align}
Performing the contour integration over $\epsilon_k$, and changing
variables into $y=\epsilon_k/\Sigma{\omega}$ and
$t = \sqrt{\omega_0}\omega/((v_F q)^{3/2})$, we obtain:
\beq
\Pi (q,0) \sim \sqrt{k_F} q^{3/2}
\eeq
We note in this regard that the non-analytic momentum dependence
of the fermion-boson static vertex also comes from bosons vibrating
near fermionic mass shell, i.e., it emerges due to the same physics as we outlined above.

\subsection{Temperature dependence of the static uniform spin susceptibility}

In this section, we show that the static uniform susceptibility is
negative at finite temperature above a ferromagnetic QCP. 
 To demonstrate this, we compute the static uniform $\Pi_A (q=0, \omega =0, T) = \Pi_A (T)$, assuming that $\xi^{-1} =0$.
 The contribution from $\Pi_B (T)$ is of the same sign and comparable in magnitude. We have
\beq \label{finiteT2}
\begin{array}{l}
{\displaystyle \Pi_A (0,T) = 16i N
 \f{m \gbar^2}{(2 \pi)^3} T \sum_{p \neq 0} \Omega_p \int d^2 q} \\ \\
{\displaystyle \times \frac{1}{q^2+\frac{\gamma |\Omega_p|}{q}} \frac{1}{\left( i\Sigma(\Omega_p) -v_F q_x -\frac{q_y^2}{2m} \right)^3}} .
\end{array}
\eeq

Since the poles in $q_x$ from the fermionic Green's functions are all in the same half-plane, one expects that $q_x^{\rm Typ} \sim (\gamma \Omega)^{1/3}$ and thus dominates over the $q_y^2$ term, which in turn can be neglected in the fermionic Green's functions. It  then becomes 
 possible to perform the integral over $q_y$, which gives
\bea \label{finiteT3}
 &&{\displaystyle \Pi}_A (0,T) =  2i N
 \f{m \gbar^2}{\pi^3 \gamma v^3_F} T \sum_{p \neq 0} \frac{\Omega_p}{|\Omega_p|} \nonumber \\
&& \int_0^{(\gamma \Omega_p)^{1/3}} d q_x~ 
\frac{q^2_x \log |q_x|}{(i {\tilde \Sigma} (\Omega_p) -v_F q_x)^3}
\eea
Integrating over the half-space where there is no triple pole,
we find that the  integral is determined by the branch cut in $\log |q_x|$. Evaluating the integral
we obtain
\beq
\int\frac{d q_x q^2_x \log |q_x|}{(i {\tilde \Sigma} (\Omega_p) -v_F q_x)^3}
= {\rm sign} \Omega_p ~ \frac{i \pi}{v_F^3} \log \left ( \frac{E_F}{|\tSigma (\Omega_p)|} \right ) \ . \eeq
Thus
\bea
 \label{finiteT31}
{\displaystyle \Pi}_A (0,T) = \frac{2 N}{3}
 \f{m \gbar^2}{\pi^2 \gamma v^3_F} T \sum_{p \neq 0}
 \log{\frac{E_F}{|\Omega_p|}}
\eea
To perform the summation over p,
 we notice that when the summand does not depend on $p$  
\beq\label{finiteT6}
T \sum_{- \Lambda/ T}^{\Lambda/T} A = 2 A \Lambda \ , \eeq 
 is independent on $T$. Then the same sum but without $p=0$ term 
will be $2 A \Lambda - AT$. 
Using this, we obtain with logarithmic accuracy 
\bea
T \sum_{p \neq 0}
 \log{\frac{E_F}{|\Omega_p|}} = - T  \log{\frac{E_F}{T}} + ...
\eea
where dots stand for  $O(T)$ terms.
Substituting this into (\ref{finiteT31}), we obtain
 the final result:
\beq \label{finiteT4}
\Pi_{A} (0,T) = - \frac{2 k_F^2}{3\pi^2} \alpha \frac{T}{E_F}
 \log \left(\frac{E_F}{T} \right) .
\eeq
Although small, because of the prefactor in $\alpha$, this term dominates at low temperature over any regular $T^2$ term. 
The sign of this $T \log E_F/T$ term is opposite to the sign of a conventional
correction to scaling HMM theory, which  comes
 from a constant part of the prefactor for the $\phi^4$ term. 
 In the HMM theory the
temperature dependence of the spin susceptibility is $b T^{d+z-2}{z}$, $b >0$, 
which in $d=2$ leads to a linear in T with positive prefactor.
The negative sign of non-analytic temperature correction in 
(\ref{finiteT4}) implies that the static spin susceptibility is negative 
right above the QCP.  This is another indication that the ferromagnetic QCP 
 is unstable.

\begin{figure}[tbp]
\centerline{\includegraphics[width=1.7in]{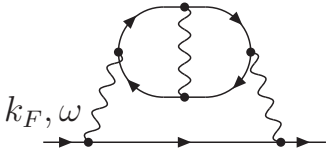}}
\caption{Three-loop contribution to the fermionic self-energy}
\label{3loopsigma}
\end{figure}

\subsection{Self-energy at the three-loop level}

Finally, we show how one can detect the instability of a ferromagnetic
QCP from an analysis of higher order self-energy diagrams.
This analysis is complimentary to the calculations hat we already
done in the previous subsections. We show
 that upon inserting the contributions from the diagrams
 presented in Fig. \ref{2loop} into the fermionic self-energy,
 we obtain  series of singular corrections in powers of $\omega_{min}/\omega$,
 where $(\gamma \omega_{min})^{1/3} =  q_{min}$, and $q_{min}$ is the scale
 at which the static susceptibility $\chi_s (q)$ becomes negative, Eq. (\ref{pi_qcp_4}).

To illustrate this, we consider one of the three-loop diagrams,
represented in Fig. \ref{3loopsigma}. In the case of a spin interaction,
we obtained a finite result after collecting the various diagrams,
so we restrict ourselves here with just one of these contributions.
The analytic form of the diagram in Fig. \ref{3loopsigma} writes:
\begin{align}
\Sigma_3 (\omega) \sim & \gbar \int d\omega_1 d^2 q_1 \frac{A(q_1,\omega_1)}{i\Sigma(\omega-\omega_1)-v_F q_{1x}-\frac{q_{1y}^2}{2m}} \nonumber \\
 & \qquad \times \left( \frac{1}{q_1^2+\frac{\gamma |\omega_1|}{q_1}} \right)^2 ,
\end{align}
where $A(q_1,\omega_1)$ is the dynamic fermionic bubble given by:
\begin{align}
A(q_1,\omega_1) & \sim N \gbar^2 \int d^2 k d\Omega \int d^2 q_2 d\omega_2 \frac{1}{q_2^2 + \frac{\gamma |\omega_2|}{q_2}} \nonumber \\
 & \times ~\frac{1}{i\Sigma(\Omega)-\epsilon_k}~\frac{1}{i\Sigma(\Omega+\omega_1+\omega_2)-\epsilon_{k+q_1+q_2}} \nonumber \\
 & \times ~\frac{1}{i\Sigma(\Omega-\omega_1)-\epsilon_{k+q_1}}~ \frac{1}{i\Sigma(\Omega+\omega_2)-\epsilon_{k+q_2}} . \nonumber
\end{align}

Approximating $A(q_1, \omega_1)$ by its singular static part $q^{3/2}_1 \sqrt{k_F}$ and
substituting into (\ref{3loopsigma}) we obtain:
\begin{align}
\Sigma_3 (\omega) \sim & \gbar \sqrt{k_F} \int d\omega_1 d^2 q_1  \left( \frac{q_1}{q_1^3+\gamma |\omega_1|} \right)^2 \nonumber \\
& \qquad \times \frac{q_1^{3/2}}{i\Sigma(\omega-\omega_1)-v_F q_{1x}-\frac{q_{1y}^2}{2m}} .
\end{align}

A simple analysis of this expression shows that the dominant contribution
to $\Sigma_3 (\omega)$ comes from $q_{1x} \sim q_{1y} \sim (\gamma \omega_2)^{1/3}$ since
the integral over $q_{1x}$ is determined by the branch cut in the bosonic propagator.
One can then safely drop the quadratic term in the fermionic propagator and integrate
over the angle $\theta$ between ${\bf k_F}$ and ${\bf q_1}$. This leads to:
\begin{align}
\Sigma_3 (\omega) \sim & \gbar \sqrt{k_F} \int_0^{\omega} d\omega_1 \int dq_1 ~\frac{q_1^{9/2}}{\left( q_1^3 +\gamma \omega_1 \right)} \nonumber \\
& \qquad \times \frac{1}{\sqrt{(v_F q_1)^2+\Sigma(\omega)^2}} \nonumber \\
\sim & \frac{\gbar \sqrt{k_F}}{v_F \sqrt{\gamma}} \int_0^{\omega} \frac{d\omega_1}{\sqrt{\omega_1}} .
\end{align}

Collecting the pre-factors, we find: \beq \label{3loopsigma3}
\Sigma_3 (\omega) \sim \left( \gbar \omega \right)^{1/2} . \eeq

We see that the  non-analyticity in the static spin susceptibility
feeds back into the fermionic self-energy leading to a contribution from
the three loop self-energy whose frequency dependence is more singular
than the $\omega^{2/3}$ dependence that we obtained assuming that the static
susceptibility is regular. Comparing these two contributions, we see that
they become comparable at a frequency:
\beq
\omega_0^{1/3} \omega_{min}^{2/3} \sim \sqrt{\gbar \omega_{min}} ~ \Longrightarrow ~ \omega_{min} \sim \frac{q^3_{min}}{\gamma} \sim \frac{E^2_F}{\gbar} .
\eeq
where $q_{min}$ is given by (\ref{pi_qcp_4}).
Parametrically, $q_{min}$ is not small since $q_{min} \sim k_F$,
and $\omega_{min} \sim E_F/\alpha$ is larger than $E_F$. However, $q_{min} \sim 0.06 k_F$
is small numerically so that $\omega_{min}$ is four orders of magnitude smaller
than $ E_F/\alpha$.

\subsection{Summary}

To summarize, we found that the Eliashberg theory for an $SU(2)$
symmetric ferromagnetic QCP has to be extended to include extra
singular terms into both the spin susceptibility and the fermionic
self-energy. These terms originate from the ``anti-Migdal''
processes in which slow bosons are vibrating near the fermionic mass
shell.  Physically, these
 extra processes originate from the dynamic long-range interaction
 between fermions, which is still present at the QCP despite the fact that fermions
 are no longer good quasi-particles.

We demonstrated that these extra non-analytic terms can
be understood in the framework of HMM $\phi^4$ theory of
quantum criticality. We showed that the prefactor for the $\phi^4$ term is
non-analytic and depends on the interplay between momentum and frequency.
The non-analytic bosonic self-energy is the feedback from the
non-analytic $\phi^4$ term to the quadratic $\phi^2$ term.

We found that these extra terms in the Eliashberg theory make a
ferromagnetic QCP unstable below a certain momentum/energy scale.
We detected the instability
 by analyzing the momentum and temperature dependence of the spin
 susceptibility, and also the fermionic self-energy at three-loop order.

\section{Static susceptibility in the non-SU$(2)$ symmetric case}

The problem of fermions interacting with bosonic collective modes
with a propagator similar to the one we considered, is quite general,
and one can wonder to what extent our analysis for a ferromagnetic case
can be extended to the case of an interaction with charge fluctuations,
a nematic QCP, or a ferromagnetic instability with Ising symmetry.

The essential difference between these cases and the ferromagnetic one
lies in the symmetry of the order parameter. In an SU$(2)$ spin-symmetric
ferromagnetic case, the order parameter (magnetization) is a three-dimensional
vector, while
 in the other cases, it is a scalar.

As we already discussed, the Eliashberg theory, and the analysis
of its validity at the two-loop level can be carried out equally for
systems with vector and with scalar order parameter: the only unessential
difference is in the numerical prefactors. On the other hand, the evaluation
of the corrections to the static susceptibility leads to different results
for scalar and vector order parameters, as we now demonstrate.

\subsection{Ising ferromagnet}

Consider first the situation of a magnetically-mediated interaction, where we
change the spin symmetry of the bosons from SU$(2)$ to Ising.
The Ising case was argued to be relevant for metamagnetic quantum critical
points \cite{millis_schoffield}.

The use of Ising spins doesn't change the expression for the Green's
functions but replaces the Pauli matrix ${\bm \sigma}$ at the
fermion-boson vertex by $\sigma^z$. As a consequence, the computations
performed for the SU$(2)$ case still hold, but the interplay between
different diagrams changes because of a change in the numerical prefactors.
In particular, instead of Eq. (\ref{gammaspin}) we now have:
\bea
\Gamma^{\rm Ising}_1 & = & \Gamma^{\rm Ising}_2 = 2, \nonumber \\
\Gamma^{\rm Ising}_3 & = & \Gamma^{\rm Ising}_4 = 0.
\eea

Under these circumstances, the non-analytic contributions from the
diagrams in Fig. \ref{2loop} cancel each other out. As a result, the
static spin susceptibility remains analytic and scales as
$\chi^{-1} (q) \propto q^2$ with a positive prefactor at the QCP.

This result can be extended to the case of a nematic instability,
following the same arguments.

\subsection{Charge channel}

For a charge vertex, one has to replace the Pauli matrices
present at the vertex by Kronecker symbols $\delta_{\alpha \beta}$. We then have:
\bea
\Gamma^{\rm Charge}_1 & =& \Gamma^{\rm Charge}_2 = \sum_{\alpha,\beta,\gamma,\delta} \delta_{\alpha \beta} \delta_{\beta \gamma} \delta_{\gamma \delta} \delta_{\delta \alpha} = 2, \nonumber \\
\Gamma^{\rm Charge}_3 & =&  \Gamma^{\rm Charge}_4 = \sum_{\alpha,\beta,\gamma,\delta,\epsilon,\zeta} \delta_{\alpha \beta}   \delta_{\beta \gamma}  \delta_{\delta \epsilon} \delta_{\gamma \alpha} \delta_{\zeta \delta} \delta_{\epsilon \zeta} = 4. \nonumber
\eea

Substituting these $\Gamma^{\rm Charge}$ into the expressions for $\Pi$,
we find that the diagrams $\Pi_1$ and $\Pi_2$, as well as $\Pi_3$ and $\Pi_4$
cancel each other out. This leaves only regular $q^2$ term in the static charge susceptibility.

\subsection{Physical arguments}

The cancellation of the non-analytic terms in the charge
susceptibility is not a direct consequence of the conservation
laws. These laws impose constraints (Ward identities) on the
behavior of the susceptibilities in the opposite limit $q=0$,
$\omega \neq 0$ ($\chi_c (q=0,\omega)$ vanishes as a uniform
perturbation cannot affect a time independent, conserved
quantity).

Instead, the absence of the non-analytic terms in the charge channel is related
to the fact that this susceptibility measures the response of the system to a change
in the chemical potential. We showed that the origin of the singular behavior of
the static spin susceptibility lies in the Landau damping term in the bosonic
propagator (see  (\ref{cflandau}), (\ref{pi_qcp_3}) ). The Landau damping does
not depend in a singular way on $k_F$ (i.e. on the density of electrons),
and therefore there is no singular response of the system to a change
in the chemical potential.

Conversely, the effect of a magnetic field on the Landau damping is singular:
for a fermionic bubble with opposite spin projections of the two fermions,
 $\Pi_{\pm} \sim |\Omega|/\sqrt{\Omega^2 + (v_F l)^2}$ 
 is replaced by $|\Omega|/\sqrt{(\Omega + 2i \mu_B H)^2 + (v_F l)^2}$ 
 in the presence of a small
magnetic field $H$. As a consequence, if the susceptibility is evaluated
 as a second derivative of the  thermodynamic potential with respect to
 $H$. Taking the second derivative of $\Pi_{\pm}$ over $H$ and setting $H=0$ later, one obtains, for $v_F l \gg \Omega$, a non-analytic 
 $d^2\Pi_{\pm}/dH^2 \sim \Omega/l^3$. This non-analyticity gives rise 
 to $q^{3/2}$ term in the static spin susceptibility~\cite{bet,ch_masl_latest}. For an Ising ferromagnet,
 this effect does not exist as there are no bubbles with opposite spin projects
 in the theory. 

The above reasoning shows that non-analyticity appears in the spin response but
not in the charge one. To further verify this argument, we
computed the subleading,  three-loop diagrams for the charge susceptibility and
found that the non-analytic contributions from individual diagrams
all cancel out. We present the calculations in Appendix
\ref{app:charge}.

The same argument holds for the self-energy at the three-loop and higher orders.
 The singular $\omega^{1/2}$ term obtained in Eq. (\ref{3loopsigma3}) appears
 in individual diagrams, but in the case of a QCP in the charge channel
 (or an Ising QCP in the spin channel) the total singular contribution cancel out.

\subsection{Summary}

To summarize, the extra singular additions to the Eliashberg
theory are specific to $SU(2)$ spin case and all cancel out
for the charge QCP, the gauge-field problem, the nematic QCP, and the spin QCP for
a scalar (Ising) order parameter.

\section{Conclusions}

We have constructed a fully controllable large-$N$ quantum
critical theory describing the interaction of fermions with gapless
long-wavelength collective bosonic modes. Our approach, similar but not identical
to the Eliashberg theory for the electron-phonon interaction, allows us
to perform detailed calculations of the fermionic self-energy and
the vertex corrections at the QCP.

We constructed a controllable expansion at the QCP as follows:
we first created a new, non Fermi-liquid ``zero-order'' theory by solving
a set of coupled equations for the fermionic and the bosonic propagators,
neglecting the vertex corrections as well as the momentum dependence
of the fermionic self-energy, and then analyzed the residual interaction
effects in a perturbative expansion around this new zero-order theory.

We have proved that this approach is justified under two
conditions: (i) the interaction ${\bar g}$ should be smaller than
the fermionic bandwidth (which we assumed for simplicity to be of
the same order as $E_F$), and (ii) either the band mass $m_B$
should be smaller than $m = p_F/v_F$, or the number of fermionic
flavors $N$ should be large. When $N = O(1)$ and $m_B \sim m$,
the corrections are of order $O(1)$. We found that the corrections
that are small in $\gbar/E_F$ come from bosons near their
resonance, as in the Eliashberg theory for the electron-phonon
interaction. The corrections small in $m_B/(Nm)$ come from bosons
for which one of the momentum component (the larger one) is near
the bosonic resonance, while the other component is close to the
fermionic mass-shell.

For an SU$(2)$-symmetric quantum critical point towards ferromagnetic
ordering, we found that there exists an extra set of singular
renormalizations which come from bosons with both momentum components
vibrating near the fermionic mass-shell. These processes
 can be understood as a consequence of an effective long-range dynamic
 interaction between quasi-particles, generated  by the Landau damping term.
These singular renormalizations are not
 small and have to be included into the Eliashberg theory. They give
 rise to a negative non-analytic $q^{3/2}$ correction to the static spin
 susceptibility, signaling that the ferromagnetic QCP is unstable.

We also demonstrated that the non-analytic $q^{3/2}$ term can be understood
in the framework of the $\phi^4$ theory of quantum-criticality. We showed how
the effective long-range dynamic interaction between fermions affects the
structure of the $\phi^4$ theory, once fermions are integrated out:
we found that the prefactors of the $\phi^3$ and $\phi^4$ terms appearing in
the effective action are non-analytic and depend on the interplay between
the typical external momentum and frequency.

We showed that the non-analytic corrections to the bosonic propagator
are specific to the SU(2)-symmetric case when the order parameter is a vector.
For systems with a scalar order parameter, like a QCP in the charge channel,
a nematic QCP, or a ferromagnetic QCP with Ising symmetry, the $q^{3/2}$
contributions from individual diagrams cancel out in the full expression of the susceptibility.

The consequences of the instability of the ferromagnetic QCP still needs
to be fully understood. There are two possible scenarios for the behavior of
the system: either the ferromagnetic transition becomes first order~\cite{belitz_rmp,braz}, or the
instability occurs at a finite $q$, leading to a second order transition towards
an incommensurate state~\cite{bkr} The full analysis of these two scenarios is clearly called on.

This work was supported by NSF-DMR 0240238 (A. V. Ch.).
J. R. and C. P. are supported by an ACI grant of the French Ministry of Research.  We acknowledge useful discussions with Ar. Abanov,  C. Castellani, A. Castro-Neto,  C. DiCastro, E. Fradkin,  M. Grilli,  D. Khveshchenko, S. Kivelson, M. Lawler, 
D. Maslov, W. Metzner, A. Millis, R. Ramazashvili, J. Schmalian and O. Starykh. 

\appendix

\section{Bosonic self-energy}

\label{app:boson}

In this section, we compute the bosonic self-energy at the one-loop level, in the case of both free and fully renormalized fermions. We prove that for an external bosonic momentum on the bosonic mass-shell, this self-energy becomes independent on the actual form of the fermionic self-energy, and reduces to the usual Landau damping term, whereas an extra term has to be included if this same external momentum is on the fermionic mass shell.

After performing the sum over spin matrices, we are left with the
following expression:
\begin{equation}
\Pi (\bq,\Omega) = 2 N \bar{g} \int \f{d^{2}k~ d\omega}{(2\pi)^3}~
G(\bk,\omega)~ G(\bk+\bq,\omega+\Omega).
\end{equation}
Introducing the angle  $\theta$ defined by
$\epsilon_{k+q}=\epsilon_k + v_F q \cos \theta$, this writes:
\begin{align}
\Pi (\bq,\Omega) & = N \f{\gbar m}{4 \pi^3} \int d\omega~ d\epsilon_k~
 d\theta ~\f{1}{i (\omega+\Sigma(\omega))-\epsilon_k} \nonumber \\ &
 \times \f{1}{i(\omega+\Omega+\Sigma(\omega+\Omega))-\epsilon_k-v_F q
 \cos \theta}.
\end{align}

Proceeding with a contour integration over $\epsilon_k$, and noticing
that $\omega$ and $\Sigma(\omega)$ have the same sign, we get:
\begin{align}
\Pi (\bq,\Omega) & =  i N \f{\bar{g} m}{2 \pi^2} \int_{- \infty}^{+
\infty} d\omega~ \int_0^{2 \pi} d\theta \left( \theta(\omega+\Omega) -
\theta(\omega) \right) \nonumber \\ & \times \f{1}{i(\Omega +
\Sigma(\Omega+\omega)-\Sigma(\omega))-v_F q \cos \theta}.
\end{align}

Performing the integration over $\theta$, and rearranging a little bit
the integration over $\Omega$, we are left with:
\begin{align}
\Pi (\bq,\Omega) & = N \f{m \bar{g}}{\pi v_F} \int_{0}^{\Omega} d\omega~\mbox{Sign}(\Omega)  \nonumber \\
 & \times \f{1}{\sqrt{(v_F q)^2+(\Omega +\Sigma(\Omega-\omega)+\Sigma(\omega))^2}}.
\end{align}

We know from direct perturbative calculation that the fermionic
self-energy goes like $\Sigma(\omega) = \omega_0^{1/3} \omega^{2/3}$,
where $\omega_0 \sim \alpha \gbar$. It follows that if the external
bosonic momentum is on the bosonic mass-shell, i.e. if
$q \sim \Omega^{1/3}$, it dominates over the frequency-dependent
term in the integral, so that:
\begin{eqnarray}
\Pi (\bq,\Omega) & = &  N \f{m \bar{g}}{\pi} \int_{0}^{\Omega} d\omega~
\f{\text{Sign}(\Omega)}{v_F q} \nonumber \\ ~ & = & N \f{m \bar{g}}{\pi v_F}~ \f{|
\Omega |}{q}. \label{ldpg}
\end{eqnarray}

We recover here the expression of the Landau damping term with a
prefactor  depending on the details of the interaction considered in
our model. This  result is independent on the fermionic self-energy
provided that $v_F q \gg (\Omega+\Sigma(\Omega))$.

However, if the external bosonic momentum is on the fermionic mass shell,
i.e. if $v_F q \sim \Sigma(\Omega)$, the frequency-dependent term can no longer
be neglected. Defining $z=\frac{\omega}{\Omega}$, one then has:
\begin{align}
\Pi (\bq,\Omega) & =  N \f{m \bar{g}}{\pi}~ \left| \f{\Omega}{\Sigma(\Omega)} \right| \nonumber \\
& \times\int_0^1 dz~\f{1}{\sqrt{\left( \frac{v_F q}{\Sigma(\Omega)} \right)^2 + \left( z^{2/3} +(1-z)^{2/3} \right)^2}}.
\label{apr25_7}
\end{align}

This formula is of limited use as it is modified by vertex corrections.
 For the calculations of the static spin susceptibility,
we will actually only
need the leading $O(1/q)$ and the subleading, $O(1/q^3)$  terms in
  $\Pi (q, \Omega)$  We will show in Appendix C  that
 these two  terms still can be evaluated without vertex
 corrections (see Eq. (\ref{mon_2}) below).
 Expanding (\ref{apr25_7}) in $1/q$ to first two orders,
we find, at $\Omega < \omega_0$,
\beq
\Pi (\bq, \Omega) = \gamma \frac{|\Omega|}{q}~\left(1 -
\frac{c^2 \Sigma^2 (\Omega)}{v^2 q^2} \right)
\label{mon_1}
\eeq
where $c \sim 1.19878$.  To simplify writing, we
  will be using the expression for the polarization operator
 with $\Omega$ dependence in the denominator, i.e., plug $\Sigma^2$ term back into denominator, i.e.,
\beq \label{specchi}
\Pi (\bq,\Omega) = N \f{m \bar{g}}{\pi}~ \f{|\Omega|}{\sqrt{(v_F q)^2+c^2 \Sigma(\Omega)^2}},
\eeq
but one should keep in mind that we will actually only need $1/q$ and $1/q^3$ terms in the calculations.

\section{Fermionic self-energy}

\label{app:fermion}

In this section, we compute the fermionic self-energy at the one,
two and three loop levels, for an external momentum taken to be
on the Fermi surface. We also analyze the momentum dependence
of the one-loop fermionic self-energy.

\subsection{One-loop}

\subsubsection{At the Fermi level}

After summing over the spin matrices, the fermionic
self-energy at the one-loop level is given by:
\begin{eqnarray}
\Sigma (\omega) & = & 3 i g^2 \int \f{d^2q~d\Omega}{(2 \pi)^3}
 G({\bf k_F}+\bq,\omega+\Omega) \chi(\bq,\Omega) \nonumber \\ ~ & = &  \f{3
 i g^2}{(2 \pi)^3} \int d\Omega~dq~d\theta ~
 \f{\chi_0}{\xi^{-2}+q^2+\gamma \frac{| \Omega |}{q}} \nonumber \\ & &
 \times \f{q}{i(\omega+\Omega+\Sigma(\omega+\Omega))-v_F q \cos \theta},
\end{eqnarray}
where we defined $\theta$ as the angle between $\bk$ and $\bq$,
and considered an external fermionic momentum $k \simeq k_F$.
The $i$ prefactor comes from the convention $G^{-1} = G_0^{-1} +
i\Sigma$.

Since the pole in $q$ from the fermionic propagator is in a definite
half-plane, the integral in $q$ is dominated by poles coming from the
bosonic Green's function, so that one can perform the integral over
the angular variable and simplify the result as follows:
\begin{align}
\Sigma (\omega) = & - \f{3 g^2}{(2 \pi)^2} \int d\Omega~dq~
\f{\chi_0}{q^3 + \gamma| \Omega |+ q \xi^{-2}} \nonumber \\ & \times
\f{q^2 \mbox{Sign}(\omega+\Omega)}{\sqrt{(v_F
q)^2+(\omega+\Omega+\Sigma(\omega+\Omega))^2}} \nonumber \\
 = & - \f{3 \bar{g}}{(2 \pi)^2}
\int_{-\infty}^{+\infty} d\Omega~ \int_0^{+\infty} dq~ \f{q^2
\mbox{Sign}(\omega+\Omega)}{v_F q} \nonumber \\ ~& \times \f{1}{q^3 +
\gamma| \Omega |+ q \xi^{-2}} .
\end{align}

Defining the new variables $z=\frac{\Omega}{\omega}$ and
$u=\frac{v_F q}{\omega}$, this leads to:
\begin{equation}
\Sigma (\omega) =  \f{3 \gbar}{2 \pi^2} ~\f{\omega^2}{\gamma v_F^3}
\int_0^1 dz~ \int_0^{+\infty} du~ \f{u}{z+a u+b u^3},
\end{equation}
where we have used $a=\f{1}{\gamma v_F \xi^2}$ and
$b=\f{\omega^2}{\gamma v_F^3}$

Let's denote by ${\cal I}$ the last double integral, and define the new
variables $y=\sqrt{\frac{b}{a^3}} z = (\omega \gamma \xi^3) z$ and
$t=\sqrt{\frac{b}{a}} u = \frac{\xi \omega}{v_F} u$. Now ${\cal I}$ reduces
to:
\begin{eqnarray}
{\cal I} & = & \left( \f{v_F}{\xi \omega} \right)^2 \int_0^{\gamma \omega
\xi^3} dy~ \int_0^{+\infty} dt~ \f{t}{t^3 + t +y} \nonumber \\ ~ & = & \left(
\f{v_F}{\xi \omega} \right)^2 h(\gamma \omega \xi^3) .
\end{eqnarray}
Substituting this back into our expression for the self-energy:
\begin{equation}
\Sigma (\omega) =  \f{3}{2 \pi m \xi^2} h(\gamma \omega \xi^3),
\end{equation}
with the following asymptotic behavior: $h(x \to 0)= \frac{\pi x}{2}$ and $h(x \to \infty) = \frac{\pi x^{2/3}}{\sqrt{3}}$.

In the two regimes we are interested in, this last result can be rewritten as:
\beq
\Sigma(\omega) = \left\{
\begin{array}{ll}
{\displaystyle \lambda \omega} & ~{\rm for}~ {\displaystyle \xi^{-1} \gg 1}, \\
{\displaystyle \omega_0^{1/3} \omega^{2/3}} & ~{\rm for}~ {\displaystyle \xi^{-1} \to 0},
\end{array}
\right.
\eeq
where the constant prefactors are defined as:
\beq
\left\{
\begin{array}{lcl}
{\dst \lambda} & = & {\dst \f{3}{4\pi} \f{\gbar \xi}{v_F} }, \\
{\dst \omega_0} & = & {\dst \f{3\sqrt{3}}{8\pi^3} \f{\gbar^3}{\gamma v_F^3} }.
\end{array}
\right.
\eeq

\subsubsection{Momentum dependence}

For definiteness, we set $\xi^{-1} =0$.  We first compute the momentum-dependent
part of the one-loop fermionic self-energy at zero external frequency:
\bea \label{sk1}
\Sigma({\bf k},\omega=0) & = & 3 i g^2 \int \f{d^2q~d\Omega}{(2 \pi)^3}
 G(\bk+\bq,\Omega) \chi(\bq,\Omega) \nonumber \\
 & = &\frac{3i  \gbar}{(2 \pi)^3} \int \frac{d\Omega d^2q}{i{\tilde \Sigma}(\Omega)-\epsilon_{k+q}} \frac{q}{\gamma |\Omega|+q^3}\ .
\eea
Expanding $\epsilon_{k+q} = \epsilon_k + v_F q \cos \theta$
 and integrating over $\theta$, we obtain
\beq \label{sk2}
\begin{array}{l}
{\displaystyle \Sigma({\bf k},0) = \frac{3  \gbar}{4 \pi^2}
\int d\Omega ~\text{Sign} (\Omega)\int \frac{q^2 dq }{q^3+\gamma |\Omega|} } \\ \\
{\displaystyle \times \frac{1}{\left((v_F q)^2 + ({\tilde \Sigma} (\Omega) + i \epsilon_k)^2\right)^{1/2}}} .
\end{array}
\eeq
At $\epsilon_k =0$, the integral vanishes by parity. Expanding to linear order
in $\epsilon_k$ we obtain:
\bea \label{sk2_1}
&& \Sigma({\bf k},0) = -3 i \epsilon_k \times  \frac{\gbar}{2\pi^2}
\int_0^\infty d\Omega ~{\tilde \Sigma} (\Omega) \times \nonumber \\
&&\int \frac{q^2 dq }{\left(q^{3} +
(\gamma |\Omega|)\right)\left((v_F q)^2 + {\tilde \Sigma} (\Omega)^2\right)^{3/2}} .
\eea

Simple estimates show that the 
 result depends on the interplay between 
$(\gamma \Omega)^{1/3}$ and ${\tilde \Sigma} (\Omega)/v_F$. 
 Introducing the scale $\omega_{\rm Max}$, defined as the frequency at
 which  $(\gamma \Omega)^{1/3} = {\tilde \Sigma} (\Omega)/v_F$ (see (\ref{wmax}), and 
 rescaling variables as $q = (\gamma \omega_{\rm Max})^{1/3} y$, $\Omega = \omega_{\rm Max} x$, we re-write (\ref{sk2_1}) as 
\beq \label{sk2_11}
 \Sigma({\bf k},0) = -3 i \epsilon_k \times  \frac{\gbar \omega_{\rm Max}}{2\pi^2 v^3_F \gamma} I
\eeq
where
\beq
I = \int_0^\infty dx \int_0^\infty dy \frac{x y^2}{(x^2+y^2)^{3/2} (x+y^3)} \simeq 1.3308
\eeq
Substituting $\omega_{\rm Max} = (\gamma v^3_F)^{1/2}$, we obtain 
(\ref{sigma2k2_2}).

To obtain the  frequency dependence of the fermionic density of states at small $\omega$, , we have to evaluate the second order
 term in $\epsilon_k$ at the mass shell, where
 $\epsilon_k = i\tSigma (\omega)$ and convert the result to real frequencies.
 We therefore will keep both $\omega$ and $\epsilon_k$ finite, and use
\beq
\Sigma({\bf k},\omega) = \frac{3 i \gbar}{(2 \pi)^3} \int d\Omega d^2q \frac{q}{\gamma |\Omega|+q^3} \frac{1}{i\tSigma (\omega+\Omega)-\epsilon_{k+q}}  .
\eeq
 Writing, as before, $\epsilon_{k+q} = \epsilon_k + v_F q \cos \theta$ and 
integrating over $\theta$,  we obtain
\beq
\Sigma({\bf k},\omega) = \frac{3 \gbar}{4 \pi^2}  
\int \frac{d\Omega q^2 d q }{\gamma |\Omega|+q^3} \frac{{\rm sign} 
(\omega + \Omega)}{\sqrt{(\tSigma (\omega+\Omega) + i \epsilon_k)^2 + 
(v_F q)^2}}
\label{may2_4}
\eeq  
  We assume and then verify that the internal $v_F q$ are still larger than $\tSigma (\omega + \Omega)$ and $\epsilon_k$, and expand
 \bea
\Sigma({\bf k},\omega) &=& - \frac{3  \gbar}{8 \pi^2 v^3_F} 
\int d\Omega {\rm sign} (\omega + \Omega) \nonumber \\
&&\int \frac{ dq}{q (\gamma |\Omega|+q^3)} \left(\tSigma (\omega+\Omega) + i \epsilon_k\right)^2
\label{may2_5}
\eea  
The lower limit of the momentum integral is $(\tSigma (\omega + \Omega) + i \epsilon_k)/v_F$. 
At the mass shell,  $i \epsilon_k = \tSigma (\omega)$. Substituting, 
we find 
 \bea
\Sigma({\bf k},\omega) &=& - \frac{3  \gbar}{8 \pi^2 v^3_F} 
\int d\Omega ~{\rm sign} (\omega + \Omega) \qquad \qquad \qquad \quad \nonumber \\
&&\int \frac{ dq}{q (\gamma |\Omega|+q^3)} \left(\tSigma (\omega+\Omega) -
\tSigma (\omega)\right)^2
\label{may2_5_1}
\eea  
We will need the contributution which is confined to 
$\Omega \sim \omega$.  The contributions from $|\Omega| \gg |\omega|$ 
 diverge in our expansion procedure, and account for the regular $O(\epsilon_k)$ and $O(\omega)$ terms in the self-energy. The last term is even smaller in $\alpha$ than the regular $O(\epsilon_k)$ term and is totally irrelevant.  As $\Omega \sim \omega$ is small, $\tSigma (\Omega) \approx 
\Sigma (\Omega) = \Omega^{2/3} \omega^{1/3}_0$. 

Because of the ${\rm sign}$ factor in the 
 numerator of (\ref{may2_4}), there are two
 distinct contributions from $\Omega \sim \omega$.  For both of 
 them, the  momentum integral is logarithnic  (this justifies
 the expansion) and yields $(1/3) \log (\omega_1/\omega)$, where 
 $\omega_1 \sim N^2 E^2_F/{\bar g}$. The first contribution comes from
 $|\Omega| \leq |\omega|$ and to logarithmical accuracy is
\bea
\Sigma({\bf k},\omega)_A &=&  - \frac{\gbar}{8 \pi^2 v^3_F \gamma} {\rm sign} (\omega) \int_{-|\omega|}^{|\omega|} \frac{d\Omega}{|\Omega|} \nonumber \\
&&  
\times \left(\Sigma (|\omega| + \Omega) - \Sigma (|\omega|)\right)^2 \log{\frac{\omega_1}{|\omega|}}
\label{may2_6}
\eea  
Rescaling the frequency, we obtain from (\ref{may2_6})
\beq
\Sigma({\bf k},\omega)_A = - \frac{\gbar I_1}{4\pi^2 v^3_F \gamma} \Sigma (\omega)|\Sigma (\omega)| \log{\frac{\omega_1}{\omega}}. 
\label{may2_7}
\eeq  
where 
\beq
I_1 = \frac{1}{2} \int_{-1}^1 \frac{dx}{|x|} \left((1+x)^{3/2} -1\right)^2 = 
0.254
\eeq
Another  comes from $|\Omega| > |\omega|$, and is 
\bea
\Sigma({\bf k},\omega)_B &=&  \frac{\gbar \Sigma (\omega)}{4 \pi^2 v^3_F \gamma} ~\log{\frac{\omega_1}{|\omega|}} \int_{|\omega|}^{\infty} \frac{d\Omega}{\Omega} \qquad \qquad \qquad \qquad  \nonumber \\
&&  
\times \left(\Sigma (|\omega| + \Omega) + \Sigma (\Omega - |\omega|) - 2 \Sigma (\Omega)\right)
\label{may2_6_1}
\eea 
Rescaling, we obtain 
\beq
\Sigma({\bf k},\omega)_B =  \frac{\gbar I_2}{4\pi^2 v^3_F \gamma} \Sigma (\omega)|\Sigma (\omega)| \log{\frac{\omega_1}{|\omega|}}. 
\label{may2_7_1}
\eeq  
where 
\bea
I_2 & = & \int_{1}^{\infty} \frac{dx}{x} \left((1+x)^{2/3} + (x-1)^{2/3}- 2 x^{2/3}\right) \nonumber \\
& = & -0.195
\eea
Combining $\Sigma({\bf k},\omega)_A$ and $\Sigma({\bf k},\omega)_B$, we obtain 
\beq
\Sigma({\bf k},\omega)_ = - \frac{0.45 \gbar}{4\pi^2 v^3_F \gamma} \Sigma (\omega) |\Sigma (\omega)| \log{\frac{\omega_1}{|\omega|}}. 
\label{may2_7_2}
\eeq  
Substituting the result for $\gamma$ in this last expression, we obtain (\ref{sigma2k3}).

\subsection{Two-loop}

We compute here 
 one of the contributions to the two-loop self-energy,  given by Fig. \ref{self}. This contribution originates from the insertion of the vertex correction into the Eliashberg self-energy.

We have:
\beq
\begin{array}{l}
{\dst \Sigma_2 (\omega) \sim \gbar^2 \int d\omega_1 d^2 q_1 \int d\omega_2 d^2 q_2 \chi ({\bf q_1},\omega_1)  \chi({\bf q_2},\omega_2)} \\ \\
{\dst \qquad \qquad \times G({\bf k_F+q_1},\omega+\omega_1) G({\bf k_F+q_2},\omega+\omega_2)  } \\ \\
{\dst  \qquad \qquad \times G({\bf k_F+q_1+q_2},\omega+\omega_1+\omega_2)} ,
\end{array}
\eeq
which gives, once we replace each propagator by its full expression:
\beq
\begin{array}{l}
{\displaystyle \Sigma_2 (\omega) \sim \gbar^2 \int d\omega_1 d^2 q_1 \int d\omega_2 d^2 q_2 \frac{q_1}{\gamma |\omega_1|+ q_1^3} \frac{q_2}{\gamma |\omega_2|+q_2^3} } \\ \\
{\displaystyle \times \frac{1}{i\tilde{\Sigma}(\omega+\omega_1)-v_F q_{1x}} \frac{1}{i\tilde{\Sigma}(\omega+\omega_2)-v_F q_{2x}-\frac{q_{2y}^2}{2m_B}}} \\ \\
{\displaystyle \times \frac{1}{i \tilde{\Sigma}(\omega+\omega_1+\omega_2)-v_F q_{1x}-v_F q_{2x}-\frac{q_{1y}^2}{2m_B}-\frac{q_{2y}^2}{2m_B}}} ,
\end{array}
\eeq
where we use the shorter notation $\tilde{\Sigma}(\omega) = \omega+\Sigma(\omega)$.

Integrating successively over $q_{1x}$ and $q_{2x}$, closing each contour on the upper half-plane, one has:
\beq
\begin{array}{l}
{\dst \Sigma_2 (\omega) \sim \f{m_B \gbar^2}{v_F^2} \int d\omega_1 \int d\omega_2 ~\Theta(\omega,\omega_1,\omega_2)} \\ \\
{\dst \qquad \int d q_{1y} \int d q_{2y}  \f{|q_{1y}|}{| q_{1y}|^3 +\gamma |\omega_1|} \f{|q_{2y}|}{|q_{2y}|^3 +\gamma |\omega_2|} } \\ \\
{\dst \qquad \qquad \times \f{1}{\left( \gamma^2 \omega_1 \omega_2  \right)^{1/3}} \f{1}{\frac{q_{1y} q_{2y}}{\left( \gamma^2 \omega_1 \omega_2  \right)^{1/3}}+i \zeta}} ,
\end{array}
\eeq
where $\zeta = m_B \f{\tilde{\Sigma}(\omega+\omega_1)+\tilde{\Sigma}(\omega+\omega_2)-\tilde{\Sigma}(\omega+\omega_1+\omega_2)}{\left( \gamma^2 \omega_1 \omega_2  \right)^{1/3}}$ and $\Theta(\omega,\omega_1,\omega_2)$ comes from the choice of a contour for the integration and is given in our case by:
\beq
\begin{array}{l}
{\dst \Theta(\omega,\omega_1,\omega_2) = \left(\theta(\omega+\omega_1) - \theta(\omega+\omega_1+\omega_2) \right) \quad} \\ \\
{\dst \quad \times \left( \theta(\omega+\omega_2) - \theta(\tSigma(\omega+\omega_1+\omega_2) -\tSigma(\omega+\omega_1))  \right)} .
\end{array}
\eeq

It is convenient at this stage to rescale the perpendicular components of the bosonic momenta. Introducing $x=q_{1y}/(\gamma |\omega_1|)^{1/3}$ and $y=q_{2y}/(\gamma |\omega_2|)^{1/3}$, we obtain:
\beq
\begin{array}{l}
{\displaystyle \Sigma_2 (\omega) \sim \frac{m \gbar^2}{v_F^2} \int d\omega_2 \int d\omega_1 \int_{0}^{\infty} dx dy~ \frac{\Theta(\omega,\omega_1,\omega_2)}{(\gamma^2 \omega_1 \omega_2)^{2/3}} }\\ \\
{\displaystyle \qquad \qquad \times \frac{i \zeta}{x^2 y^2 + \zeta^2} \frac{x y}{\left( 1+ x^3 \right) \left( 1+ y^3 \right)}}  ,
\end{array}
\eeq
where we rearranged the double integral over $x$ and $y$ to make it real.

Since all internal frequencies typically go like $\omega$, the typical value of $\zeta$ is given by the small parameter $\beta$ given in (\ref{beta}).
 Expanding the double integral for small values of $\zeta$, the leading contribution from the integral over $x$ and $y$ reads:
\beq
\int_{0}^{\infty} dx dy~\frac{1}{x^2 y^2 + \zeta^2} \frac{x y}{\left( 1+ x^3 \right) \left( 1+ y^3 \right)} \sim \log^2 \zeta .
\eeq

If one considers now free fermions, it becomes possible to reduce the expression of the two-loop self-energy to:
\bea
\Sigma_2^{\rm free} (\omega) & \sim & \frac{m_B^2 \gbar^2}{v_F^2} \int_0^{\omega} d\omega_2 \int_{\omega-\omega_2}^{\omega} d\omega_1 \frac{\log^2 \left( \frac{m_B\omega}{(\gamma^2 \omega_1 \omega_2)^{2/3}} \right)}{\gamma^2 \omega_1 \omega_2} \nonumber \\
& \sim & \f{m_B^2 \gbar^2}{\gamma v_F^2} \omega \int_0^1 dz_2 \int_{1-z_2}^1 dz_1 \f{\log^2 \left( \frac{m_B^3 \omega}{\gamma z_1 z_2}  \right)}{z_1 z_2} \nonumber \\
& \sim & \beta^2 \omega \log^2 \omega ,
\eea
where we only kept the leading contribution in the last expression, and $\beta = m_B / m N$ is one of the small parameters defined in the text.

Now, in the case of dressed fermions, we need to take $\tSigma (\omega) = \omega_0^{1/3} \omega^{2/3}$. The procedure is identical to the free fermion case, but the final expression is a bit more complicated:
\beq
\begin{array}{l}
{\dst \Sigma_2^{\rm Dressed} (\omega)  \sim  \frac{m^2 \gbar^2}{\gamma^2 v_F^2} \omega_0^{1/3} \omega^{2/3} \int_0^1 dz_2 \int_{1-z_2}^1 dz_1 \f{1}{z_1 z_2} } \\ \\
{\dst \times \left[ (1-z_1)^{2/3} + (1-z_2)^{2/3} + (z_1+z_2-1)^{2/3} \right]   } \\ \\
{\dst \times \log^2 \left( \f{(1-z_1)^{2/3} + (1-z_2)^{2/3} + (z_1+z_2-1)^{2/3}}{\frac{\gamma^{2/3}}{m \omega_0^{1/3}} z_1^{1/3} z_2^{1/3}} \right)} .
\end{array}
\eeq
Expanding the $\log^2$, one is left with a double integral that only contributes as a numerical prefactor, and the dominant term is then given by:
\beq
\Sigma_2 (\omega)  \sim \Sigma_1 (\omega)~ \beta^2 \log^2 \beta ,
\eeq
where $\Sigma_1 (\omega) = \omega_0^{1/3} \omega^{2/3}$ is the self-energy in the Eliashberg theory.

\subsection{Three loop}

We now turn to the computation of the three-loop self energy. We are only interested here in one diagram, given in fig. \ref{3loopsigma}, where we try to analyze the feedback of the non-analytic susceptibility into the the higher-order diagrams for the fermionic self-energy. For spin interaction, there is no cancellation between different diagrams for the static susceptibility, which justifies that we restrict ourselves to just one contribution.

The analytic expression for this diagram is:
\begin{align}
\Sigma_3 (\omega) \sim & \gbar \int d\omega_1 d^2 q_1 \frac{A(q_1,\omega_1)}{i\Sigma(\omega-\omega_1)-v_F q_{1x}-\frac{q_{1y}^2}{2m_B}} \nonumber \\
 & \qquad \times \left( \frac{1}{q_1^2+\frac{\gamma |\omega_1|}{q_1}} \right)^2  ,
\end{align}
where $A(q_1,\omega_1)$ is the factor from the
 fermionic bubble:
\begin{align}
A(q_1,\omega_1) & \sim N \gbar^2 \int d^2 k d\Omega \int d^2 q_2 d\omega_2 \frac{1}{q_2^2 + \frac{\gamma |\omega_2|}{q_2}} \nonumber \\
 & \times ~\frac{1}{i\Sigma(\Omega)-\epsilon_k}~\frac{1}{i\Sigma(\Omega+\omega_1+\omega_2)-\epsilon_{k+q_1+q_2}} \nonumber \\
 & \times ~\frac{1}{i\Sigma(\Omega-\omega_1)-\epsilon_{k+q_1}}~ \frac{1}{i\Sigma(\Omega+\omega_2)-\epsilon_{k+q_2}} . \nonumber
\end{align}

Approximating $A(q_1, \omega_1)$ by its singular static part $q^{3/2}_1 \sqrt{k_F}$ and substituting into the expression of $\Sigma_3$ we obtain:
\begin{align}
\Sigma_3 (\omega) \sim & \gbar \sqrt{k_F} \int d\omega_1 d^2 q_1  \left( \frac{q_1}{q_1^3+\gamma |\omega_1|} \right)^2 \nonumber \\
& \qquad \times \frac{q_1^{3/2}}{i\Sigma(\omega-\omega_1)-v_F q_{1x}-\frac{q_{1y}^2}{2m_B}} .
\end{align}
The integral over $q_{1x}$ is determined by the branch-cut in the bosonic propagator and one then expects that this very integral is dominated by $q_{1x} \sim (\gamma \omega_1)^{1/3}$. It follows that the term in $q_{1x}$ dominates inside the fermionic propagator allowing us to neglect the curvature term. Defining the angle $\theta$ between ${\bf k_F}$ and ${\bf q_1}$, and integrating over it, this leads to:
\beq
\Sigma_3 (\omega) \sim \frac{\gbar \sqrt{k_F}}{v_F} \int \f{d\omega_1 dq_1 ~ {\rm sign}(\omega-\omega_1)}{\sqrt{ q_1^2+\frac{\Sigma(\omega-\omega_1)^2}{v_F^2}}} \f{q_1^{9/2}}{(q_1^3+\gamma |\omega_1|)^2}  .
\eeq

Since the dominant contribution comes from $q_1 \sim (\gamma \omega_1)^{1/3}$, one can neglect the fermionic self-energy in the denominator. This in turn allows to simplify the frequency integral, which then writes:
\bea
\Sigma_3 (\omega) &\sim& \f{\gbar \sqrt{k_F}}{v_F} \int_0^{\omega} d\omega_1 \int dq_1 \f{q_1^{7/2}}{(q_1^3+\gamma \omega_1)^2} \nonumber \\
&\sim& \f{\gbar \sqrt{k_F}}{\sqrt{\gamma}v_F} \int_0^{\omega} \f{d \omega_1}{\sqrt{\omega_1}}  ,
\eea
where we introduced $z=q_1/(\gamma \omega_1)^{1/3}$, so that the integral over $z$ just contribute to the numerical prefactor.

Collecting prefactors, one finally has:
\beq
\Sigma_3 (\omega) \sim \sqrt{\gbar \omega} .
\eeq

\section{Vertex corrections}

\label{app:vertex}

In this section, we compute the various vertex corrections analyzed in the text.

\subsection{$q=\Omega=0$}

Consider first the simplest 3-leg vertex, with strictly zero incoming frequency $\Omega$ and momenta $q$, as presented in fig. \ref{vertices}a. Its analytic expression writes:
\bea
\left. \frac{\Delta g }{ g }\right|_{q=\Omega =0} & \sim & g^2 \int d \omega d^2 p~ G({\bf k_F},\omega)^2~ \chi({\bf p},\omega) \nonumber \\
& \sim & \gbar \int \frac{ d \omega d^2 p } { \frac{\gamma | \omega | }{  p } + p^2 } \ \frac{1}{
\left ( i \tilde{\Sigma} (\omega ) - v_F p_x - \frac{p_y^2 }{ 2m_B  }\right  )^2 }  , \nonumber
\eea
where we defined $\tilde{\Sigma}(\omega) = \omega + \Sigma(\omega)$ and we have chosen ${\bf k}_F$ along the $x$ axis.

Since both poles coming from the fermionic Green's functions are in the same half plane, the integral over $q_x$ is finite only because of the branch cut in the bosonic propagator. Since at the branch cut $p_x$ and $p_y$ are of the same order, we can drop the curvature term in the fermionic propagators and introduce polar coordinates for the internal bosonic momentum. Defining $\theta$ as the angle between ${\bf k_F}$ and ${\bf p}$, and integrating over it, one has:
\beq
\left. \frac{\Delta g }{ g }\right|_{q=\Omega =0}  \sim \gbar \int \f{d\omega dp}{p^3+\gamma|\omega|} \f{p^2 |\tSigma (\omega)|}{\left( (v_F p)^2+\tSigma(\omega)^2 \right)^{3/2}} .
\eeq

Introducing the frequency $\omega_{\rm max}$ up to which bosons are slow modes compared to fermions, i.e. up to which $(\gamma \omega)^{1/3} > \tSigma(\omega) /v_F$, one can split the frequency integral into two parts, and define in each of them the reduced momentum $z=p/{\rm Min}((\gamma \omega)^{1/3},\tSigma(\omega) /v_F)$ so that the integral over $z$ only contributes to the numerical prefactor, leading to:
\beq
\left. \frac{\Delta g }{ g }\right|_{q=\Omega =0}  \sim  \gbar  \int_{0}^{\omega_{\rm max}} d\omega \f{\tSigma(\omega)}{\gamma v_F^3 \omega} \sim
 \f{\gbar}{\gamma v_F^3} \tSigma ( \omega_{\rm max}) 
\eeq
One can easily make sure that the frequency $\omega_{\rm max}$ at which 
 $(\gamma \omega)^{1/3} = \tSigma(\omega) /v_F$ exceeds $\omega_0$, such that
 one should use $\tSigma (\omega) = \omega$ to find $\omega_{\rm max}$. Substituting, we obtain $\omega_{max} \sim (N \bar g E_F)^{1/2}$, and 
\beq
\left. \frac{\Delta g }{ g }\right|_{q=\Omega =0}  \sim \sqrt{\alpha}
\eeq

\subsection{$q=0$, $\Omega$ finite}

Considering the same vertex, now with a finite external frequency, one has:
\bea
\left. \frac{\Delta g }{ g }\right|_{q=0, \Omega} & \sim & g^2 \int d \omega d^2 p~ G({\bf k_F+p},\omega+\Omega) \nonumber \\
& & \qquad G({\bf k_F+p},\omega) \chi({\bf p},\omega) \nonumber \\
& \sim & \gbar \int \frac{ d \omega d^2 p } { \frac{\gamma | \omega | }{  p } + p^2  } \
\frac{1}{i {\tilde \Sigma} (\omega ) - v_F p_x - \frac{p_y^2 }{ 2m_B }} \nonumber  \\
& &  \qquad \times \frac{1}{i{\tilde \Sigma}( \omega + \Omega ) - v_F p_x - \frac{p_y^2}{2 m_B}  }  ,
\eea
where we chose the $x$ axis along ${\bf k_F}$.

From the pole structure in $p_x$ of this expression, one expects two contributions to this integral. A first term comes from the branch cut in the bosonic propagator, however this contribution ultimately leads to the same result as the $q=\Omega=0$ vertex, up to small corrections from the finiteness of $\Omega$. The second contribution arises from taking the poles in the fermionic propagators. At zero external frequency, these two poles were in the same half-plane of $p_x$, so we could close the integration contour over a different half-plane and only consider the contribution from the branch but in the bosonic propagator. At a finite $\Omega$, there is a range where $\omega$ and $\omega + \Omega$ have different signs, and the two poles are in different half-planes of $p_x$. The result after integration reads:
\bea
\left. \frac{\Delta g }{ g }\right|_{q=0, \Omega} & \sim &  \frac{\gbar}{v_F} \int_0^{\Omega} d\omega \int dp_y  \frac{|p_y|}{\gamma | \omega | + |p_y|^3 } \nonumber \\
& & \times \frac{1}{{\tilde \Sigma} (\Omega -\omega) + {\tilde \Sigma} (\omega)} ,
\eea
where we slightly rearranged the frequency integral.

Performing the integration over $p_y$, we are left with:
\bea
\left. \frac{\Delta g }{ g }\right|_{q=0, \Omega} & \sim &  \frac{\gbar}{\gamma^{1/3} v_F} \int_0^{\Omega} d\omega \f{\omega^{-1/3}}{\tSigma(\Omega -\omega)+\tSigma(\omega)} \nonumber \\
 & \sim & \f{\gbar}{(\omega_0 \gamma v_F^3)^{1/3}} \nonumber \\
 & \sim & {\rm Const.} ,
\eea
where we assumed that $\Omega$ is small, i.e. $\tSigma(\Omega) = \omega_0^{1/3} \Omega^{2/3}$. This vertex thus reduces to a numerical constant, that does not contain any small parameter.

\subsection{$q$ finite, $\Omega=0$}

Conversely, the same vertex taken at finite external momentum $q$, but zero external frequency writes:
\bea
\left. \frac{\Delta g }{ g }\right|_{q, \Omega=0} & \sim & g^2 \int d \omega d^2 p~ G({\bf k_F+p+q},\omega) \nonumber \\
& & \qquad G({\bf k_F+p},\omega) \chi({\bf p},\omega) \nonumber \\
& \sim & \gbar \int \frac{ d \omega d^2 p } { \frac{\gamma | \omega | }{  p } + p^2  } \
\frac{1}{i {\tilde \Sigma} (\omega ) - v_F p_x - \frac{p_y^2 }{ 2m_B }} \nonumber  \\
& &  \times \frac{1}{i{\tilde \Sigma}( \omega ) - v_F q_x - v_F p_x - \frac{p_y^2}{2 m_B} -\frac{q_y p_y}{m_B}  }  , \nonumber
\eea
where $p_x$ is the projection of ${\bf p}$ along ${\bf k_F}$.

Like its $q=0$ counterpart, this vertex is characterized by poles in $p_x$ from the fermionic propagators lying in the same half-plane. The only nonzero contribution then comes from the branch cut in the bosonic propagator. At the branch cut, $p_x \sim p_y$ which allows us to neglect the quadratic curvature terms in the fermionic Green's functions.

This makes possible a direct integration over $p_y$. This integral can be separated from the rest of the expression, and reads:
\bea
\int d p_y \f{p}{\gamma |\omega| + p^3} & = & \f{1}{|p_x|} \int_{- \infty}^{+ \infty} dz ~\f{\sqrt{1+z^2}}{(1+z^2)^{3/2}+\gamma \frac{|\omega|}{|p_x|^3}} \nonumber \\
& = & \f{2 p_x^2}{\gamma |\omega|} \int_{0}^{+ \pi/2} du ~\f{1}{\frac{|p_x|^3}{\gamma |\omega|}+ (\cos u)^3}  , \nonumber \\
\eea
where we successively defined $z=p_y/|p_x|$ and $z=\tan u$.

This last integral can be approximated by its asymptotic form, namely:
\beq \label{intchi}
\int \f{dp_y ~ p}{\gamma |\omega| + p^3} = \left\{
\begin{array}{l}
\frac{\pi}{|p_x|}, ~ {\rm if }~ |p_x|^3 \gg \gamma \omega \\
\frac{4\pi}{3\sqrt{3}} \frac{1}{(\gamma|\omega|)^{1/3}} - \frac{1}{2} \frac{p_x^2}{\gamma|\omega|} \log \frac{p_x^2}{(\gamma|\omega|)^{2/3}} \\
\ +\left(\log 2 -\frac{1}{2}\right) \frac{p_x^2}{\gamma|\omega|},~ {\rm if }~  |p_x|^3 \ll \gamma \omega
\end{array}
\right.
\eeq

The only non-vanishing contribution once we take the integral over $p_x$ comes from the $\log$ term. After expanding in $q_x$, this contribution reads:
\bea
& & \left. \frac{\Delta g }{ g }\right|_{q, \Omega=0} - \left. \frac{\Delta g }{ g }\right|_{q=\Omega=0}  \sim  \f{\gbar v_F}{\gamma} q_x \qquad \qquad \nonumber \\
& & \qquad \qquad \times \int \f{d\omega}{|\omega|} \int dp_x \f{p_x^2 \log p_x^2}{\left( i\tSigma(\omega)-v_F p_x \right)^3} .
\eea

Defining the scaled momentum $z=v_F p_x/\tSigma(\omega)$ and performing the integration over $z$, one obtains two terms, the dominant one being:
\beq
\left. \frac{\Delta g }{ g }\right|_{q, \Omega=0} - \left. \frac{\Delta g }{ g }\right|_{q=\Omega=0}  \sim  i \f{\gbar}{\gamma v_F^2} q_x \int_{|v_F q_x|} \f{d\omega}{\omega} \log [i\tSigma(\omega)] ,
\eeq
where the frequency integral runs over $|{\tilde \Sigma} (\omega)| > |v_F q_x|$ since as we expanded in $v_F q_x$, we assumed that it was smaller than $|\tSigma(\omega)|$.

Performing the remaining integral, one finds:
\beq
\left. \frac{\Delta g }{ g }\right|_{q, \Omega=0} - \left. \frac{\Delta g }{ g }\right|_{q=\Omega=0}  \sim \f{q_x}{k_F} \log |q_x|  .
\eeq

\subsection{$q,\Omega$ finite}

Finally, we consider the general vertex where both external bosonic momentum and frequency are non-zero.
In analytic form, this writes:
\bea
\left. \frac{\Delta g }{ g }\right|_{q, \Omega} & \sim & g^2 \int d \omega d^2 p~ G({\bf k_F+p+q},\omega+\Omega) \nonumber \\
& & \qquad G({\bf k_F+p},\omega) \chi({\bf p},\omega) \nonumber \\
& \sim & \gbar \int \frac{ d \omega d^2 p } { \frac{\gamma | \omega | }{  p } + p^2  } \
\frac{1}{i {\tilde \Sigma} (\omega ) - v_F p_x - \frac{p_y^2 }{ 2m_B }} \nonumber  \\
& &  \times \frac{1}{i{\tilde \Sigma}( \omega +\Omega) - v_F q_x - v_F p_x - \frac{p_y^2}{2 m_B} -\frac{q_y p_y}{m_B}  }  , \nonumber
\eea
where $p_x$ is defined as $p_x={\bf p} \cdot {\bf k_F}$.

Integrating over $p_x$ first, there are two contributions. One comes from the branch cut in the bosonic propagator, and gives similar results to the $(q=0,\Omega=0)$ and the $(q~{\rm finite,}~\Omega=0)$ vertices up to small correction from the finiteness of the external frequency. We neglect it here and focus on the other contribution which comes from the poles in the fermionic propagators:
\bea
\left. \frac{\Delta g }{ g }\right|_{q, \Omega} & \sim &  i \frac{\gbar}{v_F} \int_0^{\Omega} d\omega \int dp_y  \frac{|p_y|}{\gamma | \omega | + |p_y|^3 }  \nonumber \\
&& \hspace{-0.5cm} \times \frac{1}{i {\tilde \Sigma} (\Omega -\omega) + i {\tilde \Sigma} (\omega) -
 v_F q_x - \frac {q_y p_y }{  m_B  }} , \quad
\eea
where the simplification of the frequency integral comes from the poles in $p_x$.

This vertex correction strongly depends on the interplay between the external $q_x$, $q_y$ and $\Omega$, and is in particular quite sensitive to the momentum anisotropy. We now analyze the various possibilities.

For the generic case where $q_x \sim q_y$ (we use the notation $q$ to designate them), one can neglect the quadratic term in the fermionic dispersion, allowing to perform the integration over $p_y$, leaving us with:
\beq
\left. \frac{\Delta g }{ g }\right|_{q, \Omega} \sim  i \frac{\gbar}{v_F \gamma^{1/3}} \int_0^{\Omega} \frac{d\omega ~\omega^{-1/3}}{i {\tilde \Sigma} (\Omega -\omega) + i {\tilde \Sigma} (\omega) - v_F q} .
\eeq
Restricting ourselves to the quantum-critical regime (i.e. $\Omega \leq \omega_0$) for which $\tSigma(\omega) = \omega_0^{1/3} \omega^{2/3}$, one has:
\beq
\left. \frac{\Delta g }{ g }\right|_{q, \Omega} =  {\cal F} \left( \frac{v_F q}{\Sigma(\Omega)}  \right) ,
\label{apr26_1}
\eeq
where ${\cal F} (x) = \int_0^1 \frac{dz}{z^{1/3}} \frac{1}{(1-z)^{2/3}+z^{2/3}+ix}$, has the following asymptotic behavior:
\beq
\left\{
\begin{array}{lcl}
{\cal F}(x\ll 1) & = & O(1) \\
{\cal F}(x\gg 1) & = & O \left( \frac{1}{x} \right)
\end{array}
\right.
\eeq
If the typical $q$ is on the bosonic mass shell, then $q \sim (\gamma \Omega)^{1/3}$, and one has:
\beq
\left. \frac{\Delta g }{ g }\right|_{q, \Omega} \sim \frac{\Sigma(\Omega)}{v_F q} \sim ~\sqrt{\alpha}~ \left( \frac{\Omega}{\omega_{\rm max}} \right)^{1/3} .
\eeq

However, we encountered in previous computations (e.g. self-energies) that a strong anisotropy can be observed between the components of the bosonic momentum, with $q_y \gg q_x$. In this case, the full expression of the vertex correction is a bit complicated and we choose to present here the most relevant case for which the curvature term dominates over $v_F q_x$ in the fermionic propagator. The vertex correction then no longer depends on $q_x$ and writes:
\bea
\left. \frac{\Delta g }{ g }\right|_{q, \Omega} & \sim &  \frac{\gbar}{v_F} \int_0^{\Omega} d\omega \int_0^{\infty} dp_y  \frac{p_y}{\gamma \omega + p_y^3 }  \nonumber \\
&& \hspace{-0.5cm} \times \frac{{\tilde \Sigma} (\Omega -\omega) + {\tilde \Sigma} (\omega)}{ \left({\tilde \Sigma} (\Omega -\omega) + {\tilde \Sigma} (\omega) \right)^2 + \left( \frac {q_y p_y }{  m_B  }\right)^2}  , \quad
\eea
Defining $u=p_y/(\gamma \omega)^{1/3}$ and $z=\omega/\Omega$, it is possible to rewrite the vertex correction in this regime as:
\beq
\left. \frac{\Delta g }{ g }\right|_{q, \Omega}  \sim {\cal G} \left( \beta \frac{(\gamma \Omega)^{1/3}}{q_y}  \right)
\eeq
where $\beta = \frac{m_B}{N m}$ and ${\cal G}(x)$ is the following double integral:
\bea
{\cal G} (x) & = & \int_{0}^{\infty} \frac{du~ u}{1+u^3} \int_{0}^{1} \frac{dz ~  z^{-1/3} ~ \left((1-z)^{2/3}+z^{2/3}\right)}{\left((1-z)^{2/3}+z^{2/3}\right)^2+\frac{u^2 z^{2/3}}{x^2}} \nonumber \\
 & \sim & x^2 \log^2 x \qquad {\rm if}~ x \ll 1 .
\eea

Finally,  for the computations of the
 full dynamic polarization bubble  will also need
the vertex averaged over the directions of $q$.
 The generic structure of this vertex,
which we define as $< \Delta g/g>$ is the same as in (\ref{apr26_1}), i.e.,
\bea
&& {\displaystyle \langle \left. \frac{\Delta g }{ g } \rangle \right|_{\rm q, \Omega} = {\tilde {\cal F}}\left(\frac{v_F q}{{\tilde \Sigma} (\Omega)}\right)}, \nonumber \\
&& ~~~~{\tilde {\cal F}}(0) = O(1), ~~{\tilde {\cal F}}(x \gg 1)= O\left(\frac{1}{x}\right)
\label{mon_2}
\eea
However, it will be essential for our further analysis that
the expansion of ${\tilde {\cal F}}(x)$ at large $x$ holds in odd powers of $1/x$,
i.e., ${\tilde {\cal F}}(x\gg1) = a_1/x + a_3/x^3 + ...$. In particular,
there is no term
$O(1/x^2)$, which we found in the polarization operator without
vertex corrections (see (\ref{mon_1})).

\subsection{4-leg vertex}

In this paragraph, we compute the renormalized 4-leg vertex $\Gamma_2 (q, \Omega)$ presented in Fig~\ref{coop4leg}, which reads:
\beq
\begin{array} {l}
{\displaystyle \Gamma_2 (q,\Omega) \sim \gbar^2 \int d \omega \ \int d^2 p \ \chi_s \left(\frac{\Omega+\omega}{2},\frac{\bf p+q}{2} \right) } \\ \\
{\displaystyle \times  \chi_s \left(\frac{\Omega-\omega}{2},\frac{\bf q-p}{2} \right) G \left(\frac{\Omega + \omega}{2},{\bf k_F}+\frac{\bf p+q}{2}\right) } \\ \\
{\displaystyle \times G \left(\frac{\Omega -\omega}{2},{\bf k_F}+\frac{\bf q-p}{2} \right) } .
\end{array}
\eeq

Performing the integration over $p_x$, projection of ${\bf p}$ along ${\bf k_F}$, we obtain:
\bea
\Gamma_2 (q,\Omega) & \sim & \frac{\gbar^2}{v_F} \int_0^{\Omega} \f{d\omega~dp_y}{\gamma (\Omega-\omega)+(q^2+p_y^2-2 q_y p_y)^{3/2}} \nonumber \\
& & \times \f{1}{i\tSigma\left( \frac{\Omega+\omega}{2} \right)+i\tSigma\left( \frac{\Omega-\omega}{2} \right)-v_F q_x -\frac{q_y^2+p_y^2}{4m_B}}  \nonumber \\
& & \times \f{\sqrt{(q^2+p_y^2)^2-4q_y^2 p_y^2}}{\gamma (\Omega+\omega)+(q^2+p_y^2+2 q_y p_y)^{3/2}} .
\eea

It is convenient at this stage to define the reduced variables $z=\omega/\Omega$ and $y=p_y/|q_y|$:
\bea
& & \Gamma_2 (q,\Omega)  \sim  \gbar \frac{(\gamma \Omega)^{1/3}}{|q_y|^3} \int_0^1 \f{dz~dy}{\frac{\gamma \Omega}{|q_y|^3} (1-z)+|1-y|^3} \nonumber \\
& & \times \f{1}{i\left[(1-z)^{2/3}+(1+z)^{2/3}\right] - \frac{v_F q_x}{\Sigma(\Omega)} - \frac{1+y^2}{\beta} \left(\frac{|q_y|^3}{\gamma \Omega}\right)^{2/3}} \nonumber \\
& & \times \f{|1-y^2|}{\frac{\gamma \Omega}{|q_y|^3} (1+z)+|1+y|^3} ,
\eea
where we assumed that $\Omega \leq \omega_0$, and we used that in all our computations, the perpendicular component of the bosonic momentum is always either dominant or comparable to the parallel one, so that one has $q_y \sim q$.

Comparing this renormalized vertex with the bare one given by $\Gamma_1 (q,\Omega) \sim \frac{\gbar q_y^{-2}}{1 + \frac{\gamma |\Omega|}{|q_y|^3}}$, one has for the ratio of the two:
\bea
& & \frac{\Gamma_2}{\Gamma_1} \sim \frac{(\gamma \Omega)^{1/3}}{|q_y|} \int_0^1 \f{dz~dy}{\frac{\gamma \Omega}{|q_y|^3} (1-z)+|1-y|^3} \nonumber \\
& & \times \f{1}{i\left[(1-z)^{2/3}+(1+z)^{2/3}\right] - \frac{v_F q_x}{\Sigma(\Omega)} - \frac{1+y^2}{\beta}\frac{|q_y|^2}{(\gamma \Omega)^{2/3}}}  \nonumber \\
& & \times \f{|1-y^2|}{\frac{\gamma \Omega}{|q_y|^3} (1+z)+|1+y|^3} ,
\eea
which only depends on two parameters: the ratios $\frac{v_F q_x}{\Sigma(\Omega)}$ and $\frac{|q_y|^3}{\gamma \Omega}$.

In the generic case of an external bosonic momentum on the mass shell, i.e. $q_x \sim q_y \sim (\gamma \Omega)^{1/3}$ one has:
\beq
\frac{\Gamma_2}{\Gamma_1} \sim \frac{\Sigma(\Omega)}{v_F q_x} \sim \sqrt{\alpha} \left( \frac{\Omega}{\omega_{\rm max}}\right)^{1/3} .
\eeq

On the contrary, for a typical $q_x \sim \Sigma(\Omega)/v_F$ and $q_y \sim (\gamma \Omega)^{1/3}$:
\beq
\frac{\Gamma_2}{\Gamma_1} \sim \beta ,
\eeq
up to logarithmic factors.

\section{Static spin susceptibility}

\label{app:chi}

\subsection{Diagrams}

In this Appendix we present the details of our calculations of the
singular terms in the static spin susceptibility.
We discuss in great detail the calculation of the first two diagrams
 in Fig. \ref{2loop} (vertex and self-energy correction diagrams).
 These two diagrams can be computed explicitly both away from QCP and at QCP.
We labeled the total contribution from these two diagrams as $\Pi_A (q,0)$.
The remaining two diagrams (their total contribution is $\Pi_B (q,0)$)
 cannot be computed explicitly at QCP, and we
 compute them in an approximate scheme.

In explicit form, the first two diagrams in Fig. \ref{2loop} are given by:
\begin{align}
\Pi_{1a} (q, 0) & =  \Gamma_a \f{\gbar^2}{(2 \pi)^6} \int d^2 K d \omega
d^2 l d\Omega~ G(\omega,k)^2 \nonumber \\
 &  \times~G(\omega+\Omega,k+l) G(\omega,k+q) \chi_s (l,\Omega) \\
\Pi_{1b} (q, 0) & =  \Gamma_b \f{\gbar^2}{(2 \pi)^6} \int d^2 K d \omega
d^2 l d\Omega~ G(\omega,k) \nonumber \\
 &  \times G(\omega+\Omega,k+l) G(\omega+\Omega,k+q+l) \nonumber \\
 & \times  G(\omega,k+q) \chi_s (l,\Omega)
\end{align}
where $\Gamma_{a,b}$ are numerical prefactors coming from spin summation.
Note that for symmetry reasons, one has to count the first diagram twice,
so that the total contribution reads:
\begin{equation}
\Pi_A (q,0) = 2 \Pi_{1a} (q, 0) + \Pi_{1b} (q, 0).
\end{equation}

\subsubsection{First diagram}

To prove our point, we try to expand the products of fermionic Green's
functions into a simpler form:
\begin{align}
G(\omega,k)^2 G(\omega+\Omega,k+l) G(\omega,k+q) = \phantom{12345678} \nonumber \\
\qquad \f{G(\omega,k)^2 G(\omega,k+q)}{\kappa(l,\omega,\Omega)} - \f{G(\omega,k)
G(\omega,k+q)}{\kappa(l,\omega,\Omega)^2} \nonumber \\
\qquad + \f{G(\omega+\Omega,k+l) G(\omega,k+q)}{\kappa(l,\omega,\Omega)^2},
\end{align}
where $\kappa (l,\omega,\Omega)i(\Sigma(\omega+\Omega)-\Sigma(\omega))-v_F l_x$

The interest of such a splitting up is that one can reduce this
drastically by performing the integration over $k$. In fact:
\begin{equation}
\int d \varepsilon_k G(\omega,k) G(\omega,k+q) = 0,
\end{equation}
since all the poles in $\varepsilon_k$ are in the same half-plane.

For this reason, we are left with:
\begin{align}
\Pi_{1a} (q, 0) & =  2 \Gamma_a \f{\gbar^2}{(2 \pi)^6} \int d^2 K d \omega
d^2 l d\Omega~  \chi_s (l,\Omega) \nonumber \\
& \times ~\f{G(\omega+\Omega,k+l) G(\omega,k+q)}{\kappa(l,\omega,\Omega)^2} .
\end{align}
Let's keep this expression as it is for the moment and move on to the
second diagram.

\subsubsection{Second diagram}

Following the same path, we can rewrite the product of fermionic Green's
functions as:
\begin{align}
 & G(\omega,k) G(\omega+\Omega,k+l) G(\omega+\Omega,k+l+q) G(\omega,k+q) = \nonumber \\
 & \qquad \f{G(\omega,k)G(\omega,k+q)- G(\omega,k) G(\omega+\Omega,k+l+q)}{\kappa(l,\omega,\Omega)^2} \nonumber \\
 & \qquad + \f{G(\omega+\Omega,k+l) G(\omega+\Omega,k+l+q)}{\kappa(l,\omega,\Omega)^2} \nonumber \\
 &  \qquad - \frac{G(\omega+\Omega,k+l)G(\omega,k+q)}{\kappa(l,\omega,\Omega)^2} ,
\end{align}
with the expression of $\kappa$ defined above.

Once again, the integration over $\varepsilon_k$ may give zero if the
poles are in the same half-plane, which reduces our previous expression
to:
\begin{align}
\Pi_{1b} (q, 0) & = - \Gamma_b \f{\gbar^2}{(2 \pi)^6} \int d^2 K d \omega
d^2 l d\Omega~  \chi_s (l,\Omega) \nonumber  \\
 & \times \left(\f{G(\omega+\Omega,k+l) G(\omega,k+q)}{\kappa(l,\omega,\Omega)^2} \right. \nonumber \\
& \qquad \left. + \f{G(\omega,k) G(\omega+\Omega,k+l+q)}{\kappa(l,\omega,\Omega)^2} \right) .
\end{align}
Changing $k$ into $k-q$ in the second part of the integral, we have:
\begin{align}
\Pi_{1b} (q, 0) & = - \Gamma_b \f{\gbar^2}{(2 \pi)^6} \int d^2 K d \omega
d^2 l d\Omega~   \chi_s (l,\Omega) \nonumber \\
 & \times \left( \f{G(\omega+\Omega,k+l) G(\omega,k+q)}{\kappa(l,\omega,\Omega)^2} \right. \nonumber \\
& \left. \qquad + \f{G(\omega,k-q) G(\omega+\Omega,k+l)}{\kappa(l,\omega,\Omega)^2}  \right)  .
\end{align}
One can then notice that $\varepsilon_{k-q} = \varepsilon_k - v_F q \cos
\theta$ changes to $\varepsilon_{k+q}$ if one changes $\theta$ into
$\theta-\pi$. This finally leads to:
\begin{align}
\Pi_{1b} (q, 0) &= - 2 \Gamma_b \f{\gbar^2}{(2 \pi)^6} \int d^2 K d \omega
d^2 l d\Omega~  \chi_s (l,\Omega) \nonumber \\
& \times \f{G(\omega+\Omega,k+l) G(\omega,k+q)}{\kappa(l,\omega,\Omega)^2}.
\end{align}

>From what precedes, we have:
\begin{equation}
\Pi_A (q, 0) = 2 \left(1 - \f{\Gamma_b}{\Gamma_a} \right) \Pi_{1a}(q, 0).
\label{eq:Pi1}
\end{equation}
Spin summation prefactors can be easily computed, and are given by:
\begin{displaymath}
\left\{ \begin{array}{lclcl}
\Gamma_a & = & \sum_{\alpha,\beta,\gamma,\delta} \sigma^Z_{\alpha \beta} {\bm \sigma}_{\beta \gamma} \cdot {\bm \sigma}_{\gamma \delta} \sigma^Z_{\delta \alpha} & = & 6 \\
\Gamma_b & = & \sum_{\alpha,\beta,\gamma,\delta} \sigma^Z_{\alpha \beta} {\bm \sigma}_{\beta \gamma} \sigma^Z_{\gamma \delta} {\bm \sigma}_{\delta \alpha} & = & -2
\end{array} \right.
\end{displaymath}

This finally leads to:
\begin{equation}
\Pi_A (q, 0) = \f{8}{3} \Pi_{1a}(q, 0).
\end{equation}

\subsection{Away from the QCP}

In the Fermi-liquid regime we have
\begin{eqnarray}
\Pi_A (q, 0) & = & \frac{16 N\gbar^2}{(2\pi)^6} \int d^2 K d \omega d^2 l d\Omega~
\f{1}{(i(1+\lambda)\omega-\epsilon_k)^2} \nonumber \\
& & \times \f{1}{i(1+\lambda)(\omega+\Omega)-\epsilon_{k+l}}~ \f{1}{i(1+\lambda)\omega-\epsilon_{k+q}} \nonumber \\
 & & \times \f{1}{\xi^{-2}+l^2+\gamma \frac{| \Omega |}{l}} ,
\end{eqnarray}
where we used for the fermionic self-energy $\Sigma(\omega)= \lambda \omega$, since we are deep in the Fermi liquid phase in this case.

Defining $\cos \theta =\frac{\bf k  \cdot l}{|{\bf k}| |{\bf l}|}$ and
$\cos  \theta^{\prime} =  \frac{\bf  k  \cdot q}{|{\bf  k}|  |{\bf q}|}$, and
integrating over  $k$ and $\omega$, one has:
\begin{align}
\Pi_A (q, 0)& = i \frac{16 Nm\gbar^2}{(2\pi)^5}  \int_0^{2 \pi}
d\theta \int_0^{2 \pi}
d\theta^{\prime} \int_{-\infty}^{+\infty} d\Omega
 \nonumber \\
&  \int_{0}^{\infty} \f{dl~l}{\xi^{-2}+l^2+\frac{\gamma|\Omega|}{l}} \f{\Omega}{\left(i(1+\lambda)\Omega-v_F l \cos \theta\right)^2} \nonumber \\
 &   \times ~\f{1}{i(1+\lambda)\Omega-v_F q \cos \theta^{\prime}-v_F l \cos\theta}.
\end{align}

The integral over $\theta^{\prime}$ then gives:
\begin{align}
\Pi_A (q, 0)& = - \frac{4 Nm\gbar^2}{\pi^4}  \int_0^{\pi}
d\theta  \int_{0}^{\infty} d\Omega \int_{0}^{\infty} dl
 \nonumber \\
&  \f{l}{\xi^{-2}+l^2+\frac{\gamma\Omega}{l}} \f{\Omega}{\left((1+\lambda)\Omega+i v_F l \cos \theta\right)^2} \nonumber \\
 &   \times ~\f{1}{\sqrt{(v_F q)^2+\left(\Omega(1+\lambda)+i v_F l \cos \theta \right)^2}} .
\end{align}

It is convenient to rescale the variables at this stage, introducing $\Omega^{\prime} = \frac{(1+\lambda)\Omega}{v_F q}$ and $l^\prime = \frac{l}{q}$, so that the previous expression reduces to:
\begin{align}
\Pi_A (q, 0)& = - \frac{4 Nm\gbar^2}{\pi^4 v_F (1+\lambda)^2} |q| \int_0^{\pi}
d\theta  \int_{0}^{+\infty} d\Omega^{\prime} \int_{0}^{\infty} dl^{\prime}
 \nonumber \\
&  \qquad \f{l}{\xi^{-2}+\frac{\gamma v_F}{1+\lambda}\frac{\Omega^{\prime}}{l^{\prime}}} \f{\Omega^{\prime}}{\left(\Omega^{\prime}+i l^{\prime} \cos \theta\right)^2} \nonumber \\
 &  \qquad \times ~\f{1}{\sqrt{1+\left(\Omega^{\prime}+i l^{\prime} \cos \theta \right)^2}} ,
\end{align}
where we kept only the leading order in $q$.

Defining $z$ and $\phi$ as $z \cos \phi = l^{\prime}$ and $z \sin \phi = \Omega^{\prime}$, one is left with:
\begin{align}
\Pi_A (q, 0)& = - \frac{4 Nm\gbar^2}{\pi^4 v_F (1+\lambda)^2} |q| \int_0^{\pi}
d\theta  \int_{0}^{\infty} dz \int_{0}^{\pi/2} d\phi
 \nonumber \\
&  \qquad \f{\cos \phi}{\xi^{-2}+\frac{\gamma v_F}{1+\lambda}\tan \phi}   \f{z \sin \phi}{\left(\sin \phi+i \cos \phi \cos \theta\right)^2} \nonumber \\
 &  \qquad \times ~\f{1}{\sqrt{1+z^2\left(\sin \phi+i \cos \phi \cos \theta \right)^2}} .
\end{align}

Subtracting the constant part $\Pi_A (0,0)$ (and neglecting it),
and integrating over $z$, this leads to:
\begin{align}
\Pi_A (q, 0)& =  \frac{4 Nm\gbar^2}{\pi^4 v_F (1+\lambda)^2} |q| \int_{0}^{\pi/2} \f{d\phi~\cos \phi \sin \phi}{\xi^{-2}+\frac{\gamma v_F}{1+\lambda}\tan \phi}
 \nonumber \\
&  \qquad \times \int_0^{\pi}  d\theta \f{1}{\left(\sin \phi+i \cos \phi \cos \theta\right)^4} .
\end{align}

The angular integration over $\theta$ can be done explicitly and gives:
\beq
\int_0^{\pi} \f{d\theta}{\left(\sin \phi+i \cos \phi \cos \theta\right)^4} = \frac{\pi}{2} \sin \phi (5 \sin^2 \phi -3) .
\eeq

Substituting this into the expression of $\delta\Pi_A$, we are left with the following final result:
\beq
 \Pi_A (q, 0) = -\frac{2 \gbar}{\pi^2 v_F (1+\lambda)}~|q|~ {\cal H} \left( \f{1+\lambda}{\gamma v_F \xi^2}  \right) ,
\label{apr26_2}
\eeq
where ${\cal H}$ is defined as:
\beq
{\cal H}(x) = \int_0^{\pi/2} d\phi \frac{\cos \phi \sin^2 \phi (3-5\sin^2\phi)}{\tan \phi + x} .
\label{apr26_3}
\eeq
In the two limits,
\beq
{\cal H}(0) = \frac{1}{3},  ~~ {\cal H}(x \gg 1) \approx \frac{2}{3x^2}
\label{apr26_4}
\eeq

As one approaches the QCP, $\xi$ gets bigger, and one can take the asymptotic form of ${\cal H} (x)$ for small $x$: ${\cal H}(0) =1/3$. Rearranging the prefactor for this limit, we are left with:
\beq
 \Pi_A (q, 0) \underset{\xi\to\infty}{=} - \frac{8}{9\pi} \xi^{-1} ~ |q| .
\eeq

\subsection{At criticality}

At the QCP, we have:
\begin{eqnarray}
\Pi_A (q, 0) & = & \f{16N \gbar^2}{(2 \pi)^6} \int d^2 K d
\omega d^2 l d\Omega~ \f{1}{(i\Sigma(\omega)-\epsilon_k)^2} \nonumber \\
& & \times~\f{1}{i\Sigma(\omega+\Omega)-\epsilon_{k+l}}~\f{1}{i\Sigma(\omega)-\epsilon_{k+q}} \nonumber \\
& & \times~\f{1}{l^2+\gamma\frac{| \Omega |}{l}},
\end{eqnarray}
where we considered that, close to criticality, the self-energy dominates completely the bare $\omega$ term in the fermionic propagators.

We expand both energies as
$\epsilon_{k+l}=\epsilon_k+v_F l_x+\frac{l_y^2}{2m_B}$ and $\epsilon_{k+q}=\epsilon_k + v_F q
\cos \theta$, and perform the integration over $\epsilon_k$, leading to:
\begin{align}
& \Pi_A (q, 0)  =  i \f{16N m \gbar^2}{(2\pi)^5} \int_0^{2 \pi}
d\theta \int_{-\infty}^{+\infty} d l_x \int_{-\infty}^{+\infty} d\Omega
\int_0^{\Omega} d\omega \nonumber \\
&  \int_{-\infty}^{\infty} \f{dl_y~ l}{\gamma|\Omega|+l^3} \f{1}{\left(i\Sigma(\Omega-\omega)+i\Sigma(\omega)-v_F l_x-\frac{l_y^2}{2m_B}\right)^2} \nonumber \\
 &   \times ~\f{1}{i\Sigma(\Omega-\omega)+i\Sigma(\omega)+v_F q \cos \theta-v_F l_x-\frac{l_y^2}{2m_B}}.
\end{align}

The integration over $l_x$ brings two contributions. One comes from the poles in the fermionic propagator, and can be neglected here since both poles are in the same half-plane. The other contribution comes from the branch cut in the bosonic propagator, and since at the branch cut $q_x \sim q_y$, one can safely drop the quadratic term in the fermionic propagators. This allows us to integrate over $l_y$. Out of the terms arising from this integral, the only non-vanishing ones come from the non-analyticities of the integrated bosonic propagator defined as:
\beq
\int dl_y \chi ({\bf l},\Omega) = \int dl_y \f{1}{l^2+\frac{\gamma v_F |\Omega|}{\sqrt{(v_F l)^2+c^2 \Sigma(\Omega)^2}}} ,
\eeq
We use here the full form of the polarization operator, Eq.
(\ref{specchi}) as we will see that typical  $v_Fl_x \geq \Sigma(\Omega)$,
 and typical $l_y$ are only larger in logarithmic sense.

This integral was performed in a slightly different form in (\ref{intchi}), but the method is the same: introducing $u$ such that $\tan u = \frac{v_F l_y}{\sqrt{(v_F l_x)^2+\Sigma(\Omega)^2}}$, one has:
\bea
\int dl_y \chi ({\bf l},\Omega) & = & \frac{(v_F l_x)^2+c^2\Sigma(\Omega)^2}{\gamma v_F^2 |\Omega|} \nonumber \\
& &  \times \int_0^{\pi/2} \frac{du}{\cos^3 u - \delta~\cos^2 u + \epsilon} ,  \qquad \label{intchifull}
\eea
where we introduced $\delta=\frac{c^2 \Sigma(\Omega)^2((v_F l_x)^2+c^2\Sigma(\Omega)^2)^{1/2}}{\gamma v_F^3 |\Omega|}$ and $\epsilon= \frac{((v_F l_x)^2+c^2\Sigma(\Omega)^2)^{3/2}}{\gamma v_F^3 |\Omega|}$.

In the process of integrating over $l_x$, two non-analytic contributions arise from (\ref{intchifull}). One comes from $l_x \gtrsim (\gamma \Omega)^{1/3}$ and goes like $\frac{\pi}{(\gamma |\Omega|)^{1/3}}$, Plugging this back into
 $\Pi_A$ and subtracting (and neglecting) a constant term, we obtain
\begin{align}
& \Pi_A^{(1)} (q,0) N\sim m\gbar^2 v_F^2 q^2 \int_{-\infty}^{+\infty} d l_x \int_{-\infty}^{+\infty} d\Omega \int_0^{\Omega} d\omega \nonumber \\
&  \qquad \times \f{1}{|l_x|} \f{1}{\left(i\Sigma(\Omega-\omega)+i\Sigma(\omega)-v_F l_x\right)^5} ,
\end{align}
where we subtracted $\Pi_A(0,0)$ (hence the notation $\delta \Pi_A$) and expanded in $q$.

We further simplify the integrals, noticing that the fermionic propagator is dominated by $v_F l_x$ since $l_x\sim (\gamma \Omega^{1/3})$:
\bea
\Pi_A^{(1)} (q,0) & \sim & q^2 \frac{Nm\gbar^2}{v_F^3} \int_0^{\omega_{\rm max}} d\Omega \int d l_x \frac{\Omega}{l_x^6} \nonumber \\
& \sim & q^2 \frac{Nm\gbar^2}{v_F^3 \gamma^{5/3}} \int_0^{\omega_{\rm max}} \frac{d\Omega}{\Omega^{2/3}} \nonumber \\
& \sim & \sqrt{\alpha} q^2 ,
\eea
where we substituted $l_x\sim (\gamma \Omega^{1/3})$ in the last steps.

The other non-analytic contribution from (\ref{intchifull}) comes from typical $v_F l_x \sim \Sigma(\Omega)$. It can be seen from an expansion of (\ref{intchifull}) for small values of both $\delta$ and $\epsilon$, and goes like:
\beq
-\frac{(v_F l_x)^2+c^2\Sigma(\Omega)^2}{2 v_F^2 \gamma |\Omega|} \log \left((v_F l_x)^2+c^2\Sigma(\Omega)^2\right) .
\label{apr25_8})
\eeq
One can explicitly verify  that to get the logarithm, we only need the polarization operator $\Pi (l, \Omega)$ to order $1/l^3$. Like we said in Appendix C, to this order, polarization bubble can be evaluated with the full fermionic Green's functions but without vertex corrections.

Substituting (\ref{apr25_8}) into the expression for $\Pi_A$ and subtracting a constant part,  we obtain:
\begin{align}
\Pi_A^{(2)} (q, 0)&  =  i \f{8 N m \gbar^2}{(2\pi)^5 \gamma v_F} q \int_0^{2 \pi}
d\theta \int_{-\infty}^{+\infty} d l_x \int_{-\infty}^{+\infty} \frac{d\Omega}{|\Omega|} \nonumber \\
&  \int_0^{\Omega} d\omega \f{\cos\theta \left[(v_F l_x)^2+c^2\Sigma(\Omega)^2\right]}{\left(i\Sigma(\Omega-\omega)+i\Sigma(\omega)-v_F l_x\right)^3} \nonumber \\
 &   \times ~\f{\log \left((v_F l_x)^2+c^2\Sigma(\Omega)^2\right)}{i\Sigma(\Omega-\omega)+i\Sigma(\omega)+v_F q \cos \theta-v_F l_x}.
\end{align}

Using the $\theta \longleftrightarrow -\theta$ symmetry, and splitting the integral over $\Omega$ into two parts, one can rearrange this expression as:
\begin{align}
& \Pi_A^{(2)} (q, 0)  =  i \f{N m \gbar^2}{c \pi^5 \gamma v_F^2} q \int_0^{\pi}
d\theta \int_{-\infty}^{+\infty} d z \int_{0}^{+\infty} \frac{d\Omega}{\Omega} \nonumber \\
& \qquad \int_0^{\Omega} d\omega \f{1}{\Sigma(\Omega)} \f{\cos\theta (1+z^2)}{\left(i\frac{\Sigma(\Omega-\omega)+\Sigma(\omega)}{c\Sigma(\Omega)}-z\right)^3} \nonumber \\
 & \qquad  \times ~\f{\log \left(1+z^2\right)}{i\frac{\Sigma(\Omega-\omega)+\Sigma(\omega)}{c\Sigma(\Omega)}+\frac{v_F q \cos \theta}{c\Sigma(\Omega)}-z}, \label{piA1}
\end{align}
where we defined $z=v_F l_x/(c \Sigma(\Omega))$.

Let's now isolate the integral over $z$, given by:
\begin{equation}
J = \int_{-\infty}^{+\infty} dz \f{(1+z^2) \log (1+z^2)}{(ia-z)^3 (ia+b-z)},
\end{equation}
where $a=\frac{\Sigma(\Omega-\omega)+\Sigma(\omega)}{c\Sigma(\Omega)}$ and $b=\frac{v_F q \cos \theta}{c\Sigma(\Omega)}$, and $a \ge 0$.

Performing the contour integration in the lower half-plane (where lies the branch-cut), one gets:
\bea
J & = &-2\pi \int_1^{\infty} dy \f{1-y^2}{(y+a)^3 (y+a-ib)} \nonumber \\
& = & -2\pi \int_1^{\infty} dy \f{(1-y^2)(y+a+ib)}{(y+a)^3 \left((y+a)^2+b^2\right)} . \label{Jfin}
\eea

Once (\ref{Jfin}) is plugged back into (\ref{piA1}), only the imaginary term survives due to the symmetry of the integral in $\theta$. We are left with:
\begin{align}
& \Pi_A^{(2)} (q, 0)  =  \f{4 N m \gbar^2}{c^2 \pi^4 \gamma v_F} q^2 \int_0^{\pi/2}
d\theta \int_{1}^{+\infty} d y \int_{0}^{+\infty} d\Omega \nonumber \\
& \qquad \int_0^{1} dw \f{1}{\Sigma(\Omega)^2} \f{\cos^2\theta}{\left[c^{-1}\left((1-w)^{2/3}+w^{2/3}\right)+y\right]^3} \nonumber \\
 & \qquad  \times ~\f{1-y^2}{\left[y+c^{-1}\left((1-w)^{2/3}+w^{2/3}\right)\right]^2+\left(\frac{v_F q \cos \theta}{c\Sigma(\Omega)}\right)^2}, \label{piA2}
\end{align}
where we changed variables, defining $w=\omega/\Omega$.

Introducing the new variable $t=\left(\frac{c\Sigma(\Omega)}{v_F q \cos \theta}\right)^{3/2}$, this rewrites:
\begin{align}
& \Pi_A^{(2)} (q, 0)  =  \f{4 N m \gbar^2}{c^{3/2} \pi^4 \gamma v_F^{3/2}\omega_0^{1/2}} q^{3/2} \int_0^{\pi/2} d\theta~ (\cos \theta)^{3/2}  \nonumber \\
& \quad \times \int_{1}^{+\infty} d y \int_0^{1} dw \f{1}{\left[c^{-1}\left((1-w)^{2/3}+w^{2/3}\right)+y\right]^3} \nonumber \\
 & \quad  \times ~\int_{0}^{+\infty} dt \f{1-y^2}{1+ t^{4/3}\left[y+c^{-1}\left((1-w)^{2/3}+w^{2/3}\right)\right]^2}. \label{piA3}
\end{align}

A final change in variables leads to:
\begin{align}
& \Pi_A^{(2)} (q, 0)  =  \f{4 N m \gbar^2}{c^{3/2} \pi^4 \gamma v_F^{3/2}\omega_0^{1/2}} q^{3/2} \int_0^{\pi/2} d\theta~ (\cos \theta)^{3/2}  \nonumber \\
& \quad \times \int_{1}^{+\infty} d y \int_0^{1} dw \f{1-y^2}{\left[c^{-1}\left((1-w)^{2/3}+w^{2/3}\right)+y\right]^{9/2}} \nonumber \\
 & \quad  \times ~\int_{0}^{+\infty} dv~ \f{1}{1+ v^{4/3}}, \label{piA4}
\end{align}
where $v=t\left[y+c^{-1}\left((1-w)^{2/3}+w^{2/3}\right)\right]^{3/2}$

Performing the integral over $y$, one is left with three independent integrals contributing to the numerical prefactor:
\begin{align}
& \Pi_A^{(2)} (q, 0)  =  -\f{32 N m \gbar^2}{105 \pi^4 \gamma v_F^{3/2}\omega_0^{1/2}} q^{3/2} \int_0^{\pi/2} d\theta~ (\cos \theta)^{3/2}  \nonumber \\
& \quad \times \int_0^{1} dw ~\f{5c+2\left((1-w)^{2/3}+w^{2/3}\right)}{\left(c+(1-w)^{2/3}+w^{2/3}\right)^{5/2}} \nonumber \\
 & \quad  \times ~\int_{0}^{+\infty} dv~ \f{1}{1+ v^{4/3}} . \label{piA5}
\end{align}

These integrals can be performed separately and read:
\bea
\int_0^{\pi/2} d\theta~ (\cos \theta)^{3/2} &  = & \f{\sqrt{2} \pi^{3/2}}{6 \left[\Gamma \left(\frac{3}{4}\right)\right]^2}  \simeq  0.8740 \nonumber \\
\int_0^{+\infty} \frac{dv}{1+v^{4/3}}& = &\frac{3 \pi \sqrt{2}}{4}  \simeq  3.3322 \nonumber \\
\int_0^{1} dw ~\f{5c+2s(w)}{\left(c+s(w)\right)^{5/2}} & \simeq & 0.9438
\eea
and we used the notation $s(w)=(1-w)^{2/3}+w^{2/3}$.

Collecting all integrals, and rearranging the prefactor, the final result for the contribution of the first two diagrams then writes:
\beq
\Pi_A^{(2)} (q, 0) = -0.1053 ~\sqrt{k_F}~q^{3/2} .
\label{apr26_9}
\eeq

\subsection{Other two diagrams}

The computation of the other two diagrams in Fig. \ref{2loop}
 proceeds along the same way.
Far away from criticality, when $\gamma v_F \xi^{2}/(1 + \lambda)$ is small,
 and one can just expand perturbatively in the interaction,
 the sum of these two ``drag'' diagrams, which we label here and in the main text as $\Pi_B$, was shown in ~\cite{ch_masl_latest} to
 be equal to $\Pi_A (q,0)$ to the leading order in ${\bar g}$ (which in our model is ${\bar g}^3$, see (\ref{apr26_2} -\ref{apr26_4}).  Near criticality such simple relation no longer holds, but $\Pi_A (q, 0)$ and $\Pi_B (q,0)$ remain of the same sign and of comparable magnitude.

At criticality, we obtained for $\Pi_B (q, 0)$
\beq
\Pi_B (q,0) = \frac{\sqrt{2}}{4 3^{3/4} \pi^4}~ q^{3/2} \sqrt{k_F} I
\label{apr26_5}
\eeq
where in rescaled variables (e.g., momentum is in units of $q$)
\beq
I = \int d \Omega d x x \int_0^{2\pi} d \theta \frac{S^2 (x, \Omega,
\theta)}{S_1 (x, \Omega) S_1 (x + \cos \theta, \Omega)}
\label{apr26_6}
\eeq
and $S(x, \Omega, \theta)$ and $S_1 (x, \Omega)$ are given by
\bea
S(x, \Omega, \theta) & = & \int_0^\Omega d \omega \int_0^{2\pi} \frac{d \theta_1}{i \Sigma^* - x \cos (\theta_1)} \qquad \qquad \nonumber \\
& & \quad \times \frac{1}{i \Sigma^* - \cos (\theta + \theta_1) - x \cos \theta_1}
\label{apr26_7} \\
S_1(x, \Omega) & = & \int_0^\Omega \frac{d \omega}{x^2 + (\Sigma^*)^2}
\label{apr26_8}
\eea
where we introduced $\Sigma^* = \Sigma^* (\omega, \Omega) = (\Omega - \omega)^{2/3} + \omega^{2/3}$.

We could not evaluate this integral explicitly, and we compute it
 under two simplifying assumptions
\begin{itemize}
\item
we compute $S_1 (x, \Omega)$ by expanding to leading order in $(\Sigma^*/x)^2$, evaluating the frequency integral and plugging the result back into denominator. This way, we approximated $S_1 (x, \Omega)$ by
\beq
S_1(x, \Omega) \approx \frac{\Omega}{x^2 + (c \Omega^{2/3})^2}
\label{apr26_91}
\eeq
where $c \approx 1.2$  (see (\ref{apr25_7})
This procedure is similar to the one which led to (\ref{specchi}), but here
 we cannot  justify that only $1.x$ and $1/x^3$ terms are relevant.

\item
We replace $\Sigma^*$ by the same $c \Omega^{2/3}$ in the integrand for $S(x, \Omega, \theta)$

\item  We assume that internal momenta are larger than the external one, i.e
 $x \gg 1$ ($x$ is measured in units of $q$), neglected terms $O91)$ compared to $O(x)$ and set the lower limit of the integration over $x$ at some number $b$.

\item We choose $b$ by applying the same approximate computation scheme to $\Pi_A (q, \omega)$ and requesting that the result coincide with the exact expression, Eq. (\ref{apr26_9}).
\end{itemize}

Carrying out this calculation for $\Pi_B (q,0)$ we obtain
\beq
\Pi_A^{(2)} (q, 0) \approx -0.14 ~\sqrt{k_F}~q^{3/2} .
\label{apr26_10}
\eeq
 This is the result that we cited in the text.

\section{Two-loop renormalization of the charge susceptibility}

\label{app:charge}

In this Appendix we show that the singular contributions to the static charge susceptibility from individual diagrams cancel out in the full expression of $\chi_c (q)$. The cancellation of the singularities in the charge responce has been  extensively studied in 1D systems \cite{metzner_charge}.

\begin{widetext}
\begin{center}
\begin{figure}[hbp]
\includegraphics[angle=90,width=7in]{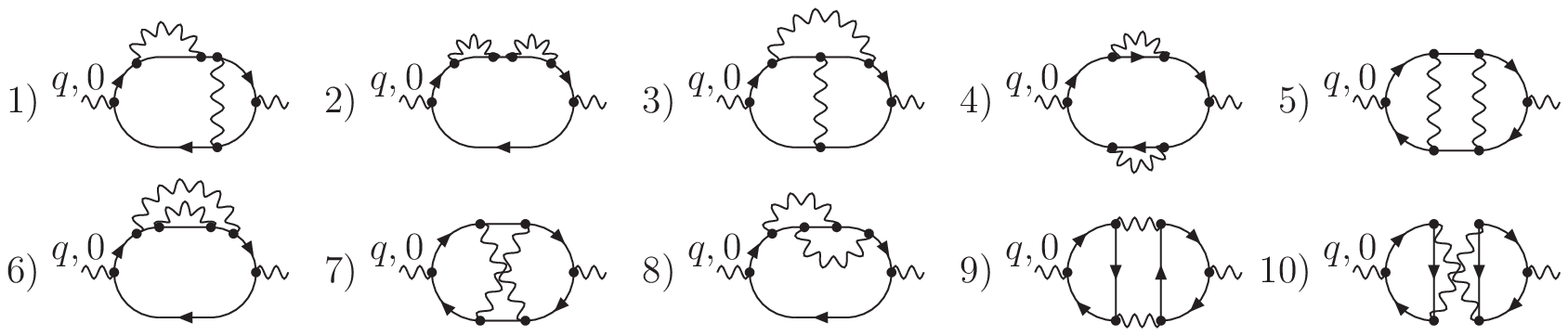}
\caption{The 10 two-loop diagrams for the charge susceptibility.}
\label{charge2loop}
\end{figure}
\end{center}
\end{widetext}

There are 10 different two-loop diagrams for the charge susceptibility, presented in Fig. \ref{charge2loop}.  The last two diagrams are identical to the ones we considered in the main text. We already argued there that these two diagrams cancel out in the case of a QCP in the charge channel.

The other eight diagrams have to be considered together. We demonstrate that the total contribution from these eight diagrams vanishes once one linearizes the dispersion of the intermediate fermions. This still leaves out the contributions from non-linear terms in the dispersion, but one can show that these contributions are regular.

To begin, consider one of these diagrams, e.g., diagram 7 in Fig. \ref{charge2loop}. In analytic form, the contribution from this diagram is:
\bea
\Pi_7 (q) &=& 2 \int d^2 q_1 d\omega_1 d^2 q_2 d\omega_2~ G_k G_{k+q} G_{k+q_1} G_{k+q_2 +q} \nonumber \\
 && \qquad \times  G_{k+ q_1 +q_2} G_{k+ q_1 + q_2 +q} \chi_{q_1} \chi_{q_2} ,
\label{m_1}
\eea
where we labeled $q_i = ({\bf q}_i, \omega_i)$, and the combinatoric factor 2 comes from the summation over spin indices.

Introduce now
\beq
G_k G_{k+q_i} = \frac{1}{\alpha_{q_i}} ~\left(G_k - G_{k+q_i}\right) ,
\label{m_2}
\eeq
where
\beq
\alpha_{q_i} = i \omega_i - q_i \cos \theta_i ,
\label{m_3}
\eeq
and $\theta_i$ is the angle between ${\bf k} \approx {\bf k}_F$ and ${\bf q}_i$. Shortening the notations further as $q_1 \equiv 1$ and $q_2 \equiv 2$, using the symbolic notation $\int_{1,2}$ for the 6-dimensional integral over $q_1$ and $q_2$, and  applying (\ref{m_2}), we obtain:
\bea
 \Pi_7 (q) & = & 2 \int_{1,2} \chi_1 \chi_2 \left[\frac{G_{k+q} G_{k +1+2}}{\alpha^2_1 \alpha^2_2}
-\frac{G_{k+q} G_{k +1+2}}{\alpha_1 \alpha_2 (\alpha_1 + \alpha_2)^2} \right. \nonumber \\
& & \left. \qquad \qquad \quad -2 \frac{G_{k+q} G_{k +1}}{\alpha^2_2 (\alpha^2_1 - \alpha^2_2)}\right]
\label{m_4}
\eea
Similarly,
\beq
 \Pi_8 (q) = 2 \int_{1,2} \chi_1 \chi_2 \left[\frac{G_{k+q} G_{k +1+2}}{\alpha_1 \alpha_2 (\alpha_1 + \alpha_2)^2} -
2 \frac{G_{k+q} G_{k +1}}{\alpha^2_1 (\alpha^2_1 - \alpha^2_2)}\right]
\label{m_5}
\eeq
and further
\beq
 \Pi_5 (q) = 2 \int_{1,2} \chi_1 \chi_2 \left[\frac{G_{k+q} G_{k +1+2}}{\alpha_1 \alpha_2 (\alpha_1 + \alpha_2)^2} +
2 \frac{G_{k+q} G_{k +1}}{\alpha^2_1 (\alpha^2_1 - \alpha^2_2)}\right]
\label{m_51}
\eeq
\bea
 \Pi_3 (q) &=& 2 \int_{1,2} \chi_1 \chi_2 \left[-\frac{G_{k+q} G_{k +1+2}}{\alpha^2_1
\alpha^2_2} \right. \nonumber \\
& & \qquad \qquad  \left. + 2 \frac{G_{k+q} G_{k +1}}{\alpha^2_1 \alpha^2_2} \left(\frac{\alpha^2_1 + \alpha^2_2}{\alpha^2_1 - \alpha^2_2}\right) \right]
\label{m_6}
\eea
\beq
 \Pi_2 (q) = 4 \int_{1,2} \chi_1 \chi_2 \frac{G_{k+q} G_{k +1}}{\alpha^2_1 \alpha^2_2}
 \left(\frac{\alpha^2_2}{\alpha^2_1 - \alpha^2_2}\right)
\label{m_7}
\eeq
\bea
 \Pi_1 (q) & = & 4 \int_{1,2} \frac{\chi_1 \chi_2}{\alpha^2_1 \alpha^2_2}~
 \left[G_{k+q} G_{k +1+2} \frac{\alpha^2_2}{ \alpha^2_1 -\alpha^2_2} \right. \nonumber \\
& & \qquad \qquad \quad \left. -  G_{k+q} G_{k +1} \frac{3\alpha^2_2 - \alpha^2_1}{\alpha^2_1 - \alpha^2_2}\right]   \label{m_8}
\eea
\beq
 \Pi_4 (q)  = \int_{1,2} \frac{\chi_1 \chi_2}{\alpha^2_1 \alpha^2_2} \left[G_{k+q} G_{k +1+2}  -2 G_{k+q} G_{k +1} \right]
\label{m_9}
\eeq
and finally,
\bea
 \Pi_6 (q) &=& \int_{1,2} \frac{\chi_1 \chi_2}{\alpha^2_1 \alpha^2_2}~
 \left[G_{k+q} G_{k +1+2} \frac{\alpha^2_1 + \alpha^2_2}{( \alpha_1 +\alpha_2)^2} \right. \nonumber \\
& & \qquad \qquad \qquad  \left. - 2 G_{k+q} G_{k +1}\right]
  \label{m_10}
\eea
Collecting the prefactors for $G_{k+q} G_{k +1+2}$ and $ G_{k+q} G_{k +1}$ from all of the eight contributions we find that they cancel out.

\section{Mass-shell singularity}
\label{app:mass_shell}

In this Appendix, we analyze in more detail the form of the self-energy near the fermionic mass shell.  The interest to the mass-shell behavior of the self-energy was triggered by recent studies of the self-energy near a mass shell in a 2D Fermi liquid~\cite{cmgg} and for 2D Dirac fermions~\cite{ch_tsvelik}. In both cases, the lowest-order self-energy diverges at the mass-shell, which forces to re-sum the perturbative series.

At first glance, the same situation holds in our analysis at the QCP. Evaluating the self-energy in a two-loop expansion {\it around free fermions} and using the fermionic dispersion with the curvature, we obtain near the mass shell~\cite{mitya_unp}
\beq
\Sigma (k, \omega) \sim \frac{1}{N^2} (i \omega - \epsilon_k)~\left[N \log \frac{i \omega - \epsilon_k}{\epsilon_k}\right]^2 .
\label{t_1}
\eeq
This result implies that the ``effective'' 
quasiparticle residue for the Eliashberg theory, 
$Z_{eff} \propto d \Sigma/d \epsilon_k$ logarithmically diverges on the mass shell of free fermions. Without the curvature of the dispersion, the divergence would be stronger than logarithm.

The issue we now have to address is whether $Z$ still diverges on the
 mass shell if we expand around the Eliashberg solution, i.e., around fermions with
\beq
G_0 (k, \omega) = \frac{1}{i {\tilde \Sigma} (\omega) - \epsilon_k} ,
\label{t_2}
\eeq
where, we remind, ${\tilde \Sigma} (\omega) = \omega +  \Sigma (\omega)$.

It turns out that this is not the case: the expansion around the Eliashberg solution leads to a finite residue  $Z$.
At the two-loop order, we obtain, instead of (\ref{t_1})
\bea
&& \Sigma (k, \omega) \sim \frac{1}{N^2} \int_0^1 dz \int_{1-z}^1 dz^\prime
\left( i \Sigma (\omega) \psi_{z,z^\prime}- \epsilon_k\right)~ \nonumber \\
&& \times \left[\log \frac{N (i \Sigma (\omega) \psi_{z,z^\prime}- \epsilon_k)}{\epsilon_k}\right]^2
\label{t_3}
\eea
where
\beq
\psi_{z,z^\prime} = (1-z)^{2/3} + (1-z^\prime)^{2/3} + (z+z^\prime -1)^{2/3} .
\label{t_4}
\eeq
For simplicity, we restricted ourselves to the quantum critical regime where ${\tilde \Sigma} (\omega) \approx \Sigma (\omega)$.  If  $\psi_{z,z^\prime}$ were equal to a constant, as it is when the system is in the Fermi liquid regime, and $\Sigma (\omega) = \lambda \omega$, $Z$ would diverge at $\omega = \epsilon/(1+ \lambda)$. However, since $\Sigma (\Omega - \omega) + \Sigma (\omega)$ does not reduce to $\Sigma (\Omega)$, we have two additional integrations over $z$ and $z^\prime$, and the logarithmic singularity is washed out. In particular, at $\epsilon_k = i \Sigma (\omega)$, i.e. at the ``Matsubara mass shell'', we have
\beq
Z_{eff} \sim \frac{1}{N^2} \left[\frac{\pi^2}{6} \log^2 N - 4.08 \log N + 2.88\right] ,
\label{t_5}
\eeq
in which case $Z_{eff}$ is just a constant. Combining this with our earlier result that the renormalization of $\epsilon_k$ is also finite, Eq. 
(\ref{sigma2k2_2}), we obtain  for the full fermionic Green's function at the smallest 
 $\omega$ and $\epsilon_k$ 
\beq
G (k, \omega) = \frac{Z_{eff}}{i {\tilde \Sigma} (\omega) - \epsilon^*_k},
\label{t_6}
\eeq
where $\epsilon^*_k$ differs from $\epsilon_k$ by a constant factor.

\end{document}